\documentclass[a4paper,11pt]{article}

\usepackage{jheppub}
\usepackage[usenames,dvipsnames,svgnames,table]{xcolor}
\usepackage{placeins}
\usepackage{booktabs}
\usepackage{ragged2e}
\usepackage{etoolbox}
\usepackage{multirow}
\apptocmd{\thebibliography}{\justifying}{}{}

\makeatletter\g@addto@macro\bfseries{\boldmath}\makeatother

\definecolor{myblue}{rgb}{0.152941176,0.549019608,0.670588235}

\hypersetup{
    unicode=false,          
    pdftoolbar=true,        
    pdfmenubar=true,        
    pdffitwindow=false,     
    pdfstartview={FitH},    
    pdftitle={Resurrecting bbh with kinematic shapes},
    pdfauthor={Christophe Grojean, Ayan Paul and Zhuoni Qian},
    pdfkeywords={Future Colliders} {bbh} {Machine Learning} {Shapley Values},
    pdfnewwindow=true,
    colorlinks=true,
    linkcolor=myblue,
    citecolor=myblue,
    filecolor=myblue,
    urlcolor=myblue,
    linktocpage=true
}

\def\equationautorefname~#1\null{Eq.\,(#1)\null}
\newcommand{\appendixref}[1]{\hyperref[#1]{appendix~\ref{#1}}}

\newcommand*{\gev}{\text{GeV}}
\newcommand*{\tev}{\text{TeV}}
\newcommand*{\fb}{\text{fb}}
\newcommand*{\iab}{\ensuremath{\text{ab}^{-1}}}

\newcommand{\inab}{\,{\rm ab}^{-1}}

\newcommand{\ee}{e^+e^-}
\newcommand{\bbh}{b\bar bh}
\newcommand{\tth}{t\bar th}
\newcommand{\ggh}{gg\to h}
\newcommand{\bbaa}{b\bar b\gamma\gamma}
\newcommand{\tkab}{\tilde{\kappa}_b}
\newcommand{\kab}{\kappa_b}

\newcommand{\rb}[1]{\rotatebox{90}{#1}}

\newcommand*{\figuretitle}[1]{%
    {\centering
    \textbf{#1}
    \par\medskip}
    \vspace{-0.1in}
}

\title{Resurrecting $b\bar{b}h$ with kinematic shapes}

\author[a,b]{Christophe Grojean,}
\author[a,b]{Ayan Paul,}
\author[a,c]{and Zhuoni Qian}

\affiliation[a]{DESY, Notkestra{\ss}e 85, D-22607 Hamburg, Germany}
\affiliation[b]{Institut f\"ur Physik, Humboldt-Universit\"at zu Berlin, D-12489 Berlin, Germany}
\affiliation[c]{Department of Physics, Shandong University, Jinan, Shandong 250100, China}

\emailAdd{christophe.grojean@desy.de}
\emailAdd{ayan.paul@desy.de}
\emailAdd{zhuoni.qian@desy.de}

\abstract{
The associated production of a $b\bar{b}$ pair with a Higgs boson could provide an important probe to both the size and the phase of the bottom-quark Yukawa coupling, $y_b$. However, the signal is shrouded by several background processes including the irreducible $Zh, Z\to b\bar{b}$ background. We show that the analysis of kinematic shapes provides us with a concrete prescription for separating the $y_b$-sensitive production modes from both the irreducible and the QCD-QED backgrounds using the $b\bar{b}\gamma\gamma$ final state. We draw a page from game theory and use Shapley values to make Boosted Decision Trees interpretable in terms of kinematic measurables and provide physics insights into the variances in the kinematic shapes of the different channels that help us complete this feat. Adding interpretability to the machine learning algorithm opens up the black-box and allows us to cherry-pick only those kinematic variables that matter most in the analysis. We resurrect the hope of constraining the size and, possibly, the phase of $y_b$ using kinematic shape studies of $b\bar{b}h$ production with the full HL-LHC data and at FCC-hh.
}
\begin{document} 

\begin{flushright}
DESY 20-175\\
HU-EP-20/28
\end{flushright}

\maketitle

\section{Introduction}
\label{sec:Intro}

After the discovery of the Higgs at the LHC~\cite{Chatrchyan:2012ufa, Aad:2012tfa}, the measurement of its couplings to the other Standard Model (SM) particles through its production and decay channels has been an ongoing quest~\cite{ATLAS:2020qdt}. The subsequent measurement of the $\tth$ signal marks the direct probe of the coupling of the Higgs to the third-generation fermions~\cite{Sirunyan:2018hoz, Aaboud:2018urx}. Despite the large branching ratio of the Higgs decaying to the bottom pair, the $ h\to b\bar b $ eluded measurement until much later through the $Vh, h\to b\bar b$ channel~\cite{Butterworth:2008iy,Aaboud:2018zhk, Sirunyan:2018kst}, bringing about sensitivity to the bottom-quark Yukawa, $y_b$, as well. The current sensitivity on $\kappa_b \sim y_b/y_b^\textrm{SM}$ is of the order of 7\% and it is expected to improve to 2.2\% at HL-LHC~\cite{Cepeda:2019klc}. Other direct probes to the Higgs couplings to the bottom quark can be envisaged through the measurements of the associated production of the Higgs with the bottom quark in the associated $bh$ or $\bbh$ production channels. 

The $\bbh$ signal is sizable at the LHC with a cross-section of about 0.5 pb. Calculations of the $\bbh$ cross-section focusing on the contribution proportional to $y_b$ has been a developing effort, with improvements from including higher order contribution, matching and resumming bottom-mass effects, evaluation of four and five-flavor matching scheme and optimal scale choices for Monte Carlo implementation with parton showers~\cite{Harlander:2003ai,Dittmaier:2003ej,Dawson:2003kb,Campbell:2004pu,Dawson:2005vi,Wiesemann:2014ioa,Forte:2015hba,Bonvini:2015pxa,Jager:2015hka,Bonvini:2016fgf,Deutschmann:2018avk,Pagani:2020rsg}. Despite the large inclusive rate, $\bbh$ production remains a challenging channel for experimental measurements. Given that the bulk of the signal contains only soft $p_T$ $b$-jets, which evade selection cuts, the signal drops by orders of magnitude after two, or even one, $b$-jets are explicitly tagged. It is even more challenging to achieve sensitivity on the $y_b$ coupling among all the amplitudes contributing to $\bbh$ production. Even with the large amount of luminosity that will be available at HL-LHC, the $y_b$ mediated diagrams contribute only a small fraction of the total cross section to the inclusive rate~\cite{Deutschmann:2018avk,Pagani:2020rsg}. 

In this work we shall appeal to kinematic shapes, machine learning and game theory to glean information on $y_b$ from the associated production of the Higgs with a bottom-quark pair. In doing so we would like to clarify the following open questions:
\begin{itemize}
    \item The irreducible background coming from $Zh$ production obfuscates the measurement of $y_b$~\cite{Pagani:2020rsg} in the $\bbh$ channel. Can multivariate analysis using kinematic shapes overcome this hurdle?
    \item Machine learning tools are infamous for providing little insight into the underlying physics. Can using high-level kinematic variables (as opposed to momenta four-vectors) be used in conjunction with interpretable machine learning tools to understand what variables are important in orchestrating a signal to background separation at HL-LHC and pave a path to measuring $y_b$ from $\bbh$ production?
    \item What are the prospects of making a far better measurement of $y_b$ from the $\bbh$ channel at FCC-hh?
    \item What can be said about the constraints on the rescaling of $y_b$, $\kappa_b$, and its phase attributed to new physics (NP), from $\bbh$ production?
\end{itemize}

Armed with these questions we proceed as follows. In \autoref{sec:bbh-col} we briefly explore the possibility of measuring $y_b$ at the future lepton colliders through $\bbh$ production before moving on to describing in detail the simulation of events for HL-LHC and FCC-hh. In \autoref{sec:kinematics} we delve into describing the machine learning algorithms that we use to explore higher dimensional kinematic shapes. We also provide an introduction to Shapley value~\cite{shapley1951notes} and measures derived from it to make machine learning interpretable. As an example of the effectiveness of Boosted Decision Trees (BDT) augmented with Shapley values, we show that the $Zh$ background can be discriminated from the signal. In \autoref{sec:hadronC} we further sharpen our analysis for HL-LHC including all irreducible backgrounds and the reducible QCD-QED $\bbaa$ dominant background. We then perform a similar analysis for FCC-hh. We round off the section with a discussion of the manner in which including systematics changes our conclusions. In \autoref{sec:YukConst} we extend our analysis to estimate bounds on $\kappa_b$ and its CP phase. The adoption of the simple $\kappa_b$ framework in presenting our results serves as a common and convenient tool in comparison with existing studies from other $y_b$ sensitive channels. Ultimately, our analysis should be included in a truly global EFT analysis to fully capture the complex structures that new physics might bring about and that could not be captured by a single parameter. We touch base with constraints from the EDM measurements to complete our analysis. Various necessary mathematical relations and the discrimination between the $y_b$- and $y_t$-driven channels are presented in the appendices. All codes necessary for reproducing our results including the simulated events are made available in a Github repository: \href{https://github.com/talismanbrandi/Interpretable-ML-bbh.git}{https://github.com/talismanbrandi/Interpretable-ML-bbh.git}.

\section{Probing the \texorpdfstring{$\bbh$}{bbh} channel at future colliders}
\label{sec:bbh-col}

In this section we will discuss the various prospects that arise for studying the $\bbh$ channel at future colliders. First, we will give a short overview of what might be possible at future lepton colliders like the FCC-ee, CEPC and muon colliders. Then we move on to a more detailed study of the prospects at HL-LHC and FCC-hh which will form the bulk of the work.

\subsection{Prospects at future lepton colliders}
\label{sec:Lepton}

At a foreseeable electron-positron collider, the $\bbh$ signal could be produced above the threshold of about 135\,GeV. Once the center-of-mass energy is sufficient for the $Zh$ Higgs-strahlung process, the dominant production of $\bbh$ is through the decay of resonant $Z$ to a bottom-quark pair, which is about five orders of magnitude larger than the subdominant $y_b$-sensitive production channel. With such a large irreducible background, the channel proportional to $y_b^2$, with its cross-section of $2.6 \times 10^{-4}\,\fb$, will be completely overwhelmed by that of the irreducible background of about 35\,$\fb$ from the Higgs-strahlung process.

Unlike the Higgs-strahlung process, for which the rate drops as inverse of the center-of-mass energy, $\sqrt{s}$, at the amplitude level, the $y_b$-sensitive channel has a weaker dependence on $\sqrt{s}$. Thus, similar to the hadron machines, the rate of $y_b$-sensitive channel benefits from going to higher $\sqrt{s}$ as compared to the Higgs-strahlung contribution. At 1\,$\tev$, the cross-section for the channel proportional to $y_b^2$ rises to about $3.5 \times 10^{-4}\,\fb$, while that for the Higgs-strahlung contribution drops to 2\,$\fb$. It is, however, still difficult to achieve competitive sensitivity on $y_b$.

Another hope arises from studying the $\bbh$ channel below the Higgs-strahlung ($Zh$) threshold. At about 160\,GeV center-of-mass energy~\cite{Ghosh:2019dmv}, with sufficient luminosity, the $\bbh$ signal could be produced and observed. The sensitivity to a possible interference term that is linearly dependent on $y_b$ is, however, difficult to achieve. The study would thus provide sensitivity on $y_b^2$ but not to the sign or phase of $y_b$.

Another recent work suggested the study at a lepton collider of the Higgs decaying to $b\bar b g$, where they focus on the region with 2 $b$-jets being collinear~\cite{Bi:2020frc}. The selection favors the kinematic region which is sensitive to a sizable interference between the $ggh$ and $\bbh$ mediated channels. When fixing the other SM values and varying only a complex $y_b$, the phase of the Yukawa was shown to be constrained to about 60$^\circ$ at CEPC.

For completeness, we estimate the possibilities of probing the $\bbh$ signal at a muon collider at 14\,TeV~\cite{Alexahin:2013ojp}. The $\bbh$ cross-section from $y_b$-mediated diagram is estimated to be about $1 \times 10^{-5}\,\fb$. As a comparison, the Higgs-strahlung induced $\bbh$ production at the proposed muon collider has a cross-section of about 0.01\,fb. Here the two contributions differ by only three orders of magnitude in total cross section. Once more, the bulk of the Higgs-strahlung process has the bottom-quark pair invariant mass peaking at around the $Z$ mass. Provided there is a sizable luminosity to produce the $y_b$-mediated channel, we expect $\bbh$ to be a clean signal at a muon collider to probe $y_b$.

We summarize the leading order (LO) cross section for the $\bbh$ signal at various future lepton colliders in \autoref{tab:lept_bbh}. The interference term between $y_b$-mediated and $Zh$ type of signal  in inclusive rate is orders of magnitude smaller than either of the contribution, and are not shown. From the results of the LO calculation, $y_b$-sensitive channel is overwhelmed by the $Zh$-mediated contribution by over three orders of magnitude across collider setups considered here. With the current projected luminosity at future $e^+e^-$ colliders, it is unforeseeable that sensitivity to $\bbh$ production driven by $y_b$ can be achieved.

\begin{table}
    \centering
    \begin{tabular}{|c|c|c|}
    \hline
        Colliders / cross section (fb) & $y_b$-mediated & $Zh$-mediated \\ \hline
        $\ee\to\bbh~( \sqrt{s}=160~\gev )$ & $1.14\times 10^{-5}$ & $1.59\times 10^{-2}$\\ \hline
        $\ee\to\bbh~( \sqrt{s}=250 ~\gev )$ & $2.6\cdot 10^{-4}$ & 35\\ \hline
        $\ee\to\bbh~( \sqrt{s}=1000 ~\gev)$ & $3.5\cdot 10^{-4}$ & 1.9 \\ \hline
        $\mu^-\mu^+\to\bbh~( \sqrt{s}=14 ~\tev)$ & $7.3\cdot 10^{-6}$ & $9.2\cdot 10^{-3}$\\ \hline
    \end{tabular}
    \caption{\it Total cross section for $\bbh$ signal at leading order at various possible future lepton colliders.}
    \label{tab:lept_bbh}
\end{table}

\subsection{Events simulation for HL-LHC and FCC-hh}
\label{sec:Sim}

\begin{figure}
    \includegraphics[width=0.32\textwidth]{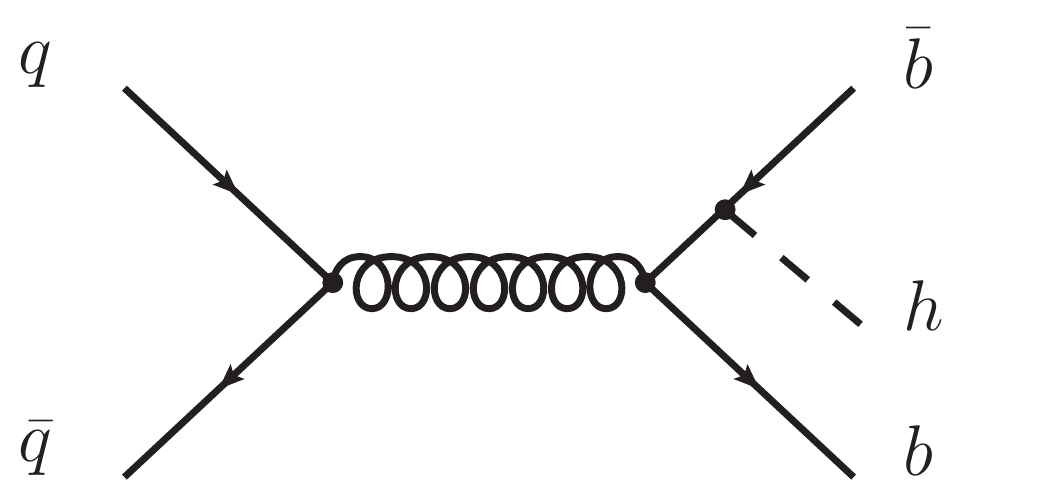}
    \includegraphics[width=0.32\textwidth]{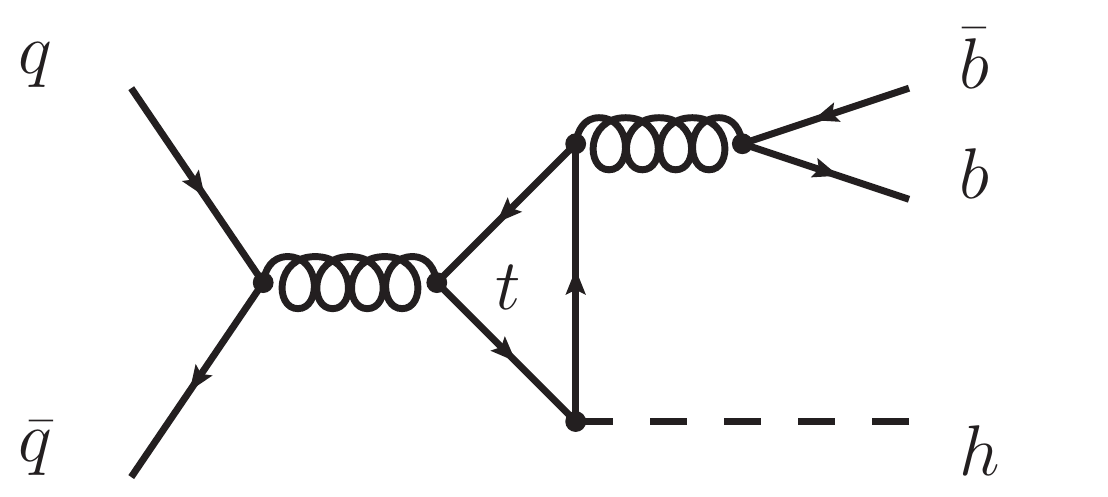}
    \includegraphics[width=0.32\textwidth]{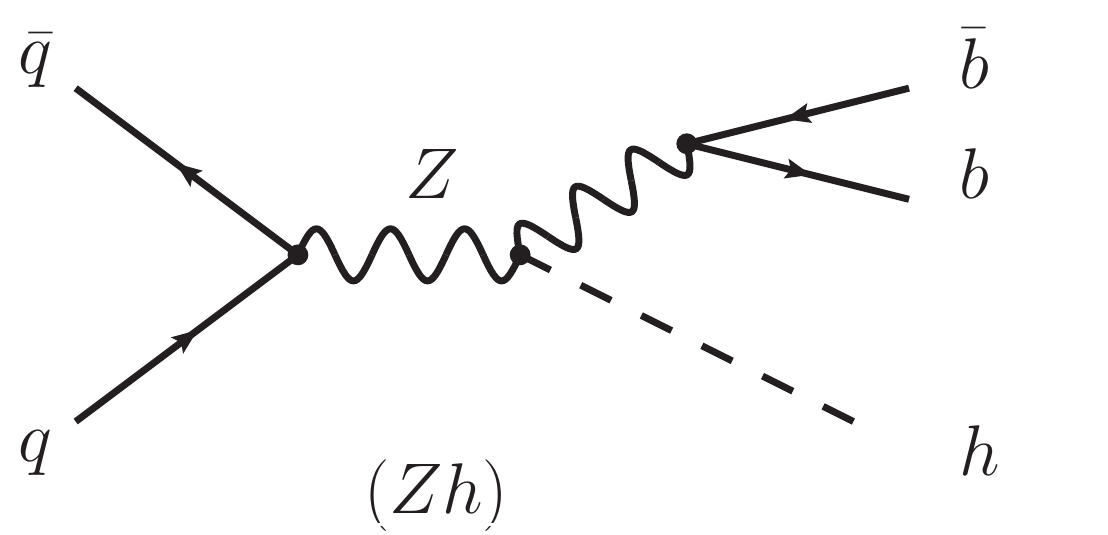}\\
    \includegraphics[width=0.32\textwidth]{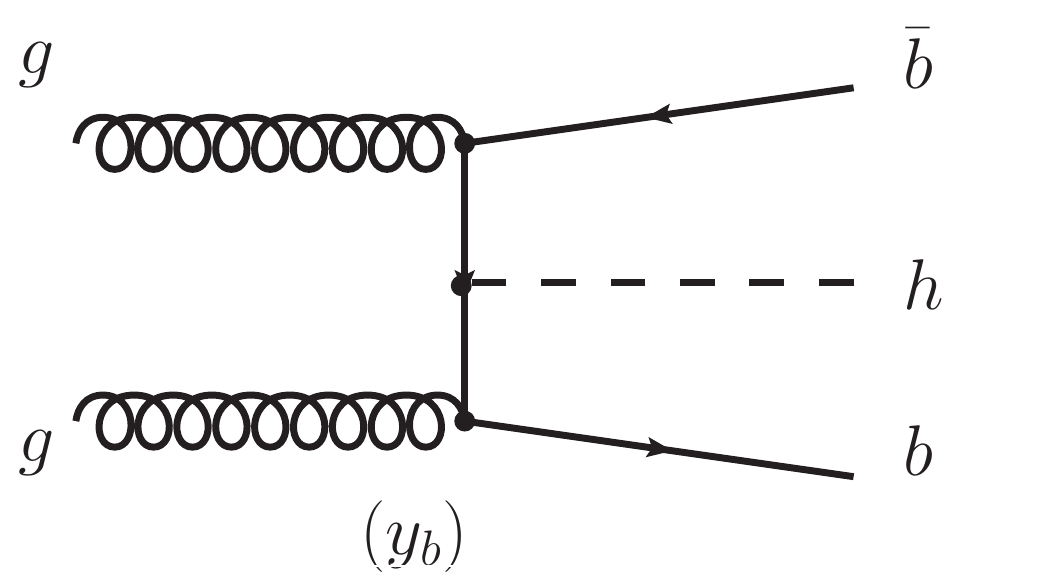}
    \includegraphics[width=0.32\textwidth]{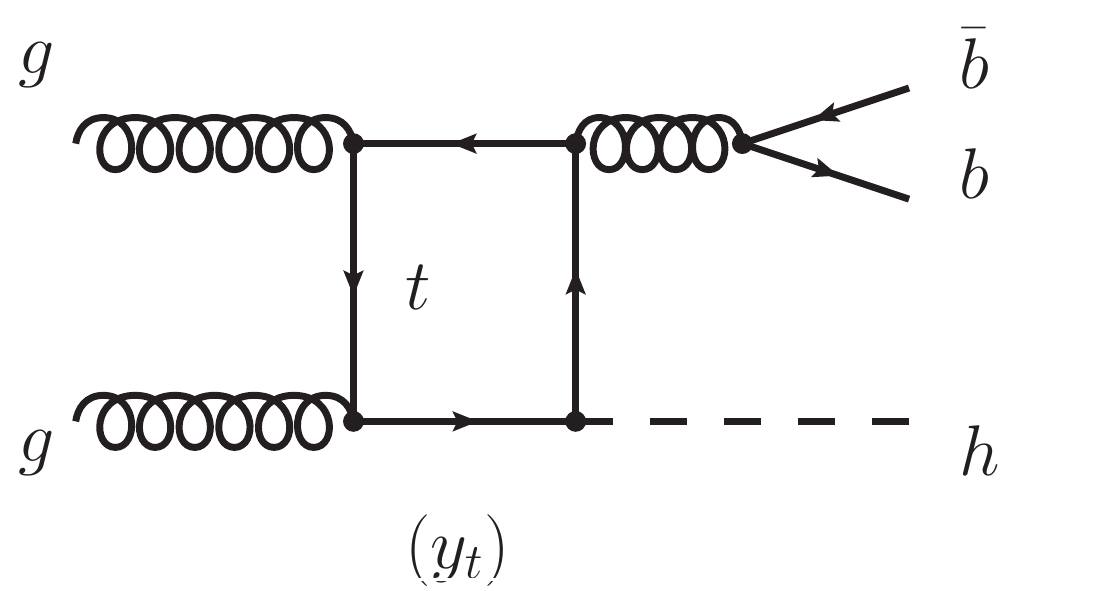}
    \caption{\it Feynman diagrams for the $\bbh$ production channels. The diagrams are grouped into amplitudes proportional to $y_b$ and $y_t$ and the amplitude generating $Zh, Z\to b\bar{b}$ production.}
    \label{fig:feyndiag}
\end{figure}

There are four types of sizable contributions at leading order (LO) that need to be considered separately for the $\bbh$ signal at hadron colliders. The first are the pure $y_b$-inducing diagrams, $(y_b)$ in \autoref{fig:feyndiag}, which contributes at the amplitude level and are proportional to $y_b$. Then there are the gluon-fusion induced Higgs production diagrams, $(y_t)$ in \autoref{fig:feyndiag}, with an additionally radiated gluon splitting to a bottom pair, which are dominantly proportional to $y_t$. Thirdly, there is a non-negligible interference between the first two types of diagrams proportional to $y_ty_b$. Lastly, there is the $Zh$ production mode with the $Z$ decaying to a pair of bottom quarks, the right-most diagram ($Zh$) in \autoref{fig:feyndiag}.

The importance of the contribution proportional to $y_t$ in the $\bbh$ signal has been recently emphasized in Ref.~\cite{Deutschmann:2018avk}, calculated to next-to-leading order (NLO) QCD in the effective $hgg$ coupling framework, and the results are presented from the fixed order calculation. Recently, the $Zh$ and vector-boson fusion (VBF) contribution were included to NLO in QCD and electroweak (EW) couplings, and found comparable to the contribution proportional to $y_b$ as well~\cite{Pagani:2020rsg}. The various contributions to the $\bbh$ signal increase, in general, the total rate and observability of the signal at the HL-LHC. But it also poses a challenge to disentangle the different contribution necessary to improve sensitivity to $y_b$, at least, when looking at 1D differential distributions from fixed order calculations presented in the aforementioned papers.

Thus, as a first attempt to estimate a ``realistic'' sensitivity to the contributions from the different channels, to the size of $y_b$ and, possibly its CP-phase, we start the study with the relatively clean and well-reconstructed channel of the Higgs decaying to two photons\footnote{A study on the other decay channels could be beneficial, such as the $h\to b\bar b$, which has its challenges in large QCD background and $b$-jet reconstruction, while having a large rate. In addition, $h\to 4\ell$ is another channel that is clean and comparable to the di-photon channel as in the single Higgs case. It, however, suffers from a even smaller signal rate and is difficult to observe at HL-LHC.}. 

In this work we include all the channels contributing to the $\bbaa$ final state and are either statistically overwhelming like the $\bbaa$ QCD-QED background, or have similar shapes, like the $Zh$ channel or are both, like the $y_t^2$-driven channel. There are some other backgrounds that are worth mentioning but are much easier to separate out.
\begin{itemize}
\setlength{\itemsep}{0pt}
    \item VBF (NLO-EW): As discussed before, this channel is particularly tricky since it can be as sizable as or larger than the $y_b$-sensitive channels and pose as a challenging background. However, as shown in Ref.~\cite{Pagani:2020rsg}, it can be brought under control by demanding an additional veto of light jets given the topology through which this channel is produced. The light-jet veto essentially kills the VBF contribution while not affecting any of the channels that we discuss above. This can be seen from table~3 when comparing the ``NLO$_3$'' contribution, dominated by VBF with the ``LO$_3$'' contribution dominated by $q\bar{q}\to Zh$. In addition, from the top-right panel of figure~3 of the reference we see the same suppression of the VBF channel due to the light-jet veto. Given that the VBF contribution can be suppressed independently, we do not consider it in our work.
    
    \item $hh$ production: The di-Higgs production can pose as a background if it decays into the $\bbaa$ final state. The cross-section for this is comparable to the $y_b$-sensitive channels both at the HL-LHC and the FCC-hh. However, since both $m_{bb}$ and $m_{\gamma\gamma}$ will be clustered around the Higgs-mass peak, the shape of the final state will be distinct enough to separate it from other channels that contribute to $bbh$.
    
    \item $gg\to Zh$: This channel has a small cross-section at HL-LHC and can be safely ignored. However, it grows rapidly with $\sqrt{s}$, about a factor of 40 from HL-LHC to FCC-hh energies for the inclusive rate. Hence, this channel will become comparable, but subdominant, to the $y_b$-sensitive channels at FCC-hh. Nevertheless, it can be distinguished from the $y_b$-sensitive channel because of the difference in the shape that it will have, akin to the case of $q\bar{q}\to Zh$. 
\end{itemize}

For the ``non-Higgs'' background, there is the dominant irreducible QCD-QED background of $b\bar b\gamma\gamma$. There are also potential backgrounds from various fakes such as $j\gamma\gamma$, $jj\gamma\gamma$, $cj\gamma\gamma$, $c\bar c\gamma\gamma$, or $b\bar bj\gamma$. The rate of light jets faking a bottom or photon is at the percentage or sub-percentage level, and makes the fake backgrounds mostly negligible after considering the dominant $\bbaa$ background.\footnote{The dominant fake contribution to $b\bar b\gamma\gamma$ comes from the $c\bar c\gamma\gamma$ channel. Indeed, it can be seen from Table 1 of \cite{ATLAS:2017muo}, with similar basic selection cuts, $c\bar c\gamma\gamma$ is about 8 times $b\bar b\gamma\gamma$. A $c\to b$ mistag rate of about $0.1$ makes it sub-dominant but comparable to the dominant $b\bar b\gamma\gamma$ background. The main argument for dropping this mistagged QCD-QED background is that they shall have similar shape to the $b\bar b\gamma\gamma$ background, and hence, should be separable by the BDT. Secondly, since $c$-tagging is still in the process of being perfected, we believe that this background might be better separated in the future, although a clear projection is missing from the literature.} Hence, we ignore these in this work. 

The total cross-section at 14\,TeV of the various contributing channels after basic cuts as defined in \autoref{eqn:bcuts} is shown in \autoref{tab:xsec14}. We calculate the production and decay of the signal in the four-flavor scheme at LO, following the results presented in Ref.~\cite{Deutschmann:2018avk} with their public code for  contributions proportional to $y_b^2$ and $y_by_t$. The interactions between Higgs and gluon are treated as point-like effective couplings $ggh$ and $gggh$ with massive quark effects included at one-loop defined as:
\begin{equation}
    \mathcal{L}\supset -\frac{1}{4}C_1 h G^a_{\mu\nu}G^{a,\mu\nu},
\end{equation}
\begin{equation}
    C_1 = -\sum_{Q=t,b}\frac{y_Q}{v}F\left[\frac{m_Q}{m_h}\right]\frac{\alpha_S}{3\pi},
\end{equation}
where $F[m_Q/m_h]$ is a function of the massive quark and Higgs mass ratio. The functions are defined in detail in \autoref{app:kgkgamma}. Input parameters are set to be the same as in Ref.~\cite{Pagani:2020rsg}, where sizable EW corrections are discussed and included to NLO level. Wherever applicable for our LO parton level simulation, we set $m_b^{\rm pole} = 4.58\,\gev$, $m_b^{\rm \overline{MS}}(m_b^{\rm \overline{MS}}) = 4.18\,\gev$, $m_h=125~\gev$, $m_t=173.34~\gev$, $m_Z=91.15348~\gev$, $\Gamma_Z=2.4946~\gev$ and $G_F=1.16639\cdot  10^{-5}~\gev^{-2}$. Bottom-quark Yukawa running effects as function of the dynamical renormalization scale are included. The central scale is chosen as $H_T/4$ where $H_T$ is the scalar sum of the transverse mass of the parton level $\bbh$ system.

\begin{table}[t]
    \centering
    \begin{tabular}{|c|c|c|c|c|}
    \specialrule{.8pt}{0pt}{0pt}
    Channel	    &LO $\sigma$ (fb)	&NLO-k-fact	&6$\inab$ [\#evt]   & 2$b$-jets[\%]  \\ 
    \specialrule{.8pt}{0pt}{0pt}
    $y_b^2$	    &0.0648	            &1.5	    &583                &7.7\%  \\ \hline
    $y_by_t$    &-0.00829	        &1.9        &-95                &4.0\%	\\ \hline
    $y_t^2$	    &0.123	            &2.5	    &1,840                &12\%	\\ \hline
    $Zh$	    &0.0827	            &1.3	    &645                &21\%   \\ \hline
    $\sum\bbh$	&0.262	            &-	        &2,970               &-      \\ \hline
    $\bbaa$	    &12.9	            &1.5	    &116,000              &14\%	\\ 
    \specialrule{.8pt}{0pt}{2pt}
    \end{tabular}
    \caption{\it SM cross-section for the main signal and background processes at 14\,$\tev$ with 6\,$\iab$ data, and number of events after the basic cuts as defined in \autoref{eqn:bcuts}. For the $\bbh$ production, the Higgs is decayed to a pair of photons.}
    \label{tab:xsec14}
\end{table}

As shown in \autoref{tab:xsec14}, the computation is first done at leading order (LO) and then an overall $k$-factor is applied on the LO $\bbh$ cross-sections according to the different production channels, based on the fixed order inclusive cross-section results provided in Ref. \cite{Deutschmann:2018avk}. The $k$-factors are relatively flat over the kinematic distributions for the $\bbh$ final states. We assume a further SM decay of the Higgs to di-photon and parton shower does not significantly affect this simplification. The branching ratio for the decay of Higgs to di-photon is further normalized as the Higgs cross-section working group recommended value~\cite{deFlorian:2016spz}. For the $\bbaa$ background, the $k$-factor is taken from the ratio between NLO and LO total cross-sections as given in Ref.~\cite{Faeh:2017fpp}. New topologies of VBF-like contribution that are not included in this work arise as NLO-EW effects. 

At the generator level, we apply the following cuts to avoid divergences arising from the $\bbaa$ background.
\begin{equation}
    \begin{aligned}
    & Xp_T^b>20\,\gev,~p_T^\gamma>20\,\gev. \\
    \textrm{generator level cuts:}\qquad& \eta_\gamma<3,~ \delta R_{b\gamma}>0.2. \\
    & 100< m_{\gamma\gamma} \,(\gev) < 150.
    \end{aligned}
\end{equation}
Here $Xp_T$ implies a minimum $p_T$ cut for at least one $b$-jet. For the PDF we use \texttt{lhaid=320500} from the \texttt{LHAPDF6} package~\cite{Buckley:2014ana} (\texttt{NNPDF31\_nlo\_as\_0118\_nf\_4}~\cite{Ball:2017nwa})\footnote{We use the default current version of the PDF set which is called \texttt{DataVersion=2}, an update from an earlier version. This causes a discrepancy of about $10\%$ in our total cross section from the study~\cite{Deutschmann:2018avk}, while we otherwise use the same calculation setup and input parameters. We, however, check the ratios between different jet selection and $p_T$ cut requirements that these two papers provide and find good agreement with their fixed order calculation results.}. Parton-showering is then performed on the generated LO event sample, with \texttt{Pythia8}~\cite{Sjostrand:2014zea}, without multi-parton interaction simulated. Detector simulation is applied next using \texttt{Delphes}~\cite{deFavereau:2013fsa}. The parameters are set according to the HL-LHC card, where the $b$-tagging rate for a central $b$-jet is about 75\%. The jets are then reconstructed with the anti-$k_T$ algorithm with $R=0.4$. We further require two photon jets~\footnote{We use the standard isolated photons definition from Delphes, and call them photon jets in the sense of reconstructed objects.} and at least one $b$-tagged jet which should satisfy: 
\begin{equation}
    \textrm{basic cuts:}\qquad
    \begin{array}{l}
    p_T^{bjet}>30\,\gev,~p_T^{\gamma jet}>20\,\gev,\\
    \eta_{bjet,\gamma jet}<2.5,~ 110< m_{\gamma\gamma}\,(\gev) < 140.
    \end{array}
\label{eqn:bcuts}
\end{equation}
With these sets of basic cuts, we get the cross-section and estimate the number of events at  HL-LHC ($6\inab$) shown in \autoref{tab:xsec14}. Note that the interference term between $y_b$- and $y_t$-types of diagrams is negative. It, in fact, contains two separate contributions which are $q\bar q$-initiated and $gg$-initiated. The former is positive and the latter, larger and negative, due to the different types of diagrams involving $s$ or $t/u$ channels with a massive bottom quark propagator. 

All this being said about HL-LHC, we move our focus to FCC-hh. Much like the $t\bar th$ process, the $\bbh$ cross-section rises much faster with the center-of-mass energy as compared to the other Higgs production channels like $ggh$, VBF or VH. While the cross-section of the latter processes see a growth of about a factor of $\sim 10-15$ when going from 14 TeV to 100 TeV at a $pp$ collider, the cross-section of $\bbh$ grows by about a factor of 22. This gives a distinct advantage to the $\bbh$ channel over the main irreducible background coming from $Zh$ rendering the measurement of $y_b$ easier at the FCC-hh even beyond a simple statistical scaling to 30 $\iab$.

We use the same parameter setting, generator level cuts, LO simulation and $k$-factor estimates as used for HL-LHC. It is known that the $k$-factors generally increase with center-of-mass energy of the collider. This increase is of the order of 10\%-20\% in the channels that we consider. If this increase is taken into consideration, the signals will actually stand to gain over the background. In this sense, our analysis for the FCC-hh is on the more conservative side. Keeping the same cuts does not affect the analysis since they are very basic cuts used to prune out the QCD-QED background as the rest of the discrimination is left to the multivariate analysis we describe in \autoref{sec:kinematics}. Likewise, we perform \texttt{Pythia} showering, hadronization, and a detector simulation at the FCC-hh using the FCC-hh working group recommended \texttt{Delphes} card where the $b$-tagging rate for a central $b$-jet is projected to be 85\%. We present the cross-section and expected number of events in \autoref{tab:xsec100} after basic cuts.

\begin{table}[]
    \centering
    \begin{tabular}{|c|c|c|c|c|}
    \hline
 Channel	&LO $\sigma$ (fb)	&NLO-k-fact	&30$\inab$ [\#evt] & 2$b$-jets[\%]  \\ \hline
$y_b^2$	& 1.64	&1.5	&73,700 	&8.7\% \\ \hline
$y_by_t$	&-0.197	&1.9	&-11,200		&4.4\%	\\ \hline
$y_t^2$	&4.12	&2.5	&309,000	&14\%	\\ \hline
 $Zh$	&0.605	&1.3	&23,600		&22\%\\ \hline
$\sum$ $\bbh$	&6.16	&-	&405,000		& - \\ \hline
$\bbaa$	&194	&1.5	&8,730,000  	&14\%	 \\ \hline
    \end{tabular}
    \caption{\it SM cross-section for the main signal and background processes at 100 $\tev$ with 30 ab$^{-1}$ integrated luminosity, and number of events after the basic cuts as defined in \autoref{eqn:bcuts}.
    }
    \label{tab:xsec100}
\end{table}

In Ref.~\cite{Pagani:2020rsg}, it was shown that non-negligible EW contribution from $Zh$ and VBF-like (NLO-EW) topologies arise and could possibly overwhelm the $y_b$-sensitive channels. In their fixed order calculation, the authors discuss distributions of $p_T$ and $\eta$ of the Higgs as well as those of the leading $b$-jet and light-jet, and the rapidity gap between the Higgs and leading $b$-jet, both at LO and adding QCD and EW (N)LO contributions. From these plots, it is not easy to discern the differences in the distributions in the bulk of the signal region. We will show that a careful selection of kinematic distributions backed with the discriminatory powers of simple machine learning algorithms can be effective in separating the $y_b$-driven channel. For example $m_{b_1h}$, the invariant mass of the leading $b$-jet and the reconstructed Higgs, which is the second most important kinematic variable for separating out $Zh$ contribution from the $y_b^2$-driven one, is not included in Ref.~\cite{Pagani:2020rsg}. 

Hence, the claim that discerning the $y_b^2$ contribution in the $\bbh$ signal will be quite hopeless at HL-LHC, where different categories with selected number of $b$-jets were studied as in Table~4 of Ref.~\cite{Pagani:2020rsg}, is somewhat premature given that a more refined statistical analysis proves otherwise. As we will see in our analysis, using decision trees and Shapley values, the number of $b$-jets is not important for distinguishing the various Higgs channels contrary to what was attempted in Ref.~\cite{Pagani:2020rsg}. Instead there are other variables and, more importantly, their correlations that come from higher dimensional distributions that will be the key to extracting a signal that has been declared as a lost cause.

\section{Exploring higher dimensional kinematic distributions}
\label{sec:kinematics}

The total $\bbh$ signal with the Higgs decaying to di-photon is a resonance-bump hunt in a background of a flat di-photon spectra, with an additional requirement of at least one tagged $b$-jet in our selection. At the HL-LHC, with 6 $\iab$ of data (ATLAS+CMS combined), and summing up the contributions from the four types of $\bbh$ signals at LO, we get a statistical significance of $6.8\sigma$ on the total SM $\bbh$ signal, with the basic cuts as defined in \autoref{eqn:bcuts}. Performing a mass window cut of $123< m_{\gamma\gamma}\, (\gev) < 127$ around the Higgs mass peak further increases the significance to about $10\sigma$. With this estimate, there is no doubt that we should be able to see a clear $\bbh$ signal on top of the dominant QCD-QED $\bbaa$ background at the HL-LHC.

The next goal is to evaluate the sensitivity to the contribution proportional to $y_b$ out of the total signal, which is the primary motivation of looking at the $\bbh$ channel in this work. We can gather from the number of events with 6$\inab$ given in \autoref{tab:xsec14}, the sensitivity to the $y_b^2$-driven channel is only about $1.6\sigma$ after basic cuts. We will try to see if this can be improved by exploring the higher dimensional kinematic shapes using multivariate analysis. 

The $Zh, Z\to b\bar b$ channel has the distinct feature that the invariant mass of the two $b$-jets can be reconstructed to the $Z$ boson mass. However, given the basic cuts, only about 20\% of $Zh$ channel has both $b$-jets tagged in our simulation. The fraction of events having two tagged $b$-jets is even smaller from other $\bbh$ channels, as shown in the last column of \autoref{tab:xsec14}. Given the limited signal statistics, especially at the HL-LHC, we do not require two tagged $b$-jets or stringent mass window cuts, to allow for more events from the $y_b^2$ and $y_by_t$ channels. Instead we stay as inclusive as possible with generous cuts, and resort to kinematic shapes and multivariate analyses to further explore the variance in shapes of the higher dimensional distributions amongst the different channels. 

As discussed before, the $y_b^2$-driven channel could be overwhelmed by the other $y_b$-independent contributions such as $Zh$ and those proportional to $y_t^2$. Despite the sizable contribution from these other $\bbh$ channels, and the similarity between the 1D distributions of the kinematic observables, the $Zh$ channel can still be separated from the $y_b^2$ channel with relative ease given information from higher dimensional distributions. We will see a nice separation from the multivariate analysis, and understand the physics ramifications brought about from the higher dimensional kinematic distributions as well. The $y_t^2$ contribution is a bit harder to disentangle, and remains as the dominant background which reduces sensitivity to $y_b$. However, we will show systematic approaches to enhance sensitivity to contribution proportional to $y_b^2$ or $y_by_t$ while suppressing $y_t^2$ and all other background contamination.

After detector simulation and jet definition, we have, for most events, a final state of two photon jets and at least one $b$-jet, where the two photons reconstruct back to a real scalar Higgs mass for all the $\bbh$ channels. We first define and evaluate a comprehensive set of kinematic observables as the following:
\begin{itemize}
\setlength{\itemsep}{0pt}
    \item $p_T^{b_1}$, $p_T^{b_2}$, $p_T^{\gamma_1}$, $p_T^{\gamma\gamma}$, 
    \item $\eta_{b_{j1}}$, $\eta_{b_{j2}}$, $\eta_{\gamma_1}$, $\eta_{\gamma\gamma}$,
     \item $n_{bjet}$, $n_{jet}$, $\delta R_{b\gamma_{1}}$, $\delta \phi_{b\gamma_{1}}$, 
    \item $m_{\gamma\gamma}$, $m_{bb}$, $m_{b_{1} h}$, $m_{b\bar b h}$, $H_T$.
\end{itemize}
$p_T^{{b/\gamma}_{1,2}}$ and $\eta^{{b/\gamma}_{1,2}}$ are the $p_T$ and rapidity for the tagged leading and sub-leading $b/\gamma$-jets (in our definition the subleading $b$-jet could be a null 4-vector since we require one $b$-jet inclusive), $n_{bj}$ is the number of tagged and passed $b$-jets. Lastly, $\delta R_{b\gamma_{1}}$ and $\delta \phi_{b\gamma_{1}}$ are the $R$-distance and the $\phi$-angle between the leading $b$-jet and the photon jet.. The remaining variables are the invariant masses, and $H_T$ is the scalar sum of the transverse mass of the system. We shall show in what follows, that it is not necessary to be very selective about the kinematic variables one chooses to use in the analysis. What is necessary is that all possibly useful kinematic variables are included. As can be seen from the list above, some of the variables seem to be interdependent and, probably, highly correlated. The beauty of using interpretable machine learning is that a hierarchy of importance for the variables will be built during the analysis using an over-complete basis of collider observables from which the most important ones can be chosen to fine tune the process.

\subsection{Boosted Decision Trees and Shapley values}
\label{sec:BDT}

Having to assess and discriminate between multiple competing signals, we face a classification problem which requires an understanding of underlying correlations and the necessity to be able to identify kinematic distributions that will optimize the extraction of the signal while suppressing the background. Hence, in the implementation of our multivariate analysis we resort to using decision trees that are very efficient at addressing multi-class classification problems. We will work with the BDT algorithm implemented in XGBoost~\cite{10.1145/2939672.2939785}, a publicly available scalable end-to-end boosting system for decision trees. We follow the normal procedures for training and testing the BDT with simulated data. To assess the performance of XGBoost, we compare it to the BDT implementation in \texttt{ROOT} and see better performance with XGBoost in terms of both speed and accuracy. As a further validation of our results, we implement a Deep Neural Network (DNN) in TensorFlow~\cite{tensorflow2015-whitepaper} to address the classification problem. We note that the performance of the DNN is comparable to and validates the BDT. For our final results we work with the BDT framework so that we can efficiently implement the analysis of variable importance with Shapley values as explained below.

The adoption of machine learning algorithms for theoretical or phenomenological particle physics analyses has been slower than for their experimental counterpart because of the perceived lack of interpretability of the models built by frameworks such as BDT and DNN to map variables onto outcomes. This has often led to a labeling of such algorithms as ``black-box'' tools. We would like to argue that this lack of interpretability is not an inherent property of machine learning algorithms and can be overcome with constructs such as the Shapley values. In conjunction with a concrete understanding of the kinematic distribution that can be attributed to the channels being studied, the use of Shapley values in interpreting machine learning algorithms can provide a far better understanding of the map between kinematic variables and how signal is extracted from dominant, and even irreducible, backgrounds than any simple cut-based analysis. In this work we show how the black-box of machine learning algorithms can be demystified leading us to an understanding of kinematic shapes in a truly multivariate manner.

In our work we choose to use the high-level kinematic variables listed in the previous section instead of low-level energy-momentum four vectors since the former provide a more concrete insight into the underlying dynamics. We test and confirm that working with high-level variables instead of low-level ones does not affect the accuracy of the classifier, be it a BDT or a DNN. One, of course, needs to build a over-complete set of high-level variables to make sure that none that is important is overlooked. With this setup we train the machine learning algorithm performing the necessary tuning to get optimal validation accuracy.

This procedure, however, provides very little insight into which kinematic distributions are paramount in discriminating between the different channels. To do this we introduce the Shapley values~\cite{shapley1951notes}. Formulated by Shapley in the mid-20$^{th}$ century, this is a measure in game theory of how important a player is in a game of $n$ players and a preset outcome defined as the objective of the game. A simple way of looking at Shapley values is to imagine what the outcome could have been without a particular player. Now, if one brings the player in the game, the outcome might be changed. The Shapley value provides a measure of the change of the outcome due to this player on an average. In essence, the Shapley value tells us how important the presence of a variable is in determining a certain class when compared to its absence from the multivariate problem being addressed. The process naturally and mathematically lends itself to studying the correlations between different variables since all possible combinations of variables can be taken out of the game to check the outcome.\footnote{More clarity on Shapley values and interpretable machine learning in general, along with their application can be found in \href{https://christophm.github.io/interpretable-ml-book/}{Interpretable Machine Learning } by Christoph Molnar~\cite{molnar2020interpretable}.}

Typically, Shapley values are very expensive to compute for most problems. Hence, we use SHAP (SHapley Additive exPlanations)~\cite{NIPS2017_7062}, which is based on Shapley values generated by the tree-explainer~\cite{2018arXiv180203888L,Lundberg:2020vt}. To determine the Shapley values, $S_v$, a game is played with all possible combinations of variables as the players and with the payout being the difference of the predicted value of the outcome from the marginalized value. The $S_v$ for the $i^{th}$ variable in determining the value of the outcome is measured from the difference between a value function that includes the $i^{th}$ variable and the value function which is marginalized over the $i^{th}$ variable. For decision trees this translates into conditional probabilities that are determined by the end-nodes that can be reached in each game. Using the additive property of $S_v$, the importance of the $i^{th}$ variable is determined by the ensemble average of the absolute Shapley values, $\overline{|S_v|}$. The higher the $\overline{|S_v|}$ for a variable, the more important is the variable in determining the outcome.~\footnote{In fact, there are several other measures of variable importance used in machine learning like Gini or permutation based measures to name a few. However, these measures are known to suffer from inconsistencies. We checked the Gini measure of variable importance and the permutation measure and found reasonable agreement in the most important variables.} To facilitate this, we use a BDT framework in our final analysis and not a DNN.

Let us assume that we perform a  binary classification of a dependent variable, $y$, using a BDT ($y\in [0,1]$, $y=f(x_i)$ and $i=1,\ldots,n$) with $\vec{x}$ being the vector of independent variables. The probability of belonging to the two classes is the output of a prediction of the BDT for each event. The $S_v$, for a class for any particular event would just represent the shift to the marginalized probability for the class: 
\begin{eqnarray}
p(y=0| \vec{x}) &=& \langle p(y=0)\rangle + \sum_i S_v(y=0|x_i)\nonumber,\\
p(y=1| \vec{x}) &=& \langle p(y=1)\rangle + \sum_i S_v(y=1|x_i),\nonumber\\
\textrm{with}\;\;p(y=0| \vec{x}) &+& p(y=1| \vec{x}) = 1\nonumber,\\
\textrm{and}\;\;\langle p(y=0)\rangle &+& \langle p(y=1)\rangle = 1.
\label{eq:shap}
\end{eqnarray}
which is just the additive property of $S_v$~\cite{shapley1951notes}.

In a multi-class classification using a BDT with the XGBoost implementation, the output of the tree ensemble is $\log(odds)$, where $odds$ is defined as:
\begin{equation}
    odds = \frac{p}{1-p},
\end{equation}

\subsection{The game of reducing the irreducible \texorpdfstring{$Zh$}{Zh} background}
\label{sec:BDTybZh}

\begin{figure}
    \centering
    \figuretitle{\large\texttt{HL-LHC}}
    \hspace{0.1cm}\includegraphics[width=5cm, height=5.1cm]{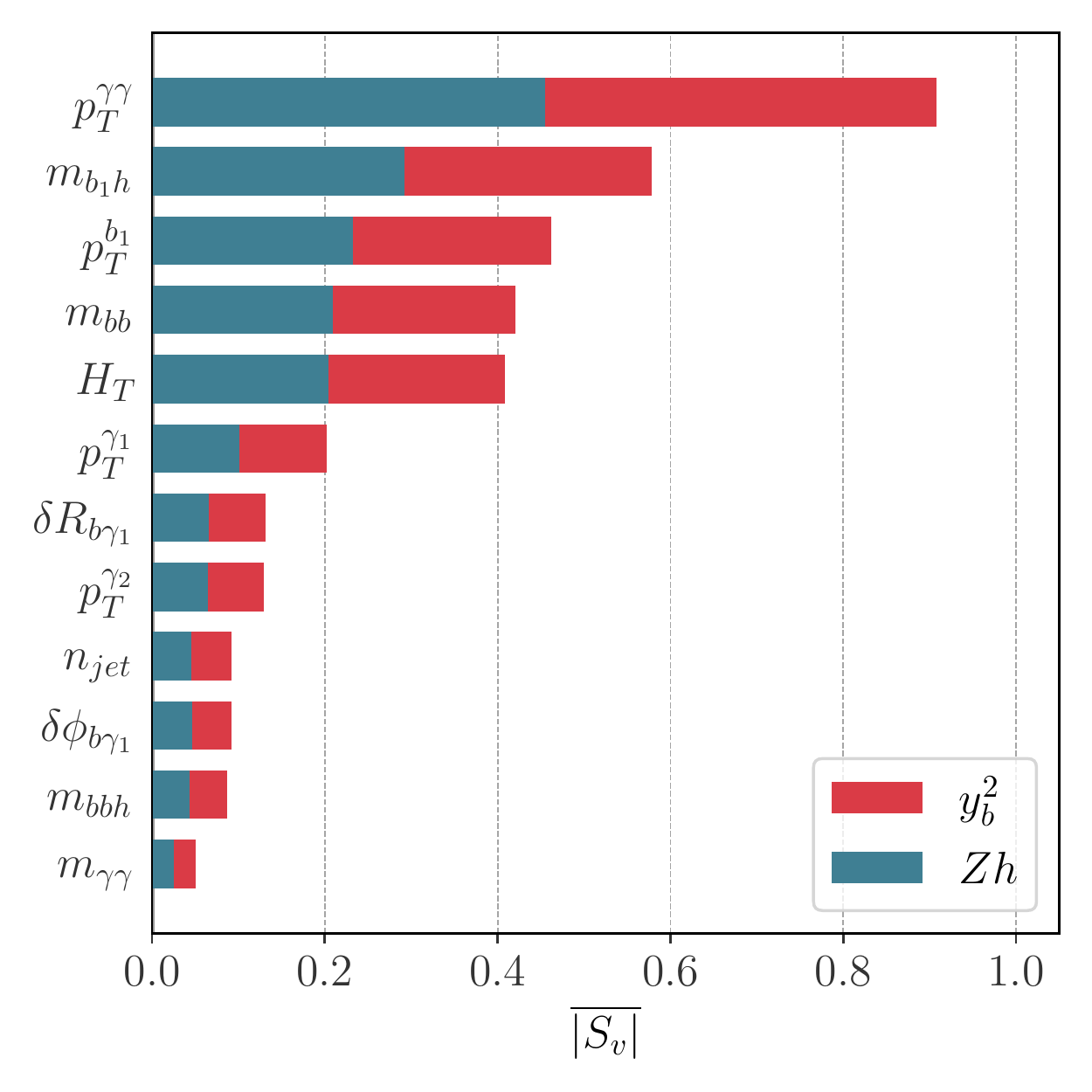}
    \includegraphics[trim=10 0 10 10,clip,width=0.32\textwidth]{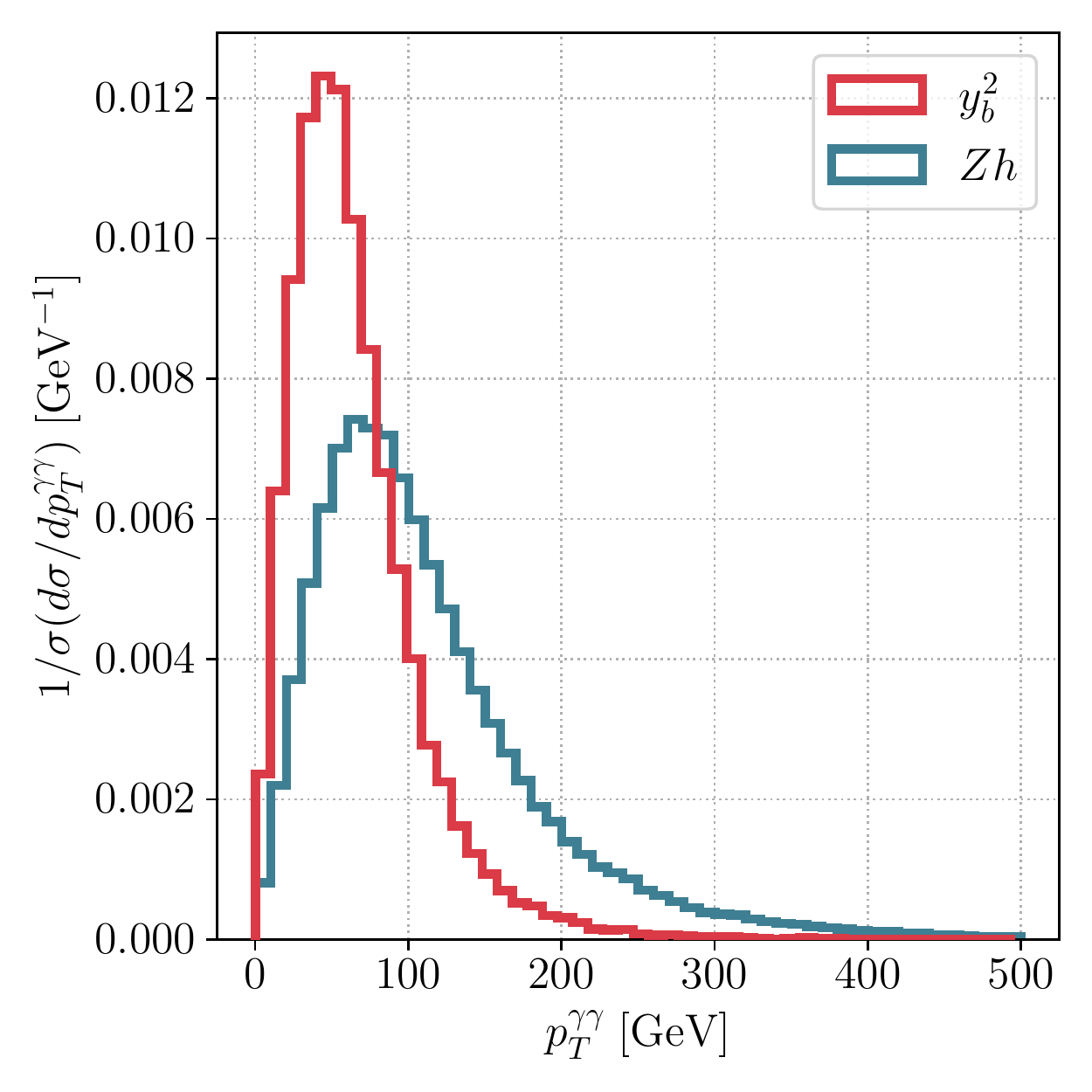}
    \includegraphics[trim=10 0 10 10,clip,width=0.32\textwidth]{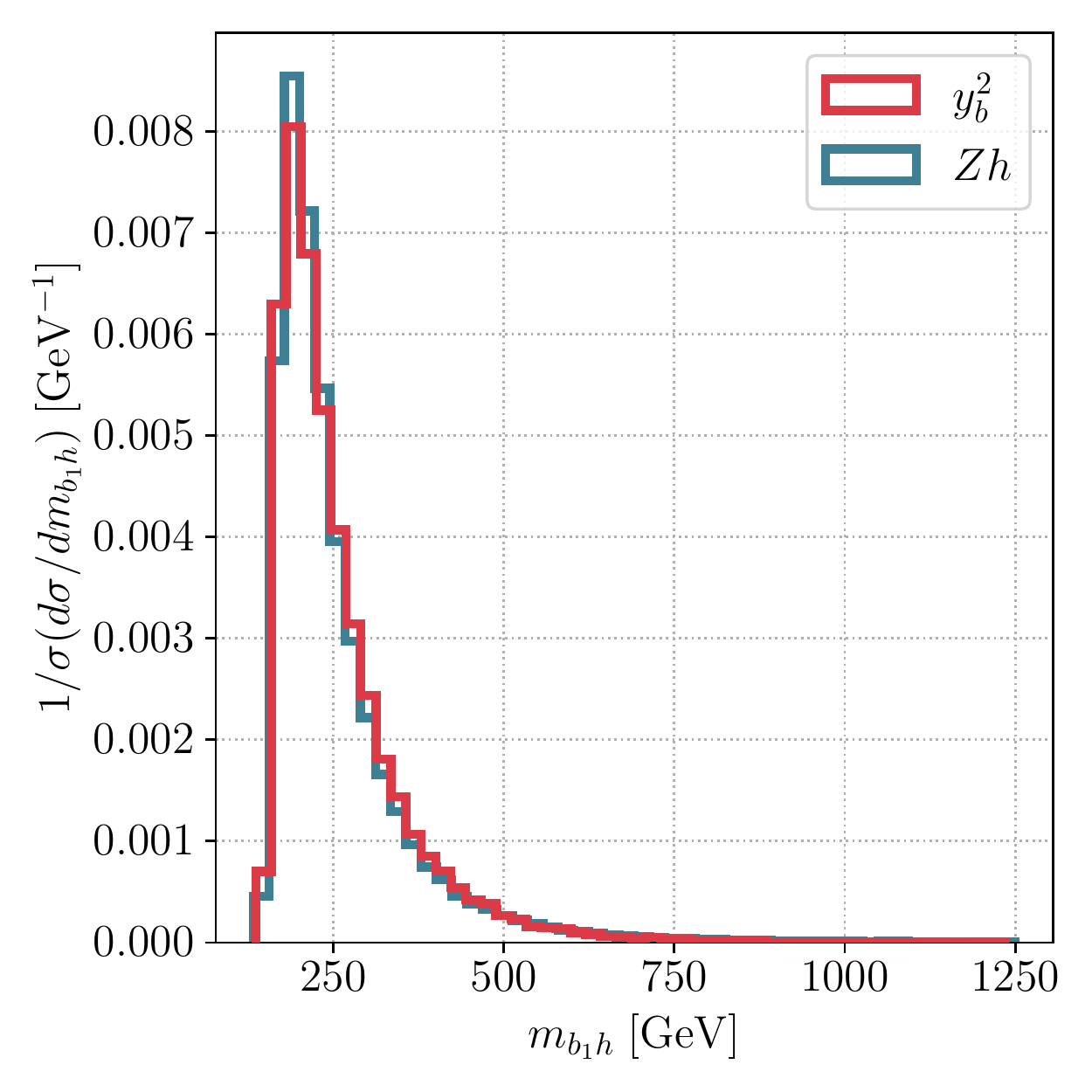}\\
    \includegraphics[trim=10 0 10 10,clip,width=0.32\textwidth]{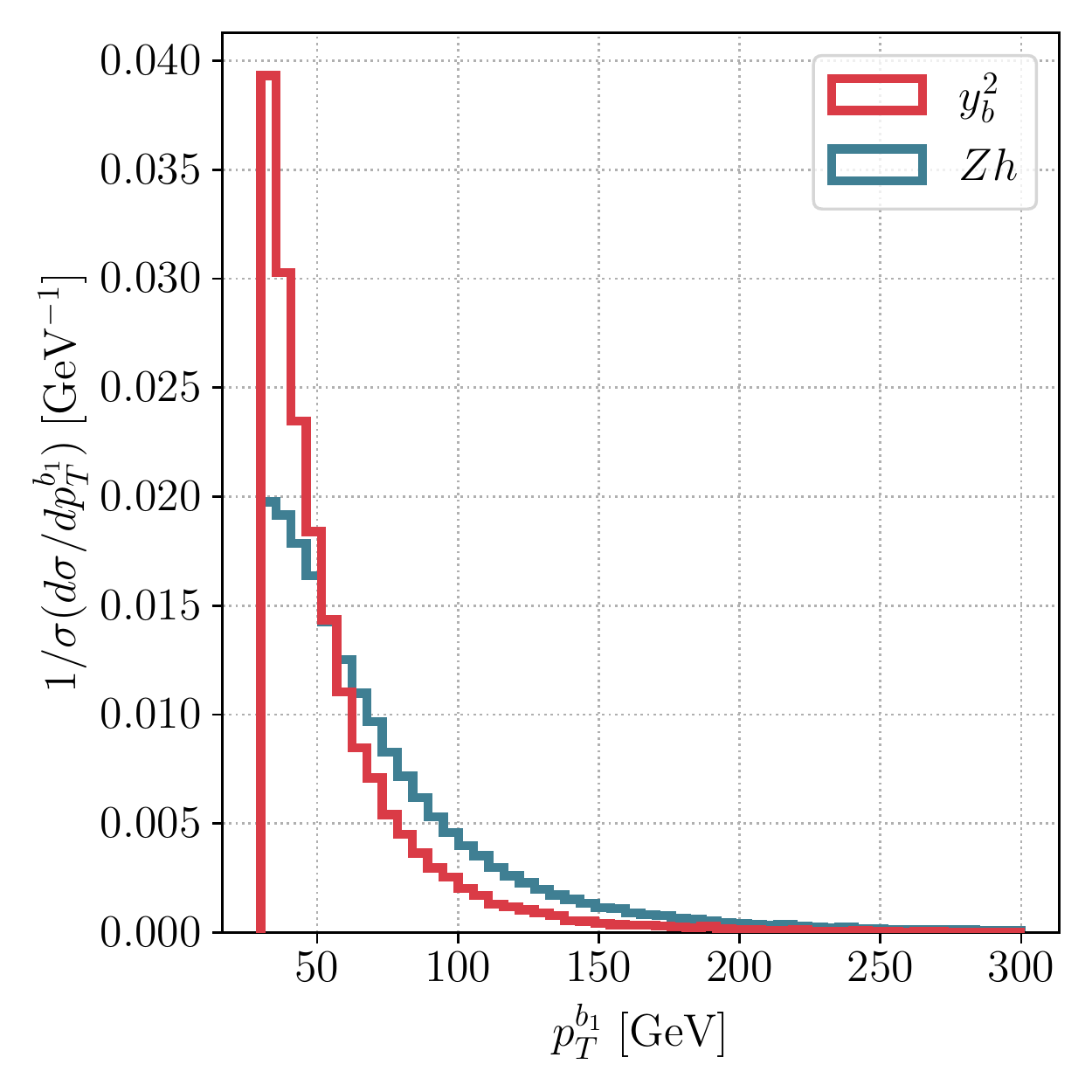}
    \includegraphics[trim=10 0 10 10,clip,width=0.32\textwidth]{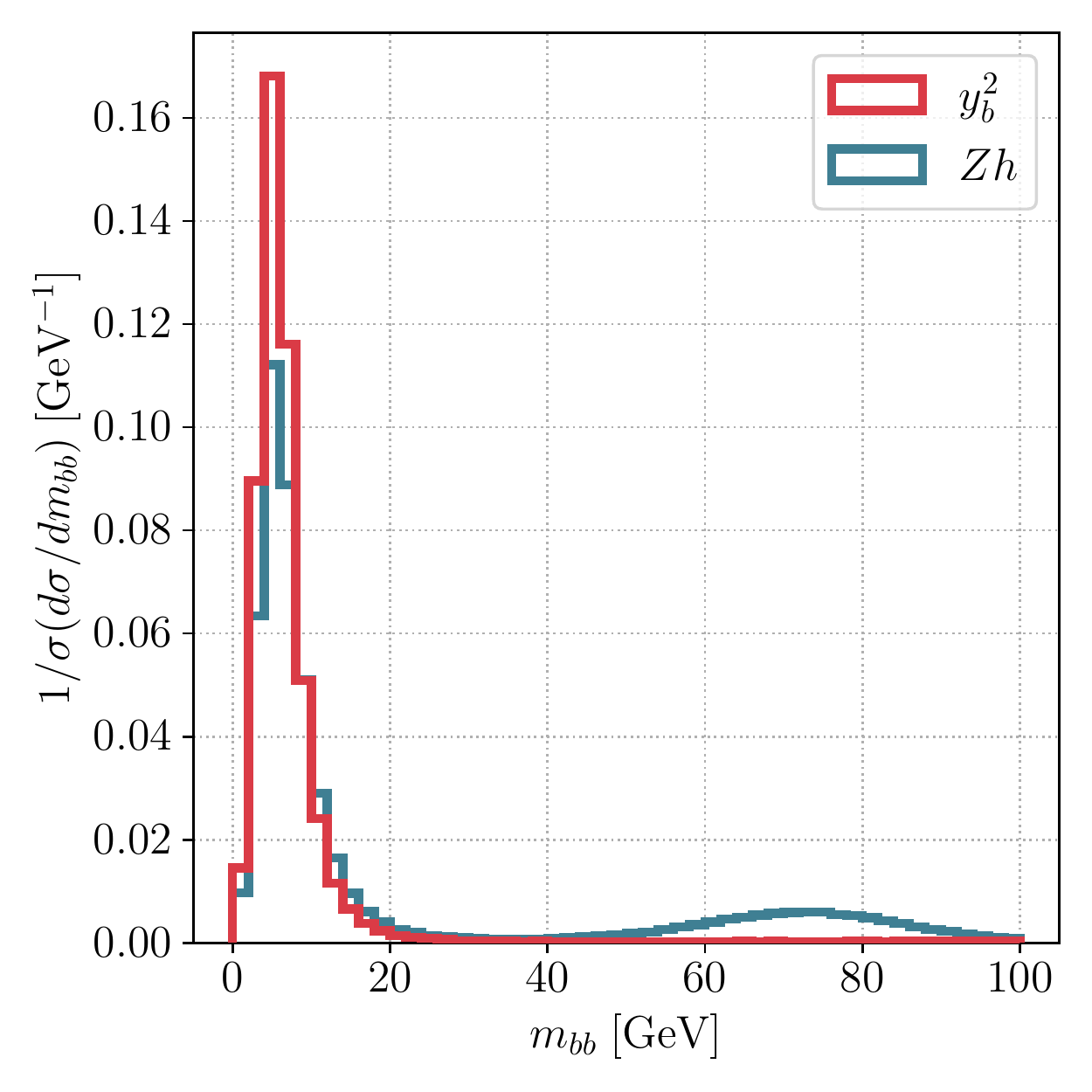}
    \includegraphics[trim=10 0 10 10,clip,width=0.32\textwidth]{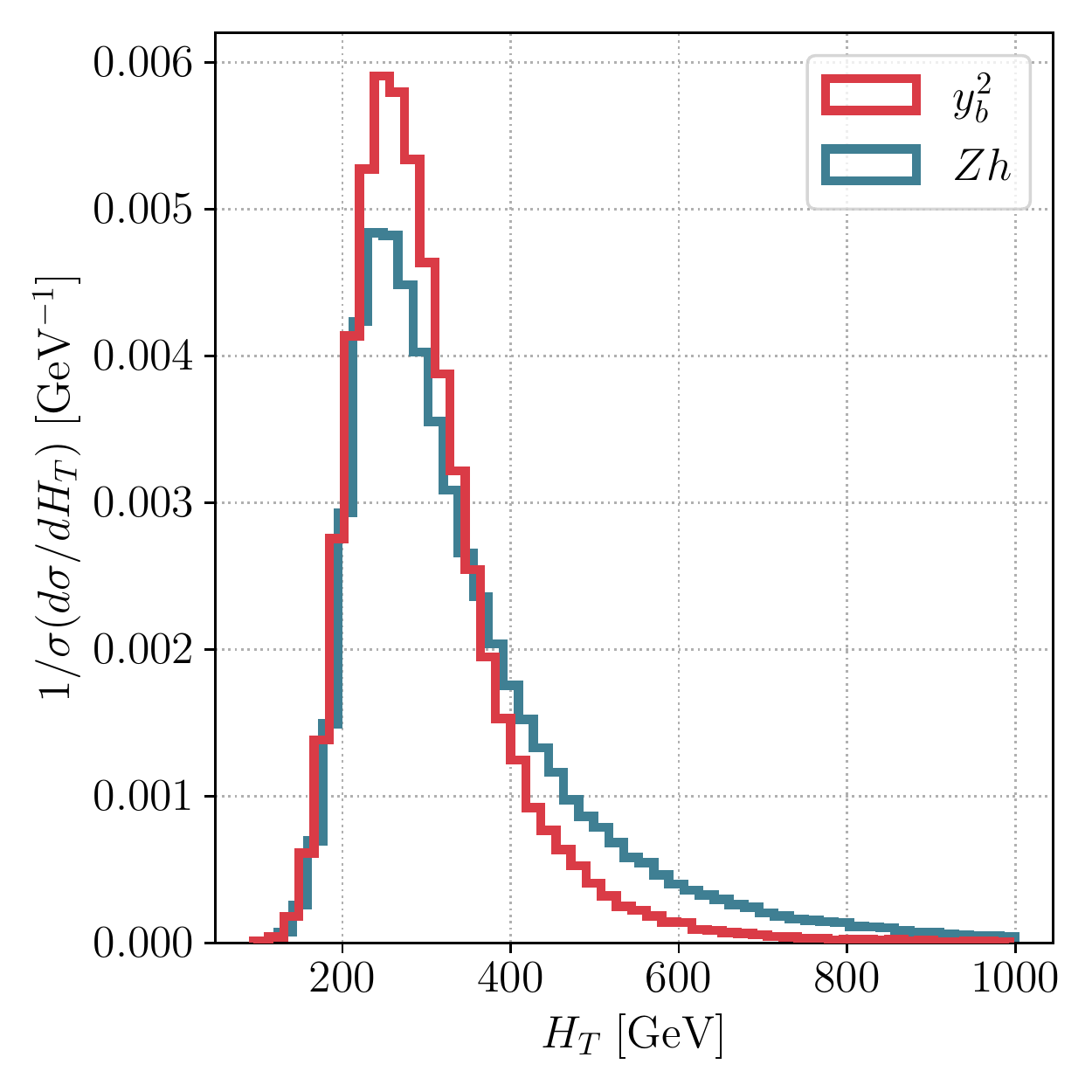}
    \caption{\it The Shapley values and differential distributions for the kinematic variables used to discriminate between the $y_b^2$ and $Zh$ contributions at HL-LHC, all \textbf{normalized to unity} for a clear comparison of shape. A SM signal is injected. \underline{Upper left corner:} The hierarchy of variable importance with increasing Shapley value denoting higher importance in the discrimination. The normalized distribution are those of the 5 most important kinematic variables as determined by their Shapley values.}
    \label{fig:zh-yb2}
\end{figure}

As shown in Ref.~\cite{Pagani:2020rsg} one of the significant hurdles in measuring $y_b$ from $\bbh$ associated production is the $Zh, Z\to b\bar b$ irreducible background. The goal of this section will be to show that this is not the case and that the $Zh$-driven channel and the contribution proportional to $y_b$ are indeed separable. This will also allow us to show how the BDT framework we use is interpreted using the Shapley values, $S_v$, thus acting as a precursor to the detailed analysis with the QCD-QED background in the next section.

We begin the analysis with the whole set of kinematic variables defined in \autoref{sec:kinematics}. After training the BDT we compute the average of the absolute Shapley values, $\overline{|S_v|}$, from a representative sample of events to assess the hierarchy of importance of the kinematic variables. From this we discard those kinematic variables that have very low $\overline{|S_v|}$ since they have a very low impact in discriminating the two channels. We check to make sure that the accuracy of the BDT is not reduced by discarding these variables. This not only reduces the training time drastically, but also provides a clearer picture of the physics behind the kinematic distributions in terms of a smaller number of variables.

The BDT is freshly retrained with the reduced set of variables to produce a final hierarchy of $\overline{|S_v|}$ which can be seen in the top left panel of \autoref{fig:zh-yb2}. From the figure it is clear that the top five kinematic variables, $p_T^{\gamma\gamma}$, $m_{b_1h}$, $p_t^{b_1}$, $m_{bb}$, and $H_T$,  are the most important ones in discriminating the two channels. We show the {\em normalized} distributions for these five kinematic variables corresponding to both the channels. It is clear that any one of those kinematic variables cannot provide a significant discrimination between the channels. Any attempt at a simple cut based analysis will clearly fail, especially since the total cross-section of the channel proportional to $y_b^2$ is lower than that of $Zh$ production. In addition, we also notice that $m_{b_1h}$ is an unlikely candidate as a discriminating variable since the 1D distributions for the two channels overlap significantly.

\begin{figure}
    \centering
    \figuretitle{\large\texttt{HL-LHC}}
    \vspace{0.1in}
    \includegraphics[width=\textwidth]{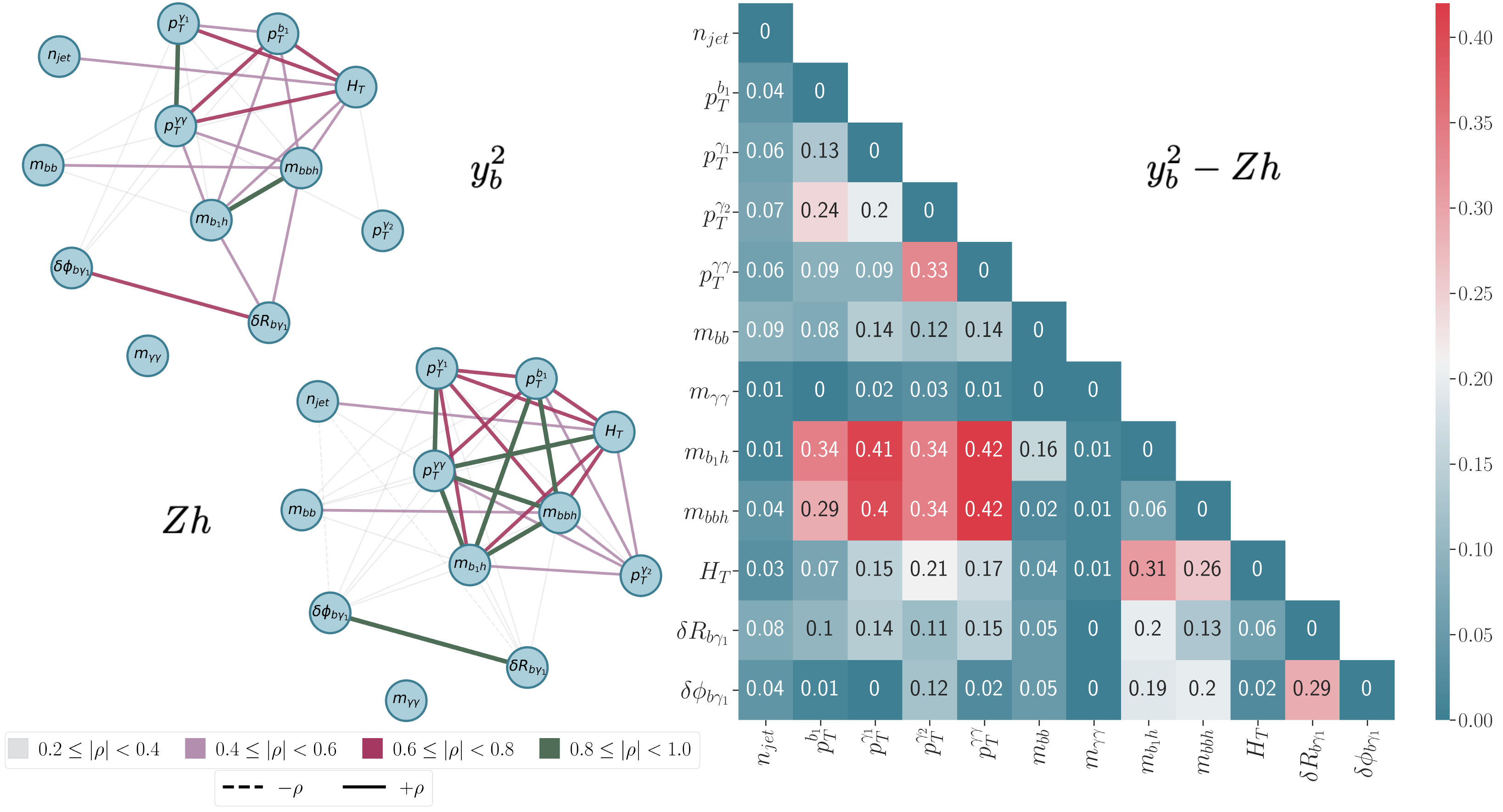}
    \caption{\it Representations of the correlations between kinematic variables for the $y_b^2$ term and the $Zh$ production channel. \underline{Left panel:} shows the correlations in the individual channels. \underline{Right panel:} shows the difference in correlations between the two channels highlighting how the BDT analysis can separate the two contributions at HL-LHC. A SM signal is injected.}
    \label{fig:corr-zh-yb2}
\end{figure}

The strength of a BDT algorithm lies in its ability to model shapes and correlations in multivariate spaces, and give us a linearized measure of the projections of these higher dimensional shapes. Therefore, to understand the hierarchical ordering of $\overline{|S_v|}$, we look at the correlation between the kinematic variables explicitly. The network plots on the left panel of \autoref{fig:corr-zh-yb2} shows the correlation between the kinematic variables in the two channels. It can be clearly seen that the correlation structure of the two channels are quite distinct hinting at variation of the shape of the multivariate kinematic distribution. The heatmap in the right panel of \autoref{fig:corr-zh-yb2} shows the {\em difference} between the correlation in the two channels. One can see that there is a misalignment of the correlation pattern for $m_{b_1h}$ with several variables, and most importantly, with $p_T^{\gamma\gamma}$. The $m_{bbh}$ variable also seems to be quite misaligned in the two channels from the heatmap, but it has a low $\overline{|S_v|}$. This can be understood by noting that $m_{bbh}$ is tightly correlated with $m_{b_1h}$ in both the channels as seen from the network plots in \autoref{fig:corr-zh-yb2} and hence, the misalignment of its correlation is inconsequential as evident from the low $\overline{|S_v|}$ it gets.

\begin{figure}
    \centering
    \figuretitle{\large\texttt{HL-LHC}}
    \vspace{0.1in}
    \includegraphics[trim= 10 10 10 20, clip, width=0.4\textwidth]{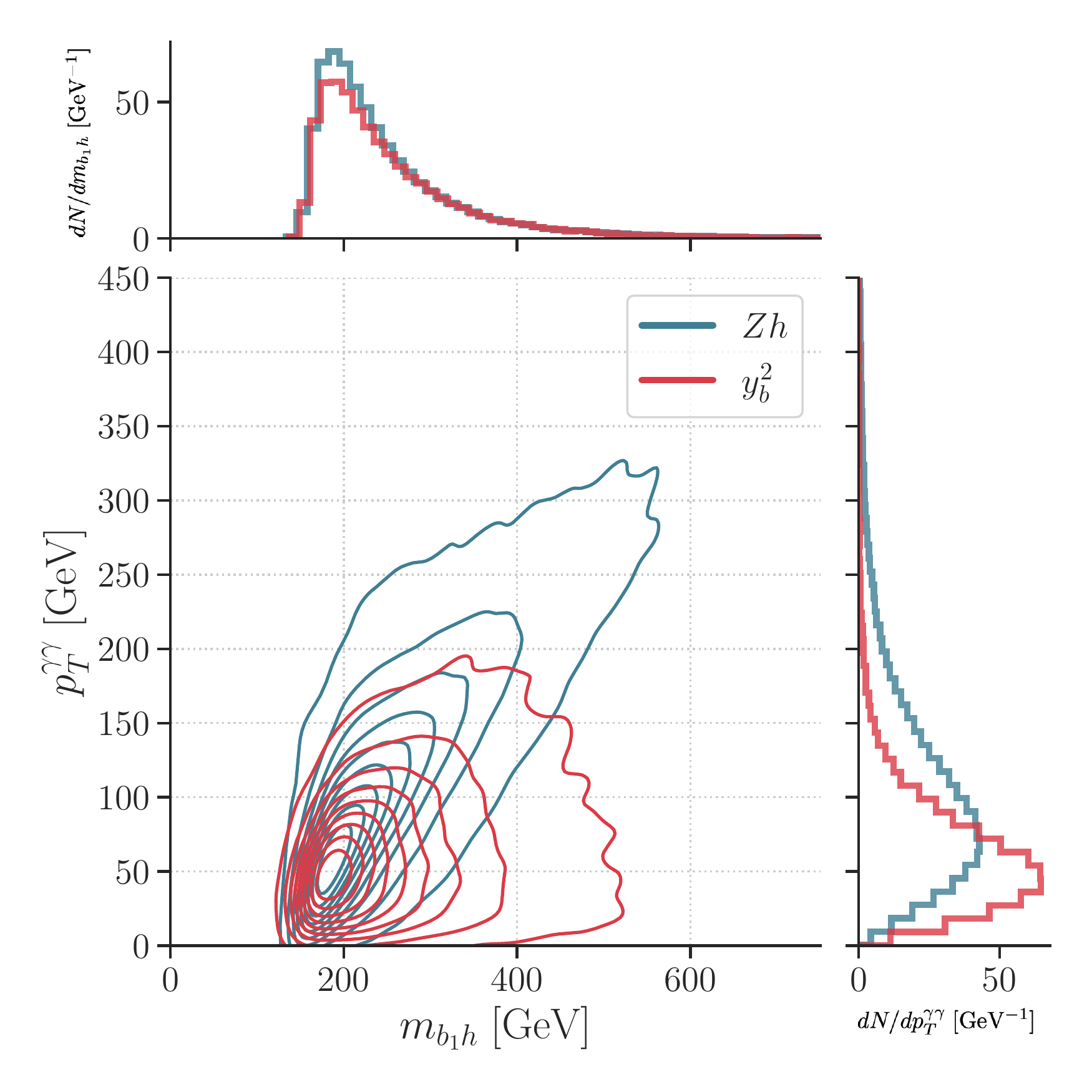}
    \includegraphics[trim= 10 10 10 20, clip, width=0.4\textwidth]{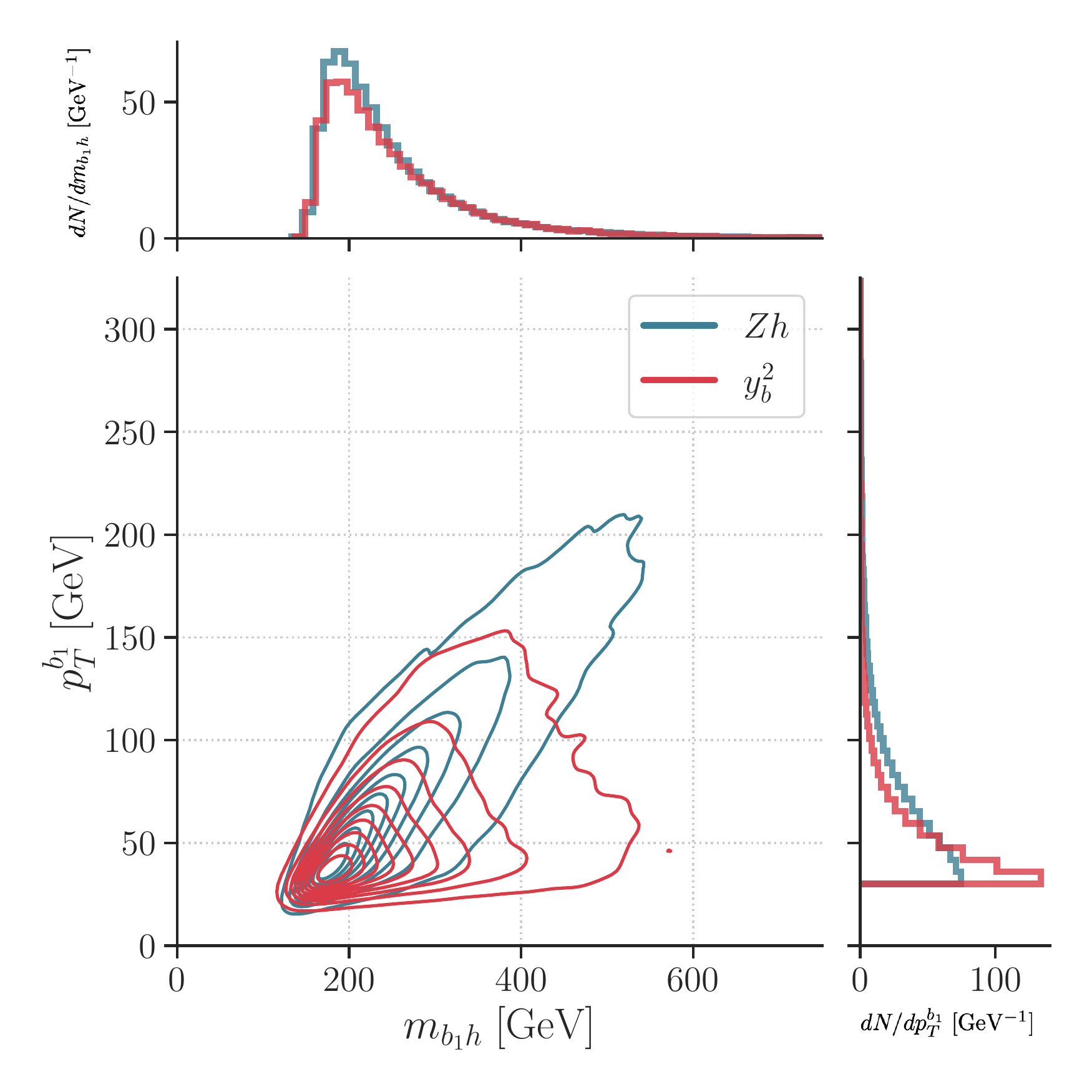}\\
    \includegraphics[trim= 10 10 10 20, clip, width=0.4\textwidth]{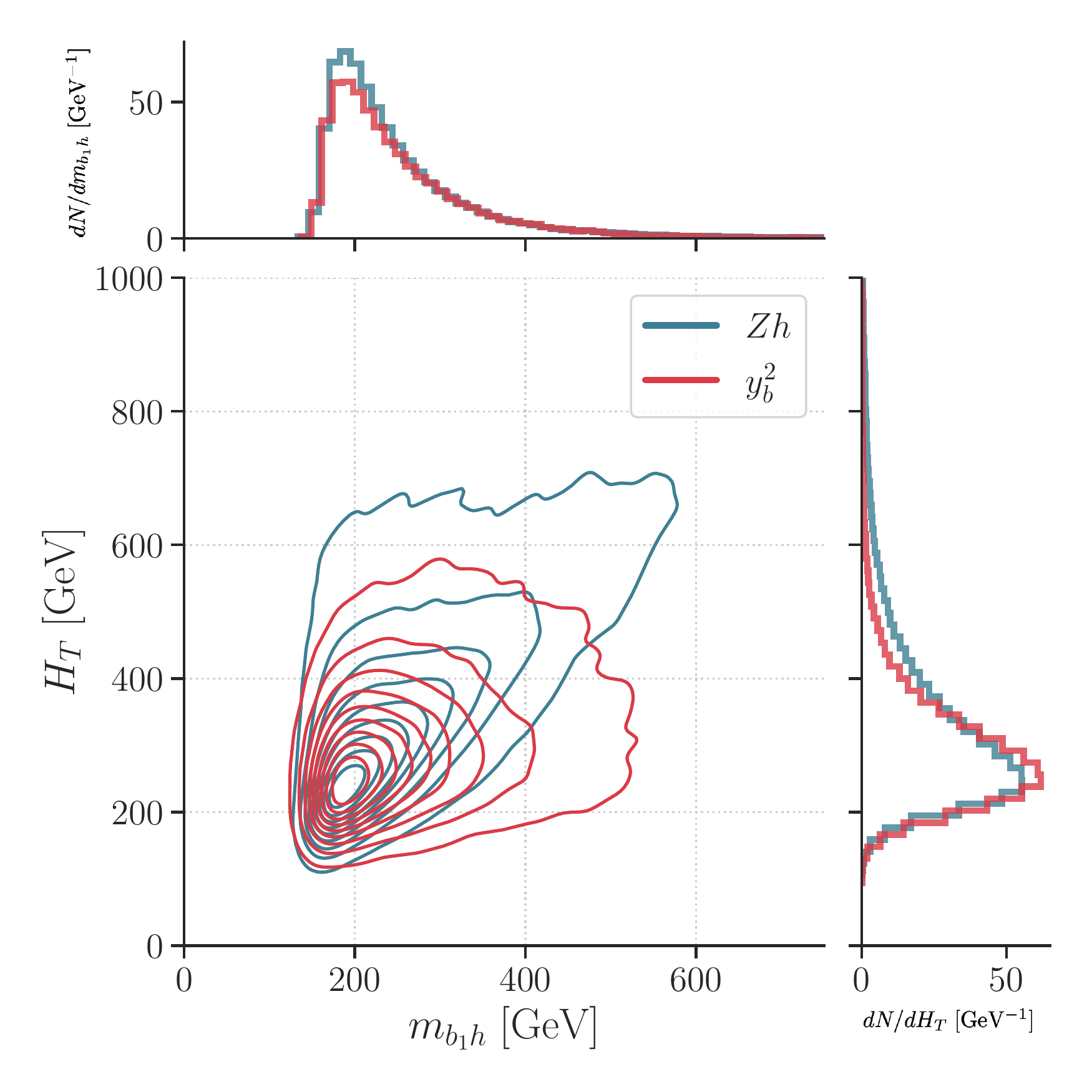}
    \includegraphics[trim= 10 10 10 20, clip, width=0.4\textwidth]{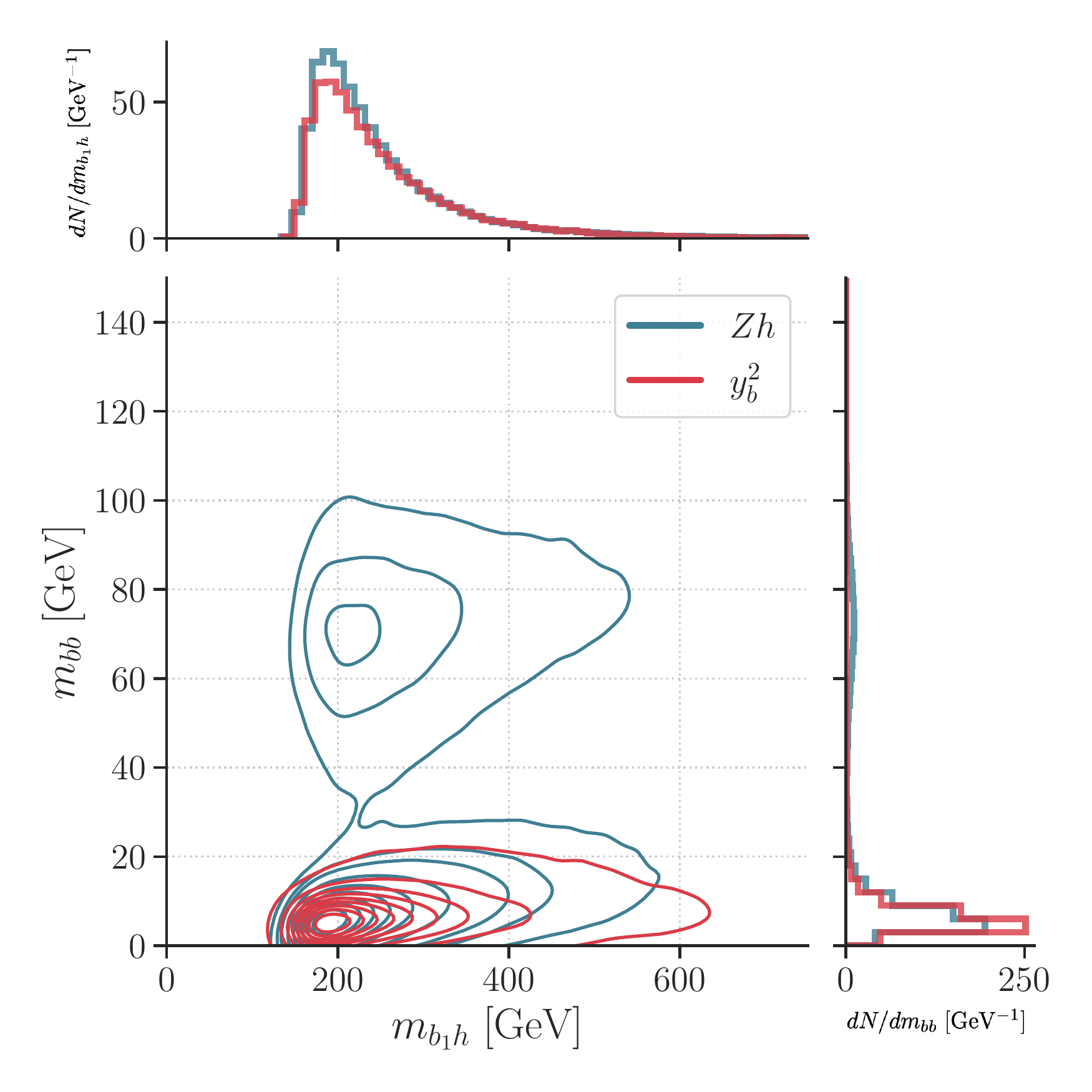}
    \caption{\it Joint \textbf{non-normalized} 1D and 2D plots, with SM signal, representing how kinematic shapes allow the BDT to discriminate between the $y_b^2$ and $Zh$ contributions at HL-LHC. While 1D distributions show that $m_{b_1h}$ by itself cannot be used to discriminate between these two contributions, its correlation with other kinematic variables creates shapes that a BDT can use to discriminate between the two contributions.}
    \label{fig:shape-zh-yb2}
\end{figure}

\begin{figure}
    \centering
    \figuretitle{\large\texttt{HL-LHC}}
    \includegraphics[width=0.53\textwidth]{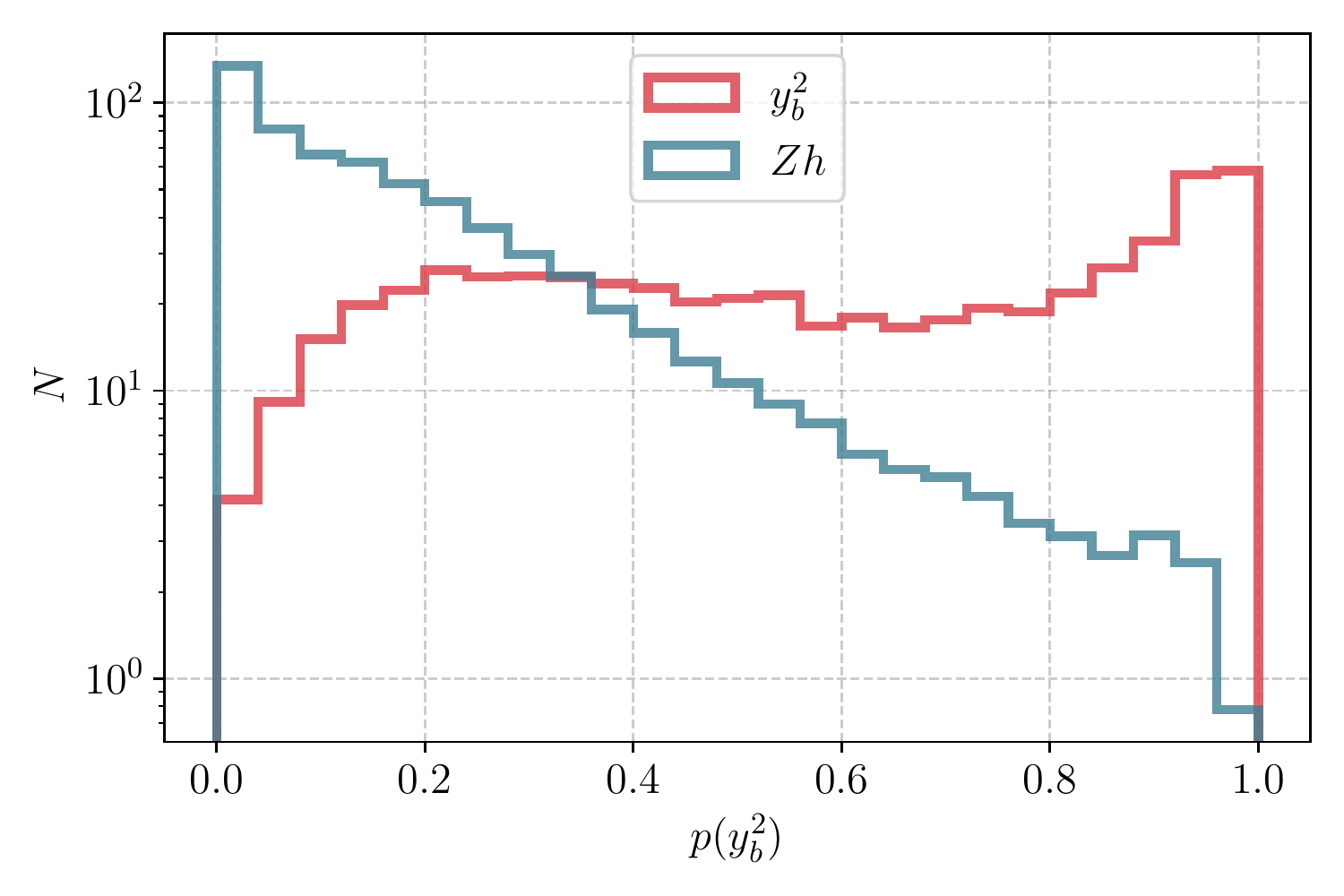}
    \caption{\it The clear discrimination between the channels proportional to $y_b^2$ and the $Zh$ channel from a multivariate BDT analysis with SM signal and 6 ab$^{-1}$ total luminosity assumed at HL-LHC.}
    \label{fig:zh-yb2-BDT}
\end{figure}

{To further understand the importance of kinematic shapes, especially in light of their relevance to the importance of a variable like $m_{b_1h}$, which would by itself be of no use in discriminating the two channels, we look at \autoref{fig:shape-zh-yb2}. In this figure, we plot the 2D distributions of $m_{b_1h}$ along with the other four of the five top variables in the $\overline{|S_v|}$ ranking. It is evident from these density plots that even when $m_{b_1h}$ does not seem like a variable worth mentioning from the 1D distributions, it becomes important in a shape based multivariate analysis. This aptly displays why a multivariate analysis performs much better in discriminating between kinematic shapes than a traditional cut-based analysis. In \autoref{fig:zh-yb2-BDT}, we show the separation of the channel proportional to $y_b^2$ and $Zh$ in the signal probability    \unskip\parfillskip 0pt \par}

\FloatBarrier

\noindent space with the former channel being the signal and the latter channel the background. The possibility to separate the two channels is indisputable. 

Similarly, we evaluate the separation between the contributions proportional to $y_b^2$ and $y_t^2$. The analysis is shown in \autoref{app:ybyt}. The separation is not as good as the previous example. Nevertheless, judging by $\overline{|S_v|}$, we find that the leading distinguishing variable is $p_T^{\gamma\gamma}$ or $p_T^h$. This comes from the requirement of a relatively hard gluon that kicks the Higgs to pick up a transverse momentum. Whereas in the case of the contribution proportional to $y_b^2$, the two bottom quarks tend to be back to back leading to a softer Higgs $p_T$. The second most important variable is $m_{b\bar b}$ where the contribution proportional to $y_t^2$ peaks towards zero and the one proportional to $y_b^2$ has a much larger $m_{b\bar b}$ value for the small percentage of events where both $b$-jets are tagged. As will be shown in the next section where we discuss the details of all the background channels at HL-LHC and FCC-hh, the dominant QCD-QED $\bbaa$ background and the irreducible contribution proportional to $y_t^2$ will still be the limiting factors to achieving sensitivity to $y_b$.

In what follows we will use the multivariate methods we discussed in this section to extract the $\bbh$ signal at the HL-LHC and FCC-hh. More importantly, we will study the possibility of isolating the contribution proportional to $y_b^2$ and $y_by_t$ to show that its a viable enterprise at the future hadron colliders.

\section{The \texorpdfstring{$\bbh$}{bbh} channel at future hadron colliders}
\label{sec:hadronC}

Having sorted out the irreducible background, let us focus on the complete analysis including the QCD-QED background. We focus on all five channels in this section and consider the signal channels as those driven by $y_b^2$ and $y_by_t$ terms since our goal is to probe $y_b$. In addition, we include the background channels driven by contributions proportional to $y_t^2$, from $Zh$ and $\bbaa$ QCD-QED production. We train a multi-class BDT with simulated data using an over-complete set of high-level variables. Then, using $\overline{|S_v|}$, we filter out the variables that are the most important without compromising the accuracy of the BDT. Finally, using these variables, we retrain the BDT to project out the possibility of measuring $y_b$ at the HL-LHC with 6 $\iab$ of total luminosity and at the FCC-hh with 30 $\iab$ of luminosity. In addition, with the help of $\overline{|S_v|}$ we point out the kinematic measurables that are the most important in these analyses to gain physics insights.

The interference term proportional to $y_by_t$ has to be dealt with considerable care. Being an interference term, it has a propensity for providing a negative contribution to the total cross-section as can be seen from \autoref{tab:xsec14}. This means that the weight of every event needs to be kept track of throughout the analysis and correctly used when projecting the significance of measuring $y_b$ from this contribution. It must be kept in mind that the BDT is agnostic to the nature of physics encoded in the simulation and is only sensitive to the kinematic shapes. Therefore, classifying a signal with a negative cross-section is not a hurdle for the BDT but rather the prediction made after the training of the BDT must be convoluted with the correct weights.

The simplicity of a binary classification is lost in a multi-class BDT classification like the one we will now describe. The prediction of a BDT comes in the form of a $n$-tuple of probabilities of an event belonging to each of the $n$ classes with the element of the $n$-tuple adding up to unity. An event is said to belong to the class corresponding the highest element in the $n$-tuple. So a plot like \autoref{fig:zh-yb2-BDT} is not very informative. To understand the results of the multi-class BDT analysis, we construct a confusion matrix. This is a $n\times n$ matrix, for $n$ classes. The sum of the elements in the $i^{th}$ row, $\sum_j N_{ij}$, gives the actual number of events of class $i$ that would be generated in a pseudo-experiment corresponding to the collider under study. The sum of the $j^{th}$ column, $\sum_i N_{ij}$, gives the number of events of the class $j$ predicted (including correct classifications and misclassifications) by the BDT in this pseudo-experiment. Hence the $(i,j)$ element of the matrix gives the number of events of the $i^{th}$ class that is classified as belonging to the $j^{th}$ class with $i\ne j$ representing a misclassification. The significance of the $j^{th}$ channel given by $signal/\sqrt{signal + background}$ can be defined as:
\begin{equation}
    \mathcal{Z}_j=\frac{|N_{jj}|}{\sqrt{\sum_i N_{ij}}},
\end{equation}
where $i$ is the row index and $j$ is the column index. Armed with these procedures and definitions we move on to examine the prospects of measurement of $y_b$ at the future colliders.

\begin{figure}[t!]
    \centering
    \figuretitle{\large\texttt{HL-LHC}}
    \includegraphics[width=5cm, height=5.1cm]{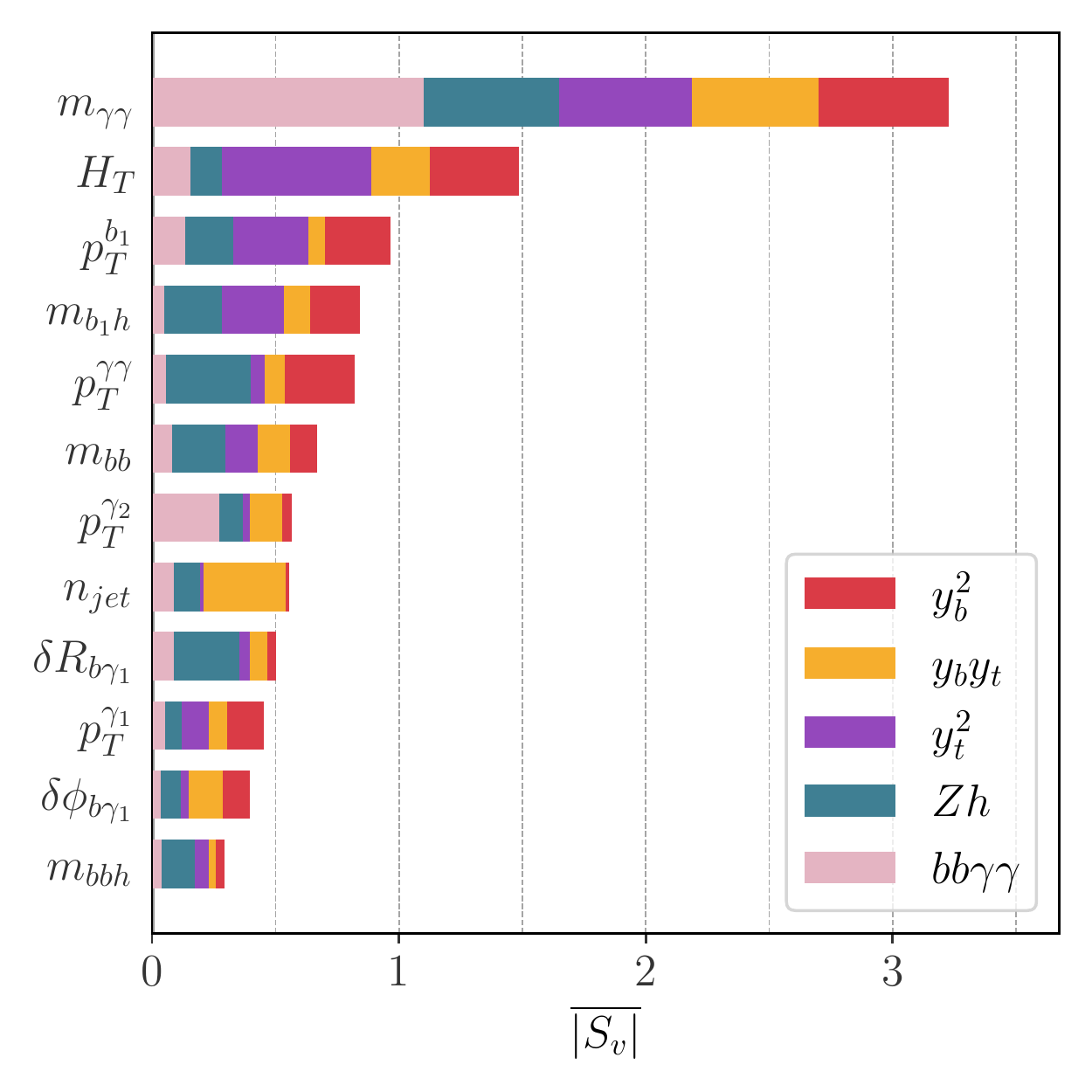}
    \includegraphics[trim=10 0 10 10,clip,width=0.32\textwidth]{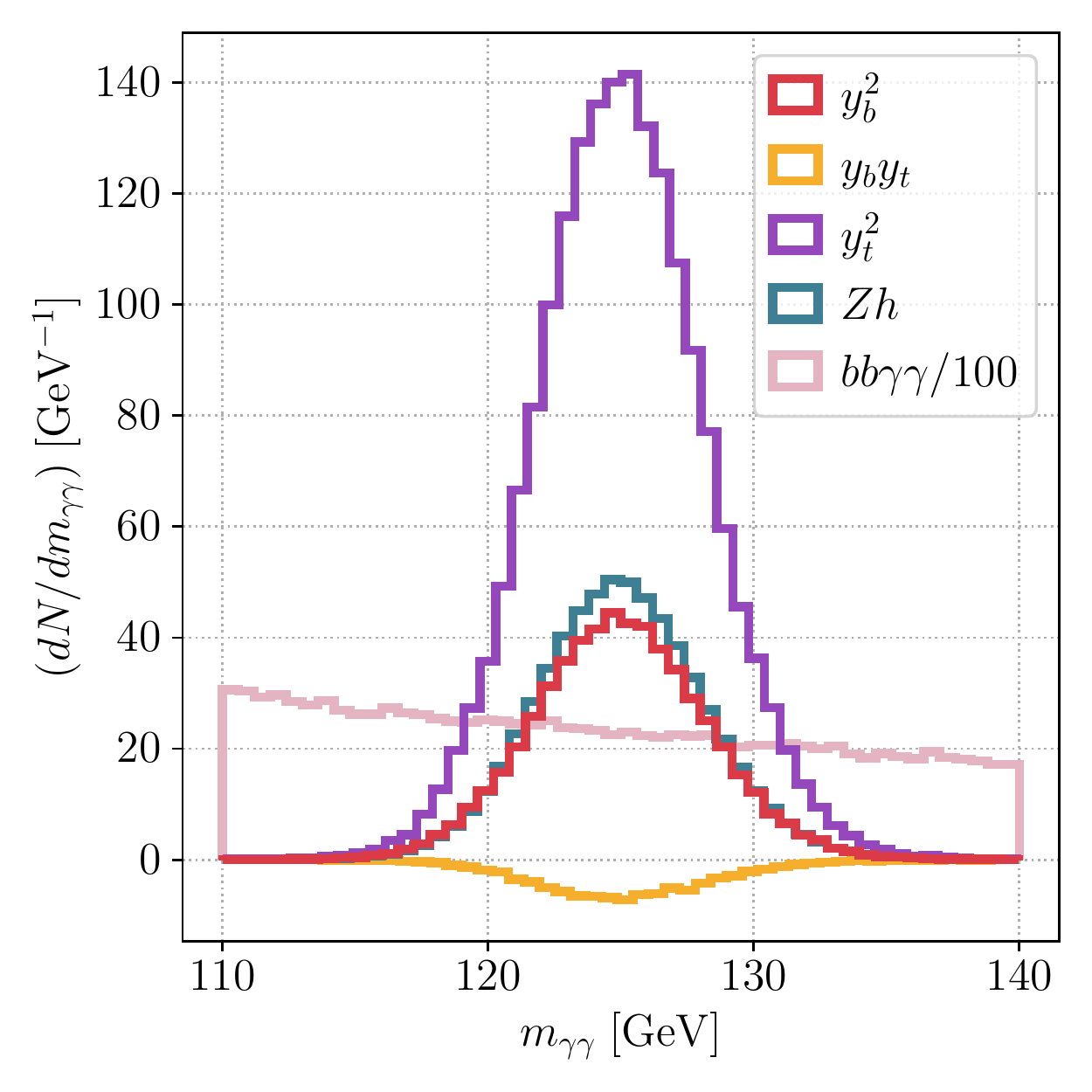}
    \includegraphics[trim=10 0 10 10,clip,width=0.32\textwidth]{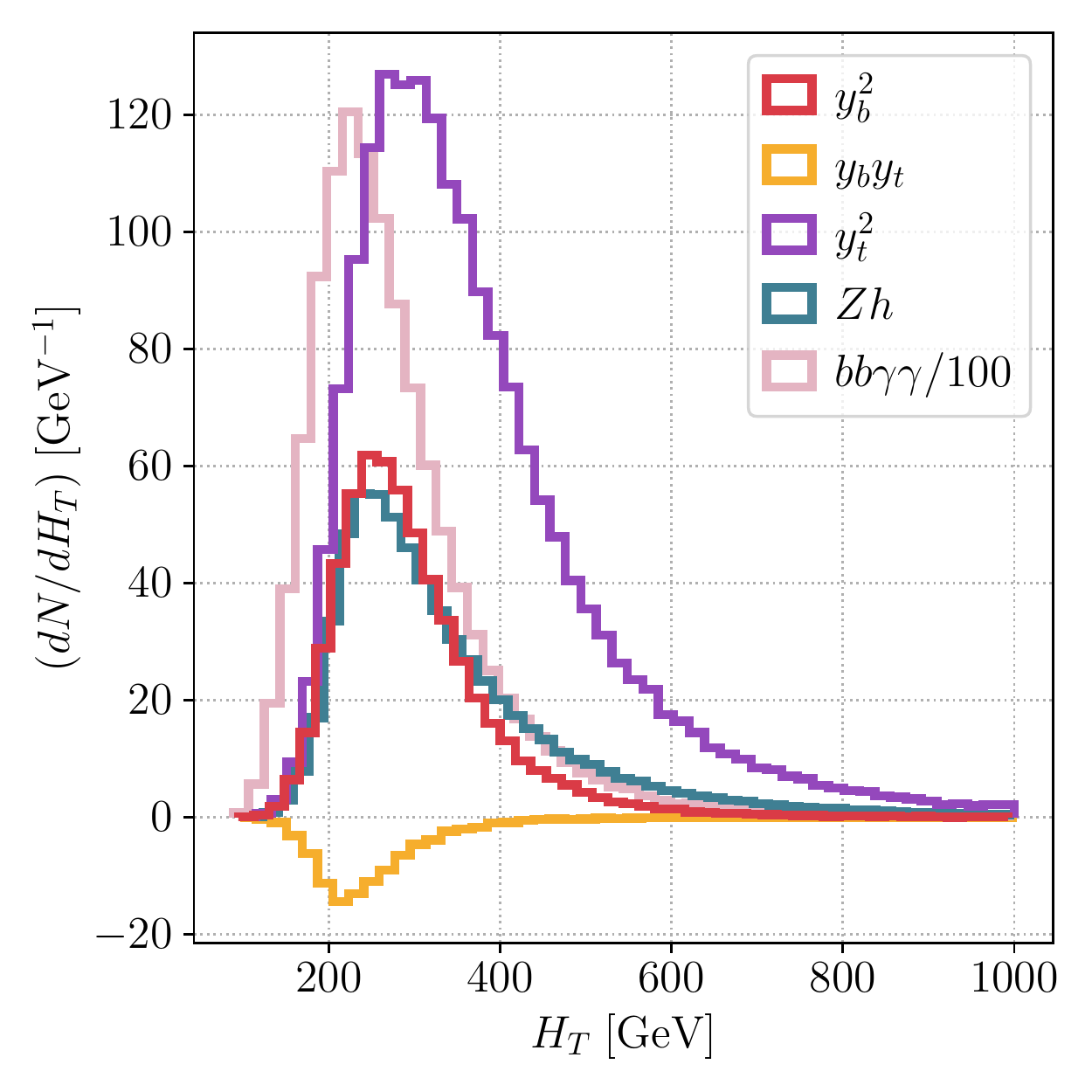}\\
    \includegraphics[trim=10 0 10 10,clip,width=0.32\textwidth]{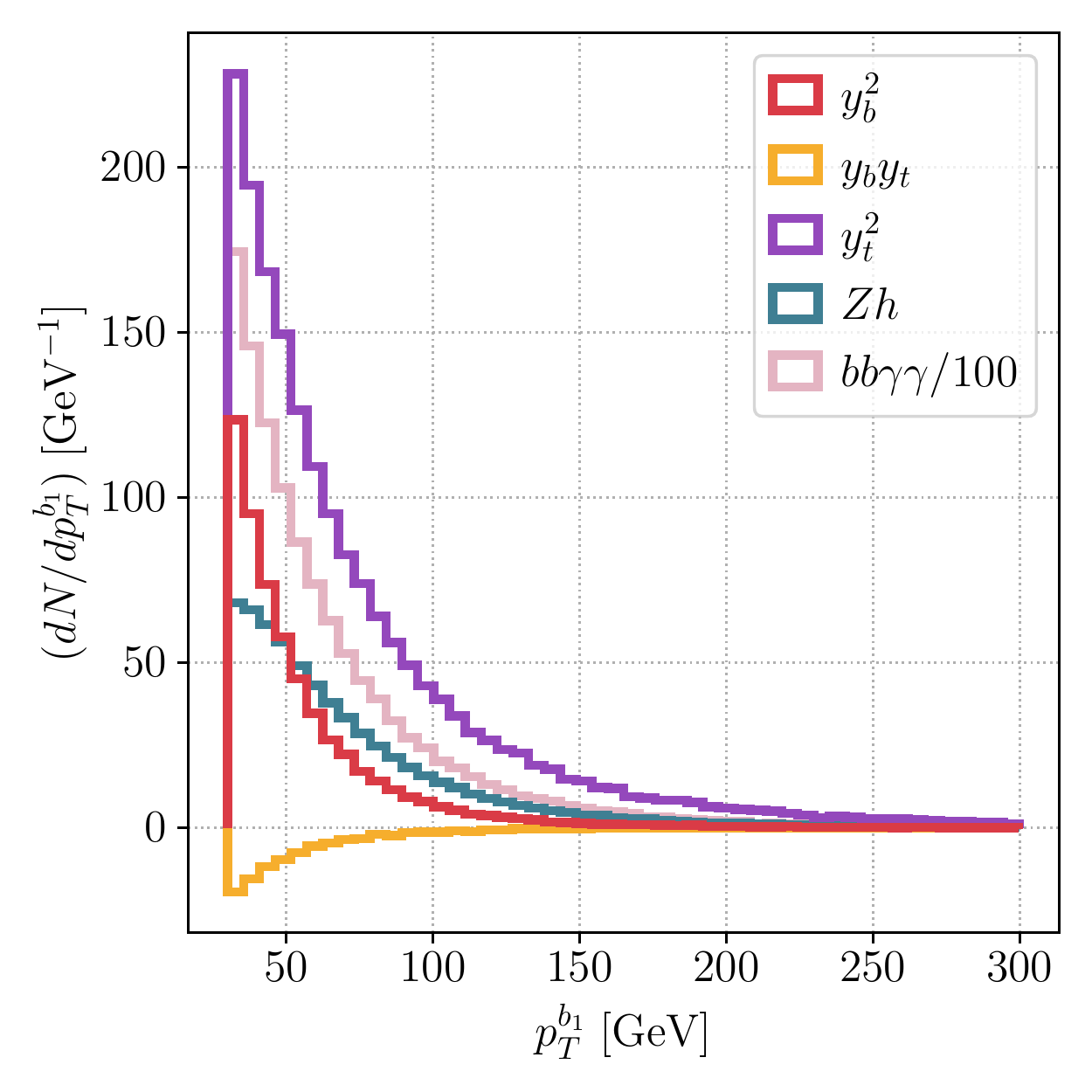}
    \includegraphics[trim=10 0 10 10,clip,width=0.32\textwidth]{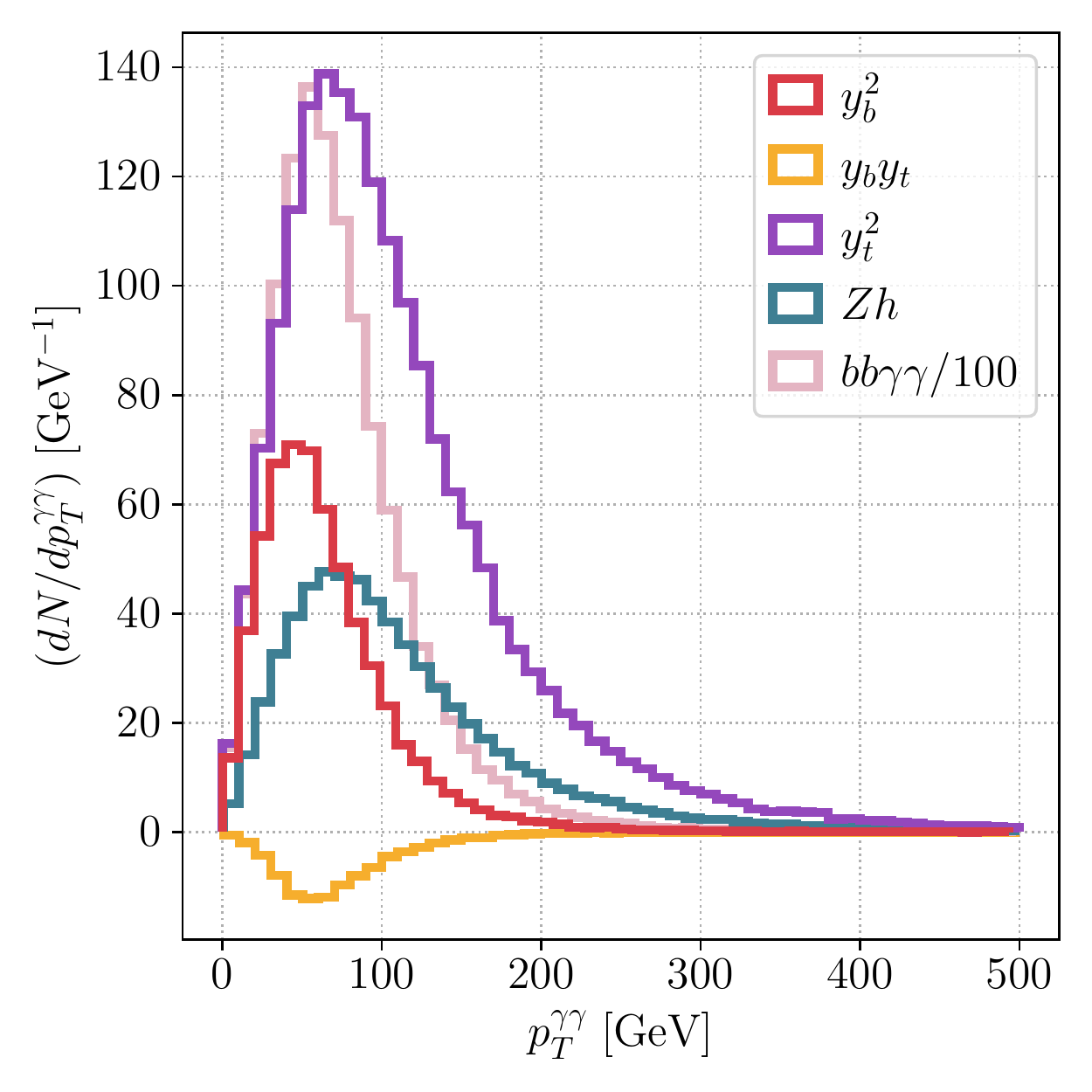}
    \includegraphics[trim=10 0 10 10,clip,width=0.32\textwidth]{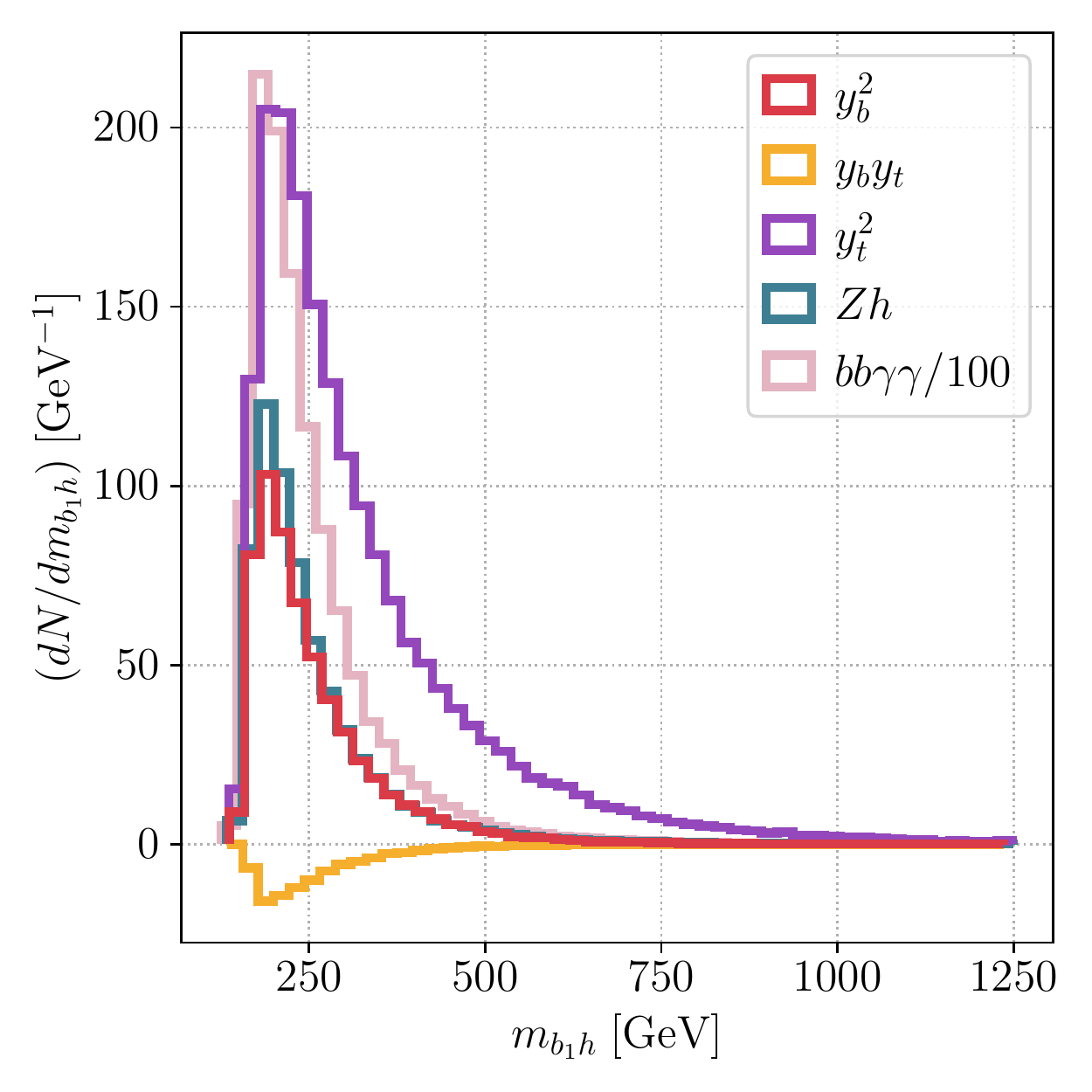}
    \caption{\it The Shapley values and differential distributions for the kinematic variables used to extract the $\bbh$ signal at HL-LHC with 6 ab$^{-1}$ luminosity (ATLAS+CMS). A SM signal is injected. \underline{Upper left corner} The hierarchy of variable importance with increasing Shapley value denoting higher importance in the discrimination. The non-normalized distribution are those of the 5 most important kinematic variables as determined by their Shapley values. The distribution for $b\bar{b}\gamma\gamma$ has been reduced by a factor of 100.}
    \label{fig:dist14}
\end{figure}

\subsection{Prospects at HL-LHC}
\label{sec:HLLHC}

The results for $\overline{|S_v|}$ from the BDT analysis are shown in the top left corner plot of \autoref{fig:dist14}. Note that the ordering of the kinematic variables according to their importance in discriminating the channels have changed from the previous example in \autoref{fig:zh-yb2}. This is not unexpected since the classes being discriminated determine what kinematic shapes are important. Secondly, unlike the $\overline{|S_v|}$ for the case of $Zh$ vs. $y_b^2$, different channels have different $\overline{|S_v|}$ for each kinematic variable. From this, one can understand the importance of each kinematic variable in discriminating each channel from the others. 

For example, $m_{\gamma\gamma}$ has the largest discriminating power for separating out the $\bbaa$ background (pink part) from the $y_b^2$-driven channel (red part) as can be seen from the $\overline{|S_v|}$ plot in \autoref{fig:dist14}. The blue, purple and yellow parts corresponding to $Zh$, $y_t^2$-driven and $y_by_t$-driven channels respectively are of the same length as the red part of the bar corresponding to the $y_b^2$-driven channel as they should be because they all have the same shape as is clear for the upper middle plot of \autoref{fig:dist14}. The distribution of $\bbaa$ is flat in $m_{\gamma\gamma}$, while the others are peaked around $m_h$ since the $\gamma$ pair comes from a Higgs decay. On the other hand the separation of the $y_t^2$-driven channel from the $y_b^2$-driven channel is orchestrated by $H_T$, which is second in the hierarchy of kinematic variables as judged by $\overline{|S_v|}$.

The matrix shown in \autoref{tab:HL-LHC-confusion} shows the confusion matrix produced from a pseudo-experiment for HL-LHC assuming 6 ab$^{-1}$ of data. The right-most column gives the actual number of events produced in each channel. The bottom-most row gives the significance of each channel. Note that this is purely a statistical significance and no systematics has been rolled into this estimate. We leave the discussion of systematics in \autoref{sec:errors}. Following these procedures, we get a resulting statistical significance of the $y_b^2$-driven channel of 3.33$\sigma$. The sensitivity for $y_by_t$ channel is 0.47$\sigma$.
\begin{table}[]
    \centering
    \begin{tabular}{ll|rrrrr|r}
    \multirow{7}{*}{\rb{\bf Actual no. of events\hspace{0.6cm}}} & \multicolumn{7}{c}{\bf Predicted no. of events at HL-LHC}\\
    \cmidrule[\heavyrulewidth]{2-8}
    &Channel	        &      $y_b^2$ &     $y_by_t$ &      $y_t^2$ &        $Zh$ &  $bb\gamma\gamma$ &     total \\
    \cline{2-8}
    &$y_b^2$          &   170 &    54 &    51 &   122 &        189 &     586 \\
    &$y_by_t$         &    -7 &   -24 &    -4 &   -20 &        -40 &     -95 \\
    &$y_t^2$          &   238 &   112 &   452 &   546 &        487 &    1,835 \\
    &$Zh$             &    22 &    28 &    21 &   416 &        161 &     648 \\
    &$bb\gamma\gamma$ & 2,183 & 2,450 &  1510  & 8,045 &    101,591 &  115,779 \\
    \cline{2-8}
    &$\mathcal{Z}_j$  &  3.33 &  0.47 &  10.  &  4.36 &        317 &         \\
    \cmidrule[\heavyrulewidth]{2-8}
    \end{tabular}
    \caption{\it Trained BDT classification (confusion matrix) of the five channel contributions at HL-LHC with 6 ab$^{-1}$ luminosity (ATLAS+CMS), assuming SM signal injection. The right-most column gives the total number of events expected in each channel in the SM.}
    \label{tab:HL-LHC-confusion}
\end{table}
We shall see in \autoref{sec:kappab} how the projection from this analysis translates into bounds on the effective rescaling of $y_b$.

\subsection{Prospects at FCC-hh}
\label{sec:100tev}

Isolating the $y_b$-sensitive channels at the FCC-hh is a much easier task not only due to the much larger luminosity but also because of the disproportionately larger growth of $\bbh$ production compared to $Zh$ production as explained in \autoref{sec:Sim}. However, keeping in mind that a multivariate analysis will still outperform a cut-based analysis, we use the combination of BDT and $S_v$ to fathom the prospects of measuring $y_b$ at FCC-hh. Since the importance of the irreducible $Zh$ background falls, and the only other backgrounds that need to be discriminated from are the contributions proportional to $y_t^2$ and the QCD-QED $\bbaa$ background, an analysis with $\overline{|S_v|}$ rightly shows that $m_{\gamma\gamma}$ is by far the most important kinematic variable in discriminating the $y_b^2$ signal from the dominant background. The results of the BDT analysis is given in \autoref{tab:FCC-confusion}. Noting that the significance quoted in the table pertains to those obtained from assuming the presence of statistical uncertainties only, we see that the contribution proportional to $y_b^2$ can be isolated at about 64$\sigma$ significance while the interference term proportional to $y_by_t$ can be isolated to about 10$\sigma$ significance. 
\begin{table}[]
    \centering
    \begin{tabular}{ll|rrrrr|r}
    \multirow{7}{*}{\rb{\bf Actual no. of events\hspace{0.6cm}}} & \multicolumn{7}{c}{\bf Predicted no. of events at FCC-hh}\\
    \cmidrule[\heavyrulewidth]{2-8}
    &{} &        $y_b^2$ &       $y_by_t$ &        $y_t^2$ &           $Zh$ &  $bb\gamma\gamma$ &      total \\
    \cline{2-8}
    &$y_b^2$          &   32,074 &   15,112  &   10,966  &    6,579  &      8,959    &    73,690 \\
    &$y_by_t$         &     -964 &   -6,815  &     -907  &     -583  &     -1,820    &   -11,089 \\
    &$y_t^2$          &   48,772 &   45,751  &  148,669  &   39,598  &      26,484   &   309,274 \\
    &$Zh$             &    1,860 &    4,498  &    2,280  &   12,661  &      2,282    &    23,581 \\
    &$bb\gamma\gamma$ &  172,088 &  373,436  &  106,335  &  126,429  &     7,952,834 & 8,731,122 \\
    \cline{2-8}
    &$\mathcal{Z}_j$ &      63.7 &     10.4 &      288 &     29.4 &      2,813    &          \\
    \cmidrule[\heavyrulewidth]{2-8}
    \end{tabular}
    \caption{\it Trained BDT classification (confusion matrix) of the five channel contributions at FCC-hh. A SM signal is injected. The right-most column gives the total number of events expected in each channel in the SM.}
    \label{tab:FCC-confusion}
\end{table}

\subsection{Discussion of theoretical and systematic uncertainties}
\label{sec:errors}

As discussed above, compared to the $Zh$ channel and the QCD-QED $\bbaa$ backgrounds, the $\bbh$ channels gain a relative enhancement on statistics at the FCC-hh. This results in a much better sensitivity to the measurement of $y_b$ that scales roughly with the square-root of statistical gain. Till now, we have not included any experimental or theory systematic uncertainties, which might not be negligible at HL-LHC or FCC-hh. The current theory uncertainty from fixed order NLO calculation ranges from $20\%$ to $50\%$ for the channels proportional to $y_b^2$, $y_by_t$, $y_t^2$~\cite{Deutschmann:2018avk}, and are the dominant sources of uncertainty. No projection of the errors at future colliders is available, but, extrapolating from the HL-LHC analyses~\cite{Cepeda:2019klc}, we would expect that in the future similar or smaller theory uncertainty as compared with statistical uncertainty would be reached. 

As for the experimental systematic uncertainties, it is expected to be important especially when low statistical errors are reached, like at the FCC-hh. We thus include an approximate estimate of the significances including experimental systematic at the level of 0.5\%, 1\% and 5\% stemming from the dominant background $\bbaa$. In practice we take the number of events classified as $\bbaa$ in the confusion matrix, multiplied by the different systematic error estimate, and added in quadrature to the statistical error, the latter being the square-root of the total signal and background events. The results are shown in \autoref{tab:systematics}. We see that even $\mathcal{O}(1\%)$ experimental systematic uncertainty from the dominant background reduces the significance of the signal considerably. Furthermore, the errors in the $\bbh$ measurements are mostly statistics dominated at the HL-LHC but can easily get systematics dominated at the FCC-hh. Thus, a good control over the systematic uncertainties will be necessary to draw the full strength of the statistics allowed at the FCC-hh.

\begin{table}[]
    \centering
\begin{tabular}{|l|cc|cc|}
\hline
\multirow{2}{*}{systematics} & \multicolumn{2}{c|}{HL-LHC (6 ab$^{-1}$)} & \multicolumn{2}{c|}{FCC-hh (30 ab$^{-1}$)}\\
\cline{2-5}
            &      $y_b^2$  &     $y_by_t$  &     $y_b^2$ & $y_by_t$  \\
\hline
0\%         &  3.33         &   0.47       & 63.7        & 10.4     \\
0.5\%       &  3.26         &   0.46       & 32.2        & 3.44      \\
1\%       &  3.06         &   0.42       & 17.9        & 1.80      \\
5\%       &  1.41         &   0.18       & 3.72        & 0.36      \\

\hline
\end{tabular}
    \caption{\it Significance from $\bbh$ analysis including systematics estimates at HL-LHC and FCC-hh.}
    \label{tab:systematics}
\end{table}

\section{Constraints on the effective rescaling of \texorpdfstring{$y_b$}{yb}}
\label{sec:YukConst}

In this section we will take a look at the rescaling of the effective Higgs couplings to the bottom quark. To account for possible CP violation in the Yukawa sector, it is natural to go to the more general complex Yukawa assumption and constrain the 2D space of the complex Yukawa. We will address the bounds on the CP-even coupling, $\kappa_b$, and its CP odd counterpart, $\tilde{\kappa}_b$ or equivalently, the magnitude, $|\kappa_b|$, and the phase $\phi_b$ of the complex coupling. The bounds we explore from $\bbh$ come indirectly from the combined contribution of CP-even and odd effects in the inclusive (interference) contribution. \footnote{Its worth mentioning here that model parameters can also be extracted using the Fisher Information that sets a lower bound to the covariance matrix of the parameters according to the Fr\'echet-Cram\'er-Rao bound~\cite{Diehl:1993br}. This concept has been revived recently in Ref.~\cite{Brehmer:2016nyr} with later works using machine learning to learn the likelihood function. Since we build the machine learning classifier in a way that the $y_b$ (or $\kappa_b$) dependence can be isolated, a simple rescaling allows us to set bounds on the parameter within certain assumptions (c.f. Footnote~9).}
Exclusive CP-odd observables give more direct and model-independent constraints on the CP violation for various couplings. 

In the $\kappa$-framework, the indirect bounds are currently more stringent compared to bounds from exclusive measurements such as those for $y_t$, when comparing bounds from inclusive Higgs rate with bounds from $\tth$ kinematic observable studies~\cite{Bahl:2020wee,Bortolato:2020zcg,Cao:2020hhb,Ma:2018ott,Hou:2018uvr,Goncalves:2018agy,Mileo:2016mxg,Gritsan:2016hjl,AmorDosSantos:2017ayi,Demartin:2014fia}. Exclusive CP measurements will benefit greatly from more statistics at HL-LHC and especially at FCC-hh. In comparison, the sensitivity to the CP-violating phase in $y_b$ (or strictly speaking, the relative phase between $y_b$ and $g^{\rm eff}_{ggh}$) is derived from the sensitivity to the $y_b y_t$ channel in the $b\bar bh$ production process. It is in the same spirit as for the sensitivity to the CP-phase of $y_t$ from inclusive $t h$~\cite{Demartin:2015uha} production and the $th+V$~\cite{Demartin:2016axk} process, where sizeable and measurable interference between $y_t$- and electroweak coupling $g_V$-driven diagrams are expected at future colliders making it possible to probe the relative phase. Separately, there are exclusive studies of CP-sensitive observable in the $t\bar th$ channel, where the decay topology of the massive top quark provides additional information on the top quark spin. Hence, CP-sensitive observable can be constructed from the decay final states. The equivalent of these CP-sensitive observable are lacking in the $b\bar bh$ process since the decay topology is buried in the $b$-jet. Constructing observable which involve both jet-substructure and the rest of the event shape might be helpful and worth exploring in future works. Hence, an exclusive study of $\bbh$ with $h\to 4\ell, \gamma\gamma$  to directly probe possible CP-violating effects in the process is left for a future work. Assuming NP effects in $y_b$, we can interpret experimental constraints on the complex rescaling of $y_b$ including our $\bbh$ study. A global fit with a relevant set of EFT operators including corrections order by order, or interpretation of a motivated NP scenario is left for a future work too.

\subsection{Constraints on a real bottom-quark Yukawa \texorpdfstring{$\kappa_b$}{kappab}}
\label{sec:kappab}

Let us first take a look at a real rescaling of $y_b$. In the $\kappa$-framework, modifying only the Higgs coupling to the bottom quarks, we have
\begin{equation}
     \mathcal{L}\supset -\kappa_b \frac{m_b}{v}  \bar b b  h,
\end{equation}
where $m_b^{\rm pole} = 4.58\,\gev$ or $m_b^{\rm \overline{MS}}(m_b^{\rm \overline{MS}}) = 4.18\,\gev$, depending on the choice of scheme, is the bottom quark mass and $v$ is the vacuum expectation value of the Higgs field normalized to 246 GeV. The rest of the couplings in the SM Lagrangian are assumed to remain unmodified. The results from the previous section can now be used to put bounds on $\kappa_b$. The contribution proportional to $y_b^2$ scales at $\kappa_b^2$, while the interference term scales with $\kappa_b$ and $\kappa_g$, the effective coupling for $\ggh$. Here we need to assume that the additional gluon in the $y_t^2$-driven contribution is not too far off-shell and the $\ggh$ is effectively energy independent\footnote{The interference term is thus proportional to $\kappa_b (1.05-0.05\kappa_b)$ which can be derived using the SM parameters as defined in \autoref{sec:Sim} in \autoref{eq:kgkb} of \autoref{app:kgkgamma}. A small quadratic piece from the inclusion of bottom-quark effects in the $\ggh$ coupling is present.}. This assumption also makes it independent of the center-of-mass energy. The rescaling of a general CP-violating $\ggh$ coupling, $\kappa_g$, as function of a CP-violating complex $\kappa_b$ is given in \autoref{app:kgkgamma} and applies for both HL-LHC and FCC-hh. From that, one can reduce to the real $\kappa_b$ case.

\begin{figure}
    \centering
    \includegraphics[width=0.49\textwidth]{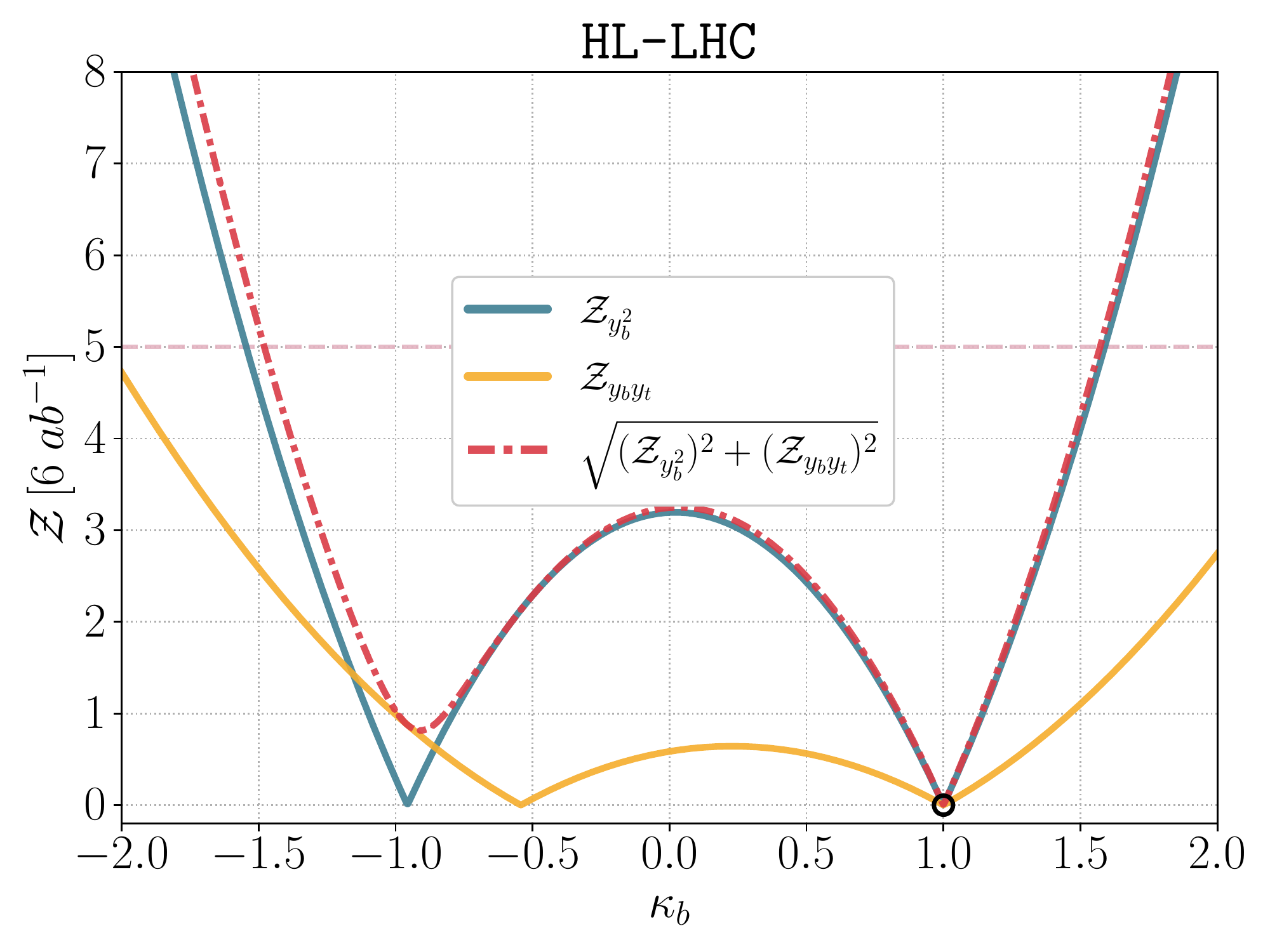}
    \includegraphics[width=0.49\textwidth]{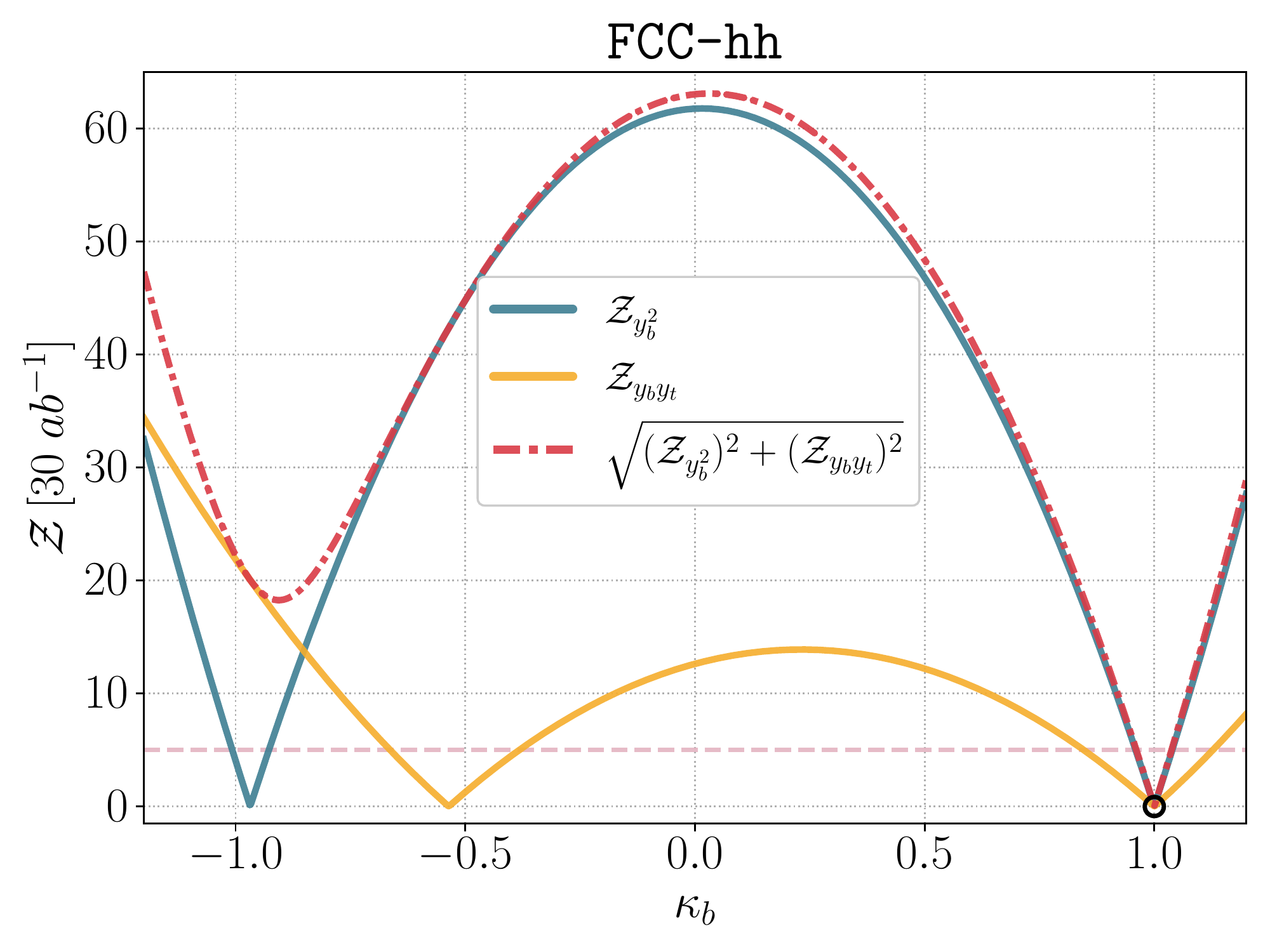}
    \caption{\it Significance, $\mathcal{Z}$, as a function of $\kappa_b$ at HL-LHC (ATLAS+CMS combined, 6 ab$^{-1}$) and FCC-hh (30 ab$^{-1}$). A SM signal is injected.}
    \label{fig:resbkappa}
\end{figure}

From the confusion matrices given in \autoref{tab:HL-LHC-confusion} for HL-LHC and \autoref{tab:FCC-confusion} for FCC-hh we can get the number of signal and background events for both the $y^2_b$-driven part and the interference term. This can be translated into a significance with which a certain value of $\kappa_b$ can be excluded by assuming a SM signal injection. The significance is given by:

\begin{eqnarray}
    \mathcal{Z}_j = \frac{1}{\sqrt{\sum_i N_{ij}}}\left| N_{1j}\kappa_b^2 + N_{2j}\kappa_b(1.05 - 0.05\kappa_b) - (N_{1j} + N_{2j})\right|,
    \label{eq:kappa_b_confusion}
\end{eqnarray}
for $j=1,2$ representing the $y_b^2$ and $y_by_t$ channels respectively in the confusion matrix. The index $i$ sums over all five channels with $i$ and $j$ being row and column indices respectively. The significance for $y_b^2$ and $y_by_t$ and their statistical combination as functions of $\kappa_b$ are shown in \autoref{fig:resbkappa}. From the term proportional to $\kappa_b^2$ there is a degeneracy in the determination of the sign of $\kappa_b$. This degeneracy is broken in the determination of $\kappa_b$ from the interference term. Hence, combining the two channels lifts the degeneracy in the determination of $\kappa_b$. While at the HL-LHC, the exclusion of the negative parameter space for $\kappa_b$ is not possible with high significance due to the limited sensitivity to the interference term, it is possible at FCC-hh with the negative solution being excluded at about 19$\sigma$. This translates to a $1\sigma$ confidence limits on $\kappa_b$ of [-0.99,-0.82] $\cup$ [0.84,1.14] at HL-LHC and [0.99,1.01] at FCC-hh. In comparison, the direct bound on $|\kappa_b|$ from $h\to b\bar{b}$ is 2.2\% at HL-LHC and 0.49\% at FCC-hh. However, the sign of $\kappa_b$ cannot be resolved with $h \to b\bar{b}$ measurements alone as discussed in discourse that follows.

\subsection{Constraints on a 2D CP-violating bottom-quark Yukawa \texorpdfstring{\{$\kappa_b,\tilde{\kappa}_b$\}}{kappab, kappabtilde} space}
\label{sed;twoparam}

Extending our analysis to a CP-violating rescaling of $y_b$ with \{$\kappa_b,\tilde{\kappa}_b$\}, we get an additional term in the SM Yukawa Lagrangian since in the SM $\tilde{\kappa}_b=0$:
\begin{equation}
    \mathcal{L}\supset -\frac{m_b}{v} (\kappa_b \bar b b + i \tilde{\kappa}_b \bar b \gamma_5 b ) h.
\end{equation}
The CP violating phase is thus $\phi_b = \arctan(\tilde{\kappa}_b)/(\kappa_b)$, with $|\kab|=1$ and $\phi_b=0$ being the SM prediction. All other couplings are fixed to their SM values in this analysis. The use of this simple $\{\kappa_b, \tilde{\kappa}_b\}$  framework has to be understood as a proxy to compare the sensitivity of the different channels and cannot be a replacement of a dedicated global analysis considering simultaneous deformations of the SM as expect in any realistic BSM scenario.

To translate the constraints from the studied $\bbh$ channels in this work, we replace $\kappa_b^2$ in the real $y_b$ scenario discussed in \autoref{eq:kappa_b_confusion} by $\kappa_b^2+\tilde{\kappa}_b^2$ \footnote{In principle, the $\kab$ and $\tkab$ contribution are different at the amplitude level and thus produce different kinematic shapes, and this naive extrapolation should not hold. In practice, the threshold behavior in squared-amplitudes of the CP-even and odd contributions differ by a factor of $\beta_b^2 = (1-4m_b^2/s)$, where $s\gtrsim m_h^2$ at threshold. The difference is thus suppressed by three orders of magnitude and negligible.}. The $y_b^2$-driven contribution scales as $\kappa_b^2+\tilde\kappa_b^2$ when including the non-zero CP-odd component. On the other hand, the $y_by_t$-driven amplitude scales linearly with $\kappa_b$. From \autoref{fig:feyndiag} it can be seen that the amplitudes proportional to $y_t$ generated by the top-quark loop, can also be generated by a bottom-quark loop. This can be done by expressing $\kappa_g$, the rescaling of the $\ggh$ coupling, as a function of $\kappa_b$. Hence, the interference term is proportional to $Re[\kappa_g]\kappa_b+Re[\tilde\kappa_g]\tkab$ and is given by:
\begin{equation}
    1.05 \kab - 0.05 \kab^2 -0.06 \tkab^2.
\end{equation}
Note that $\kab$ and $\tkab$ are real by definition. Additional details about the functional form can be found in \autoref{app:kgkgamma}. 

Since we shall attempt to draw a comparison between the constraints on the rescaling of $y_b$ from $\bbh$ production with other modes such as $h\to b\bar b$, $\ggh$ and $h\to\gamma\gamma$, it is worthwhile to give a brief overview of the current status of $h\to b\bar b$ measurements and the projected bounds from measurements of all the channels at HL-LHC and FCC-hh. The recently measured channel of $h \to b\bar b$ by ATLAS~\cite{Aaboud:2018zhk} and CMS~\cite{Sirunyan:2018kst} reports bounds on the signal strengths of:
\begin{eqnarray}
\mu_{h\to b\bar b}^{\textrm ATLAS} &=& 1.01 \pm 0.12^{+0.16}_{-0.15},\nonumber\\
\mu_{h\to b\bar b}^{\textrm CMS} &=& 1.04 \pm 0.14 \pm 0.14,
\end{eqnarray}
with the first and the second errors being statistical and systematic respectively. This translates to a bound on $\kappa_b$ of about 7\% from the combined experiments. We will see that the decay, $h\to b\bar b$, will remain the most sensitive measurement for the absolute value of $y_b$ at the HL-LHC and FCC-hh. 

Additionally, indirect constraints from various inclusive Higgs production and decay measurements are important for the determination of a CP-violating Yukawa component, especially from the large gluon-fusion production and the di-photon decay of the Higgs. For completeness, we show in \autoref{app:kgkgamma} the dependence of $\kappa_{g,\gamma}$ and $\tilde\kappa_{g,\gamma}$ on $\kappa_{b}$ and $\tilde\kappa_{b}$ after integrating out the one-loop quark contribution with the external legs on-shell. With these assumptions the projected constraints on $\kappa_g$ and $\kappa_\gamma$ from the inclusive production and decay rate can be directly mapped to bounds on $\kappa_b$.

The projected 1$\sigma$ bounds at HL-LHC (ATLAS+CMS combined, 6 ab$^{-1}$, including systematics) for $\kappa_g$ ($0.8\%$), $\kappa_\gamma$ ($1.3\%$) and $\kappa_b$ ($2.2\%$) are derived from the projected bounds on the production cross-section and decay branching ratios $\sigma_{ggh}$ ($1.6\%$), BR$(h\to\gamma\gamma)$ ($2.6\%$) and BR$(h\to b\bar b)$ ($4.4\%$) from Figs.~28-29 of the HL-LHC projections' study~\cite{Cepeda:2019klc}, where no sources of CPV are assumed. The CP-phase contribution to the total rate comes in through $|\kappa_{g,\gamma,b}|^2 + |\tilde\kappa_{g,\gamma,b}|^2$. This is true when the bounds on $\kappa_{g,\gamma,b}$ are dominated by inclusive Higgs measurements which have no additional Yukawa coupling or NP dependence other than the $\kappa$'s being constrained. 

For the $\bbh$ channel, we do not include systematics since no estimates are available other than the naive scaling we showed in \autoref{sec:errors}. It should be noted that while the error estimates for $\kappa_g$, $\kappa_\gamma$ and $\kappa_b$ from $h \to b\bar{b}$ are expected to be systematics dominated at HL-LHC, the error estimate for the $\bbh$ channel can be argued to be statistics dominated because of the low statistics for this channel at HL-LHC. This can also be seen from our naive estimate of systematics in \autoref{tab:systematics}. We proceed under this assumption to perform the combined fit.

At FCC-hh, the expected sensitivity from a global $\kappa$-fit to $\kappa_g$, $\kappa_\gamma$ and $\kappa_b$ are about $0.49\%$, $0.29\%$ and $0.43\%$, estimated, for example in Table.~3 of Ref.~\cite{deBlas:2019rxi} including experimental and theory uncertainties. This is somewhat different from what we use as projections for HL-LHC since the numbers are taken from a global $\kappa$-framework fit for FCC-hh rather than from the fits to the individual channels\footnote{The bounds given in Ref.~\cite{deBlas:2019rxi} are for FCC-ee/FCC-eh/FCC-hh. For the channels considered here, FCC-hh will, by far, dominates the combination.}. Given that the aim of this section is to make a quantitative comparison between the  channels widely advocated for constraining the size and phase for $y_b$, vis-\`a-vis what can possibly be added by $\bbh$, we feel this is a reasonable approach. As for HL-LHC, we do not include systematic errors for $\bbh$ measurement at FCC-hh for lack of clarity on such estimates.

\begin{figure}
    \centering
    \includegraphics[width=0.45\textwidth]{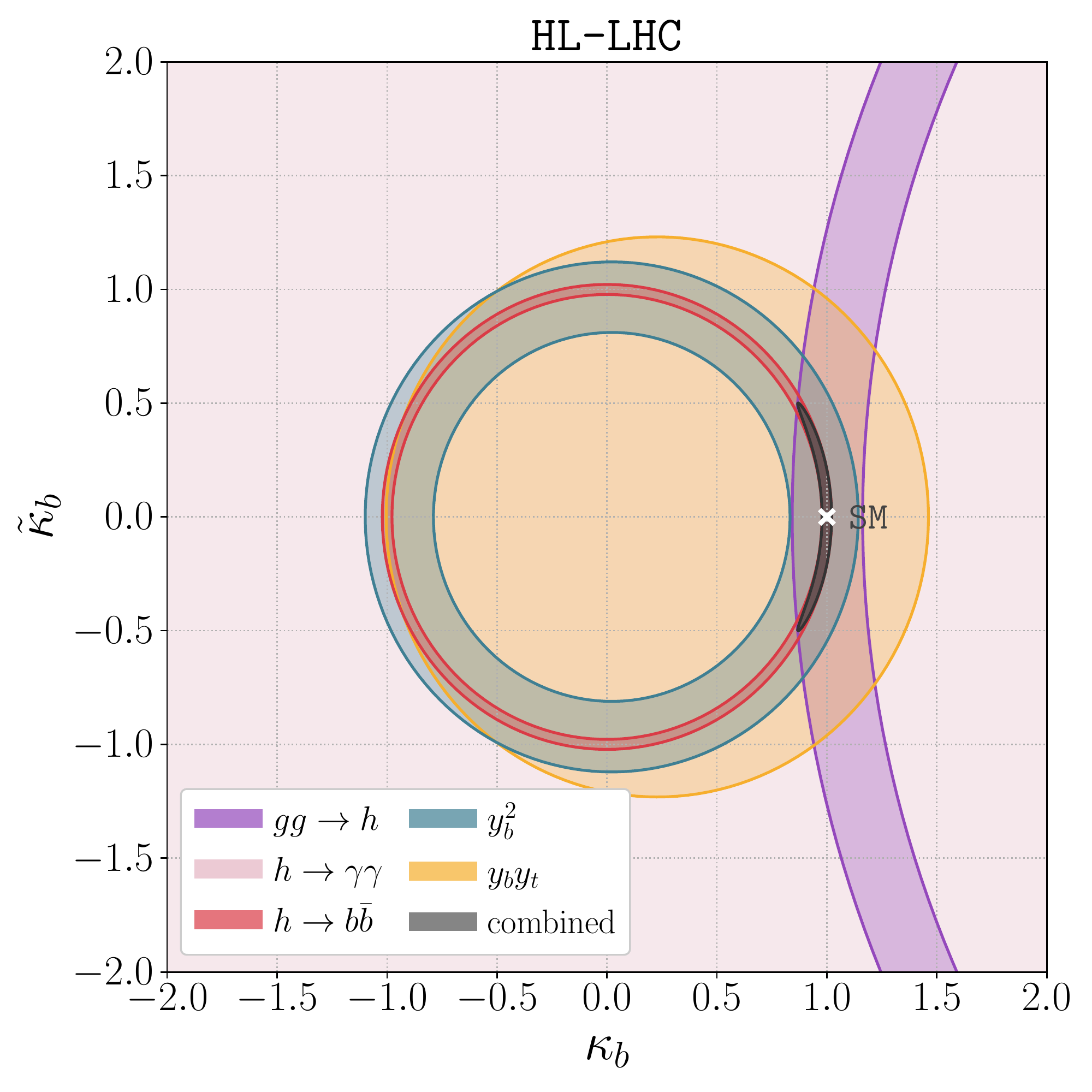}
    \includegraphics[width=0.45\textwidth]{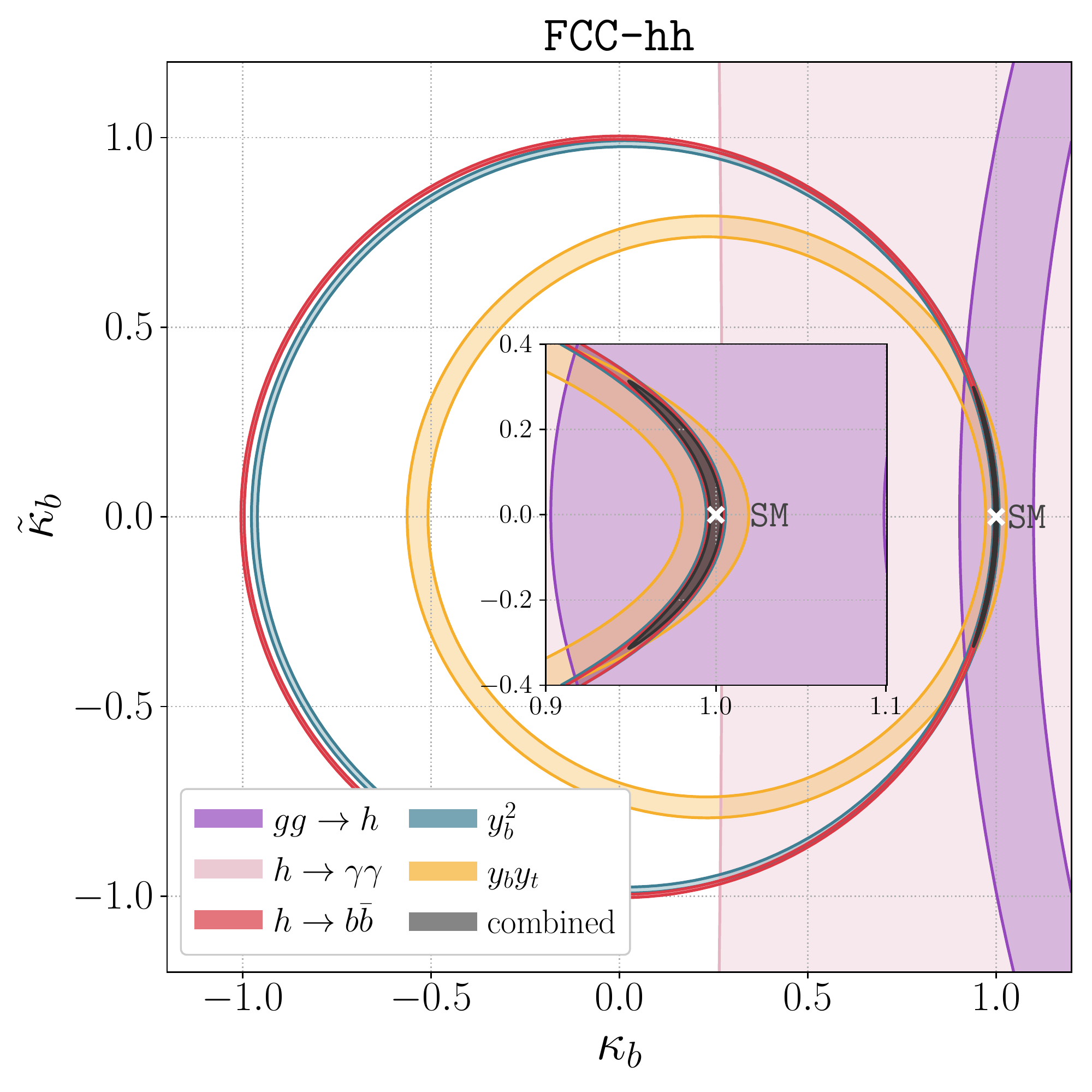}\\
    \includegraphics[width=0.45\textwidth]{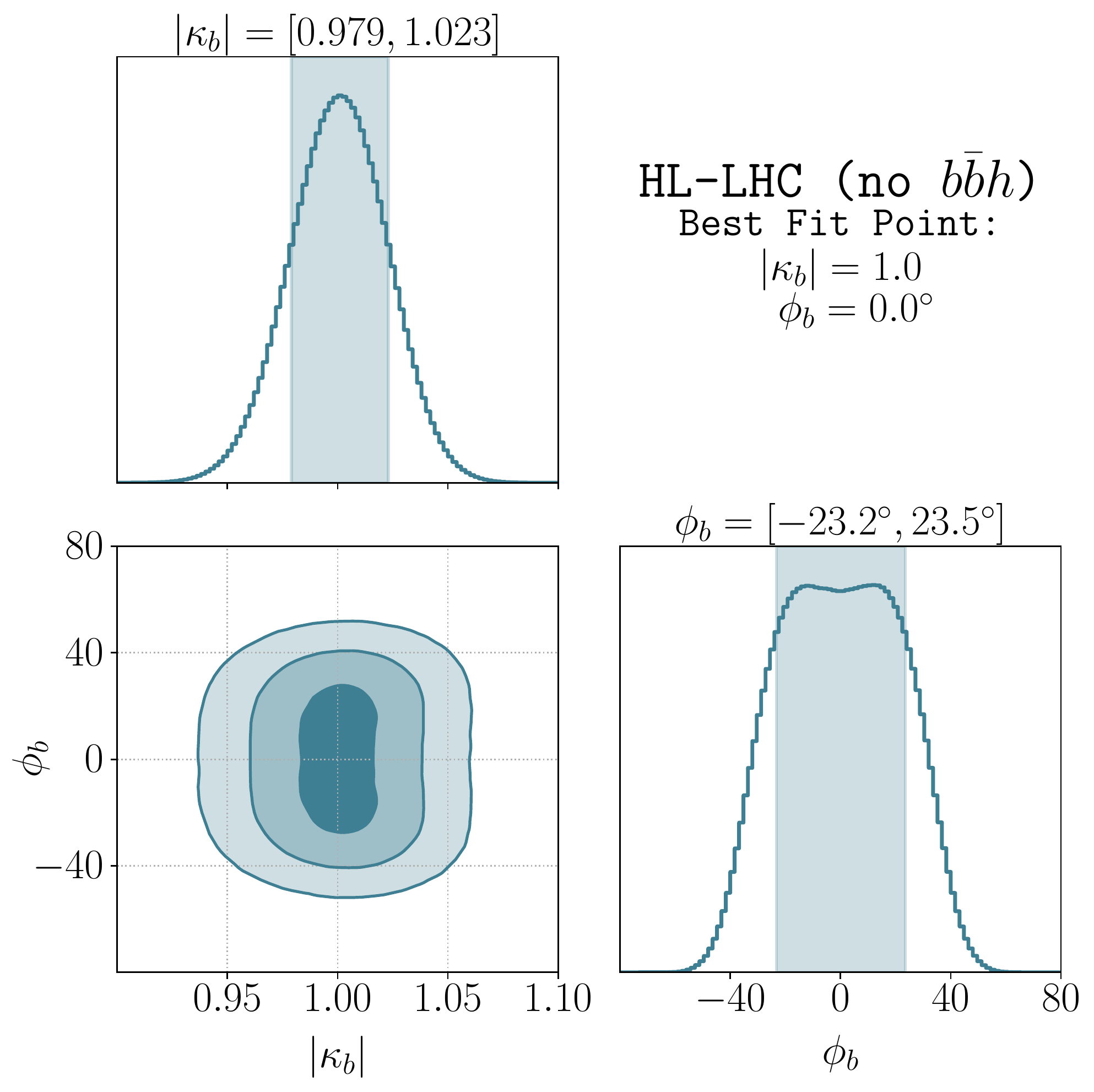}
    \includegraphics[width=0.45\textwidth]{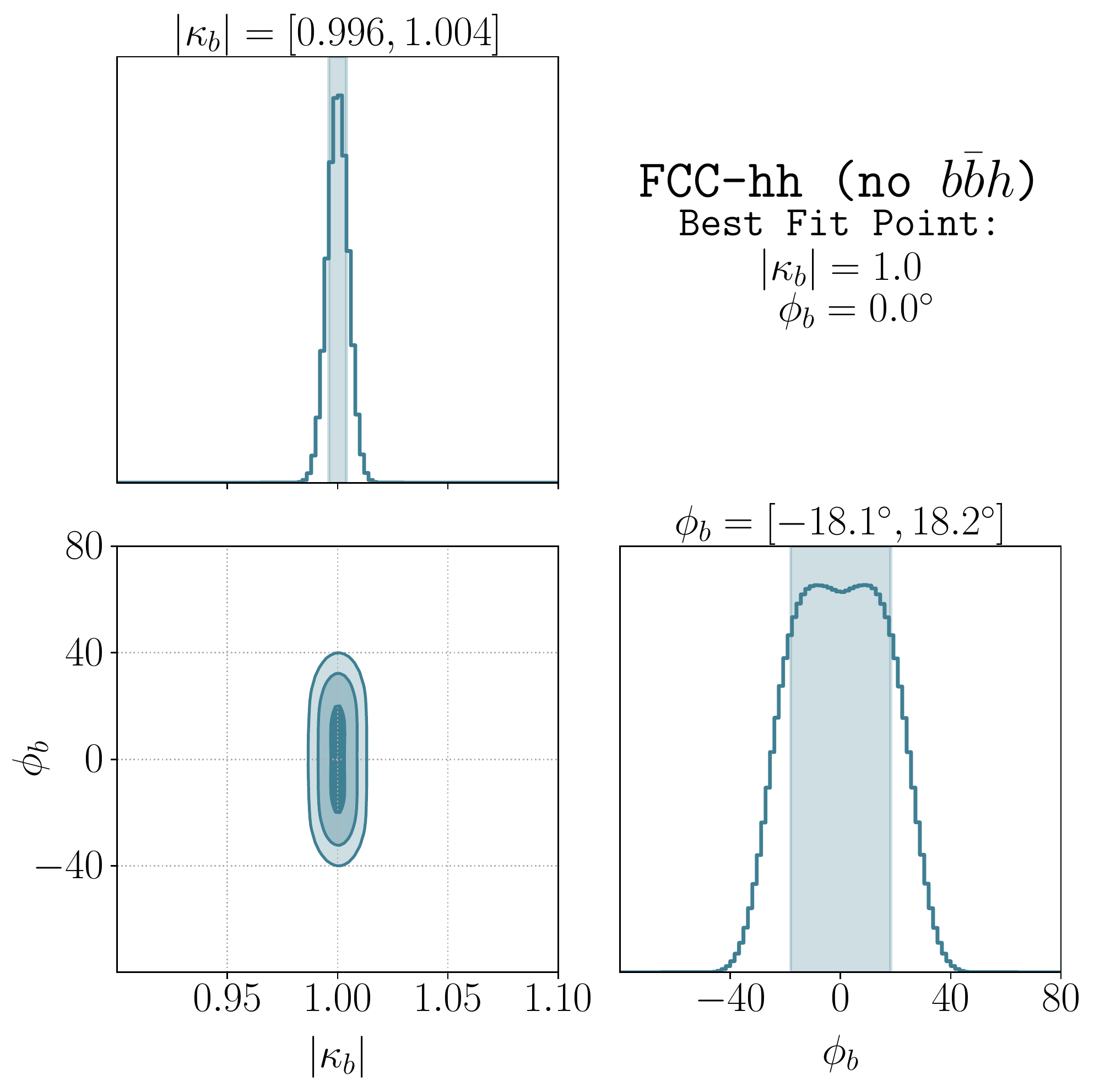}
    \includegraphics[width=0.45\textwidth]{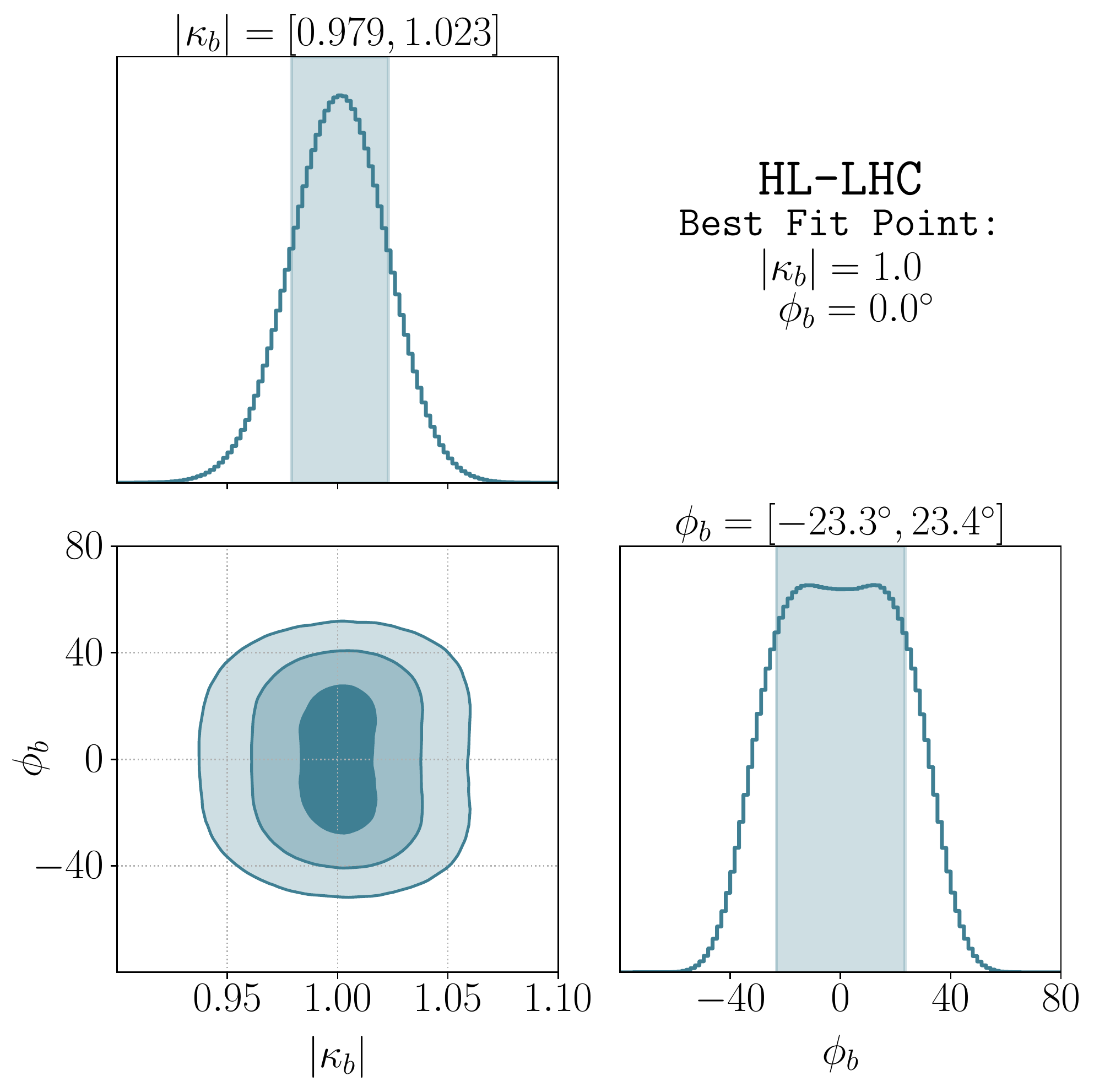}
    \includegraphics[width=0.45\textwidth]{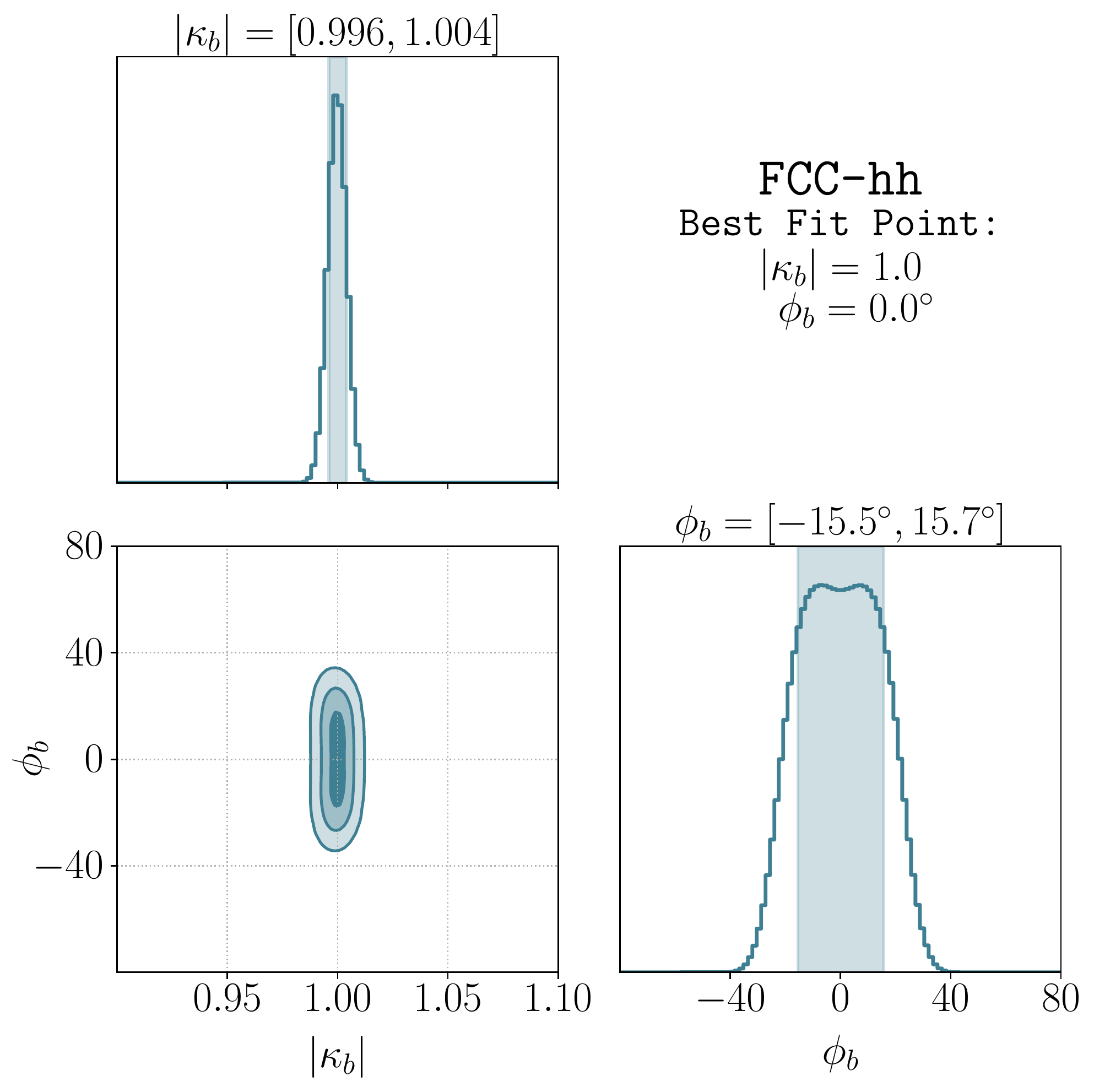}
    \caption{\it \underline{Upper panels:} $1\sigma$ sensitivity contour in the complex $y_b$ space while fixing all other parameters to SM values for HL-LHC (ATLAS+CMS, 6 ab$^{-1}$) and FCC-hh (30 ab$^{-1}$). The blue and yellow $\bbh$ constraints are from interpretations of the $\sigma_{y_b^2}$ and $\sigma_{y_by_t}$ respectively. \underline{Middle and Lower panels:} Bayesian MCMC fits of $|\kappa_b|$  and $\phi_b$ for HL-LHC (ATLAS+CMS, 6 ab$^{-1}$) and FCC-hh (30 ab$^{-1}$) showing 1D posterior distributions and 2D correlation plots without (middle) and with (lower) $\bbh$ measurements.}
    \label{fig:resbcp}
\end{figure}

Collecting all these numbers we perform a fit to all the constraints from HL-LHC and FCC-hh. In the top two panels of \autoref{fig:resbcp} we show the constraints from measurements at HL-LHC and FCC-hh of all the channels we just discussed. The weakest constraint, understandably, comes from $h\to \gamma\gamma$ (pink region) since the fermionic loops generate subdominant contributions with the bulk of the contribution coming from the top loop. However, due to a constant shift in the amplitude dependent on $y_b$, the $h\to\gamma\gamma$ channel can break the degeneracy in the sign of $\kappa_b$ at FCC-hh. The same argument holds for $gg\to h$ production since the bottom-loop contribution is sub-leading and the constraint on $\kappa_b$ is shifted due to the larger constant term coming from the top-loop contribution. However, the constraints are much stronger for $gg\to h$ both at the HL-LHC and the FCC-hh (purple region) than for $h\to \gamma\gamma$. The most stringent constraint comes from the measurement of the branching fraction $h\to b\bar b$ (red region). Being proportional to $|\kappa_{b}|^2 + |\tilde\kappa_{b}|^2$, it does not allow for the determination of the CP-violating phase $\phi_b$. The constraints from the contributions proportional to $y_b^2$ and $y_by_t$ are shown in blue and yellow respectively. They are not competitive with the constraint from $h\to b\bar b$ at  HL-LHC while being quite comparable at FCC-hh. Since there is a linear dependence of $y_b$ in the interference term, these two constraints are shifted from being centered at zero.

In the plots in the middle of \autoref{fig:resbcp} we show the constraints on the $|\kappa_b|$ and $\phi_b$ parameter space from all measurements other than $b\bar b h$ associated production. These fits are done using a Bayesian MCMC implemented in \texttt{PyMC3}~\cite{Salvatier2016}. The 1$\sigma$ high-density intervals are given for each parameter in the corner plots. 

{At the HL-LHC, $h\to b \bar{b}$ and $gg\to h$ completely determine $|\kab|$ and $\phi_b$ as evident from the three left panels of \autoref{fig:resbcp}. The contribution from $\bbh$ does not leave a mark. The picture is quite different for FCC-hh. Taking a careful look at the inset in the top right panel of \autoref{fig:resbcp}, one can notice that the phase, $\phi_b$ is constrained by an interplay between $h \to b\bar{b}$, and the bounds from the $y_by_t$- and $y^2_b$-driven contributions of $\bbh$. Indeed, 
\unskip\parfillskip 0pt \par}

\FloatBarrier

\noindent it is the misalignment of the interference term that improves the bounds on the phase by about 15\% over what is possible without $\bbh$ measurements as can be seen from comparing the middle-right and the bottom-right corner plots. This clearly shows that a more comprehensive study of the $\bbh$ channel is necessary for reducing the theoretical errors and estimating the systematics for future colliders.

\subsection{Bounds from EDM}
\label{sec:EDM}

We feel that a discussion on the current bounds from EDM on $\tilde{\kappa}_b$ is necessary since they are quite constraining. Recent update on electron EDM measurement from ACME~\cite{Andreev:2018ayy} gives stringent bound on $T$-violating (or CP-violation assuming CPT symmetry) effects in the system. Hadronic EDMs such as those from neutron or mercury EDM measurements give complementary information on hadronic CP-violating effect and do not rely on the existence of leptonic Yukawa couplings. However, they usually suffer from sizable theory uncertainties. Recent discussions and review on EDM measurements and theory constraints can be found in Refs.~\cite{Yamanaka:2017mef,Safronova:2017xyt,Chupp:2017rkp}. Future proposals to measure both leptonic and hadrnoic EDM~\cite{Vutha:2017pej,Cairncross:2017fip,Vutha:2018tsz,Hutzler:2020lmj,Strategy:2019vxc} aim to improve these bounds by one or more orders of magnitude at these small scale experiments in the near future. When focusing on the CP-violating terms in the Yukawa sector, EDM measurements are interpreted as bounds on the CP-violating phase in the Yukawa couplings~\cite{Egana-Ugrinovic:2018fpy,Brod:2018pli,Brod:2018lbf}.

When not relying on assumption of a SM electron Yukawa, Neutron EDM currently gives the strongest bound on the CP-violating coupling, $\tilde{\kappa}_b\lesssim 5$. Assuming a SM electron Yukawa coupling, the most stringent bound on $\tilde{\kappa}_b$ comes from electron EDM with $\tilde{\kappa}_b\lesssim 0.5$~\cite{Brod:2018pli}. The constraints on $\tilde{\kappa}_b$ from the neutron (or other hadronic) EDM can also be diluted or evaded, if more than one Yukawa CP-phases from different quark flavors are present and cancel among themselves. In this case, studies of CP-violating observable or interference terms in the $\tth$, $\bbh$ or $h\to\tau\tau$ processes at colliders would become valuable and provide complimentary information to pin down the individual Yukawa couplings and their CP-phases.

\section{Summary}
\label{sec:Sum}

Gleaning tiny and obfuscated signals from dominating backgrounds has been the boon and the bane of particle physics, in both the theoretical and experimental realms. The evolution of methods used to do so has gone through several stages with multivariate analysis being the most evolved. The strength of multivariate analyses in the form of machine learning methods like decision trees and neural networks have dominated experimental analyses for quite a few years now. In this work we augment phenomenological analyses by bringing forth to it the strength of interpretable machine learning.

The extraction of $y^2_b$ from the associated production of the Higgs with a $b\bar b$ pair has been written off due to the presence of dominant irreducible backgrounds, especially from $Zh$ production~\cite{Pagani:2020rsg}. The arguments were based on the fact that $Zh, Z\to b\bar b$ and the $y^2_b$-driven contribution have very similar kinematic signatures making it difficult to distinguish the weaker contribution proportional to $y_b^2$, from the irreducible background. In this work we show that, the dismissal of the measurability of the $y_b^2$-driven contribution in an attempt to isolate $y_b$ is premature. The use of multivariate analysis in the form of BDT can still allow for the separation of the hidden signal.

We go a step further and add interpretability to the machine learning method by appealing to a measure derived from game theory called the Shapley values. This facilitates us in working with high-level kinematic variables, instead of momenta four-vectors, and gives us insight into the hierarchy of importance of these kinematic variables\footnote{Without the use of Shapley values as a variable importance measure it would not be possible to choose the minimal set of kinematic variable required to separate the signal from the background and we would be left with a very large set of kinematic variables to deal with that would drastically reduce the interpretability of the procedure}. This, in turn, allows us to narrow down and focus on those kinematic variables that are the most important, providing a way of understanding the physics underlying the possibilities of separating the $y_b^2$-driven contribution from the backgrounds.

In summary, this work provides the following innovations and insights:
\begin{itemize}
    \item An irreducible background like $Zh$ production can be tamed with multivariate methods that can be better understood with interpretable machine learning tools. Kinematic shapes are the key to distinguishing small signals that cut-based analyses miserably fail at.
    
    \item Machine learning algorithms do not need to be black-boxes~\cite{Rudin2019}. While it is difficult to avoid this for several machine learning algorithms, much progress is being made to make machine learning interpretable. In this work we show that decision trees can be made interpretable and certain nuances of the distributions that they probe can be understood with metrics such as the Shapley values.
    
    \item At the FCC-hh, the associated production of $\bbh$ stands to gain as its production rate grows much faster than the $Zh$ production rate and the dominant $\bbaa$ background with rising energies. This can be exploited to get a good measurement of the magnitude and sign of $y_b$ from $\bbh$ production which can be comparable to that from a $h\to b\bar b$ measurements.
    
    \item Since constraint from $h\to b\bar b$ is very stringent on the magnitude of $\kappa_b$ and $gg\to h$ can fix the phase, $\phi_b$, in combination with $h\to b\bar b$, probing $\bbh$ does not add to these constraints significantly at HL-LHC (\autoref{fig:resbcp}). While at the FCC-hh, $\bbh$ production does not add much to the constraint on $|\kappa_b|$, the constraint on $\phi_b$ can gain by about 15\% from $\bbh$ measurements at FCC-hh (\autoref{fig:resbcp}) through the interplay between the $y_b^2$- and $y_by_t$-driven contributions along with $h\to b\bar{b}$, resulting in the $\bbh$ channel proving to be more significant than the $gg\to h$ in constraining $\phi_b$.
\end{itemize}

The emergence of machine learning as an effective tool-set to address regression and classification problems has led to a wave of applications in particle physics often at the expense of overshadowing the importance of the physics case. However, our objective lies in showing how interpretable machine learning can augment phenomenological analyses. The primary message of this work lies in the display of a method that can be instrumental in future phenomenological analyses focused on using kinematic shapes as the primary tool to probe for tiny signatures. In the precision era, this will be an important component that provides maximal statistical advantage along with a clear understanding of the underlying dynamics with interpretability. We clearly show this by resurrecting the possibility of measurement of $y_b$ from $\bbh$ production after tackling rather complex irreducible backgrounds. The robustness of our analysis brings the hope of extension of the methods developed in this work to extract  difficult signatures in other phenomenological analyses.

\acknowledgments

We would like to thank Marius Wiesemann and Marco Zaro for clarifications on their work in Ref.~\cite{Deutschmann:2018avk}. Our special thanks to Davide Pagani for clarifying quite a few key points related to Ref.~\cite{Pagani:2020rsg}. We would also like to thank Laura Reina for discussion during various stages of preparing the draft and Hyunju Kim for discussions on Interpretable Machine Learning. This work benefited from support by the Deutsche Forschungsgemeinschaft under Germany's Excellence Strategy  EXC 2121 ``Quantum Universe" -- 390833306. This research was supported in part through the Maxwell computational resources operated at DESY, Hamburg, Germany.

\appendix

\section{Discriminating between the terms proportional to \texorpdfstring{$y_b^2$}{Y-b2} and \texorpdfstring{$y^2_t$}{y-t2}}
\label{app:ybyt}

\begin{figure}
    \centering
    \figuretitle{\large\texttt{HL-LHC}}
    \hspace{0.1cm}\includegraphics[width=5cm, height=5.1cm]{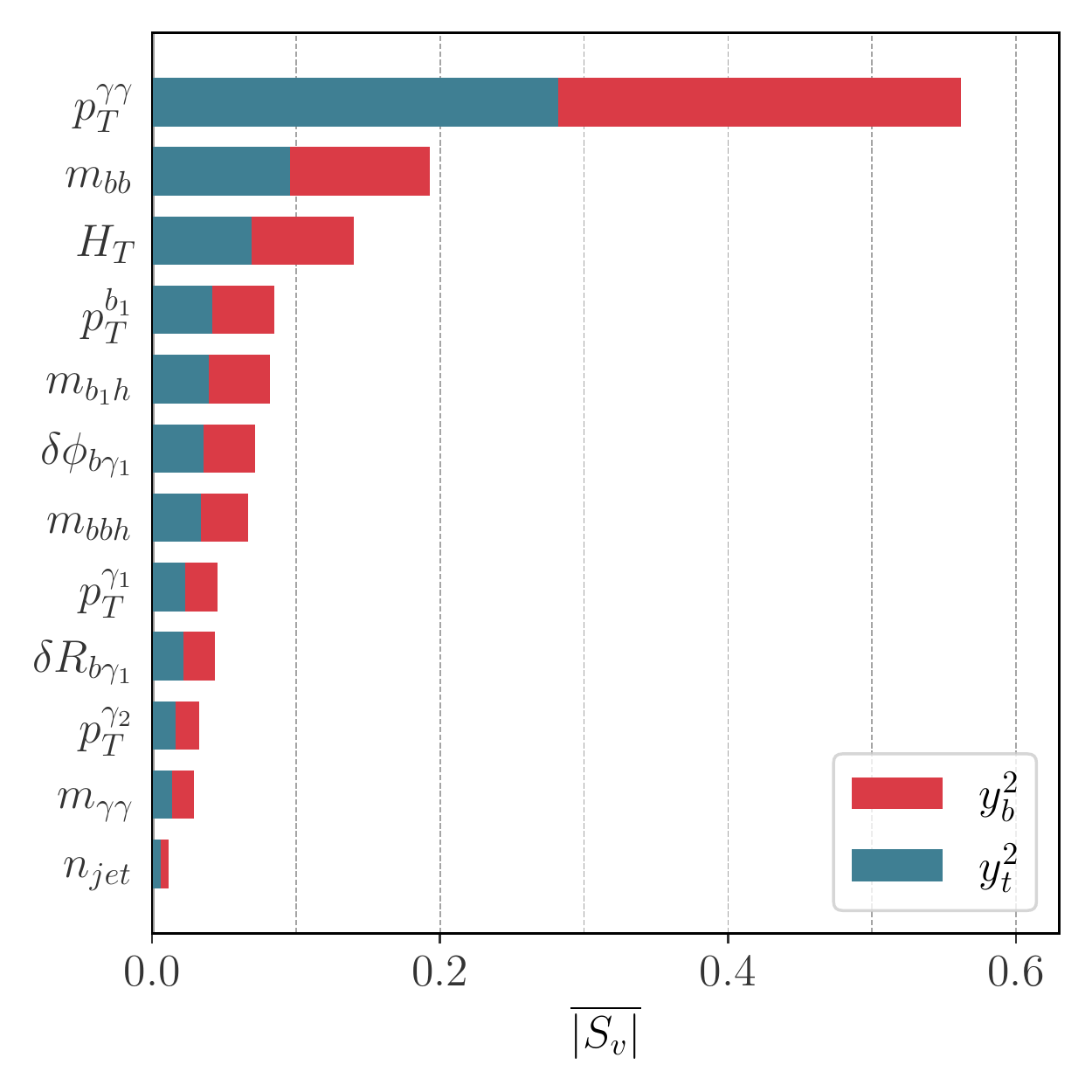}
    \includegraphics[trim=10 0 10 10,clip,width=0.32\textwidth]{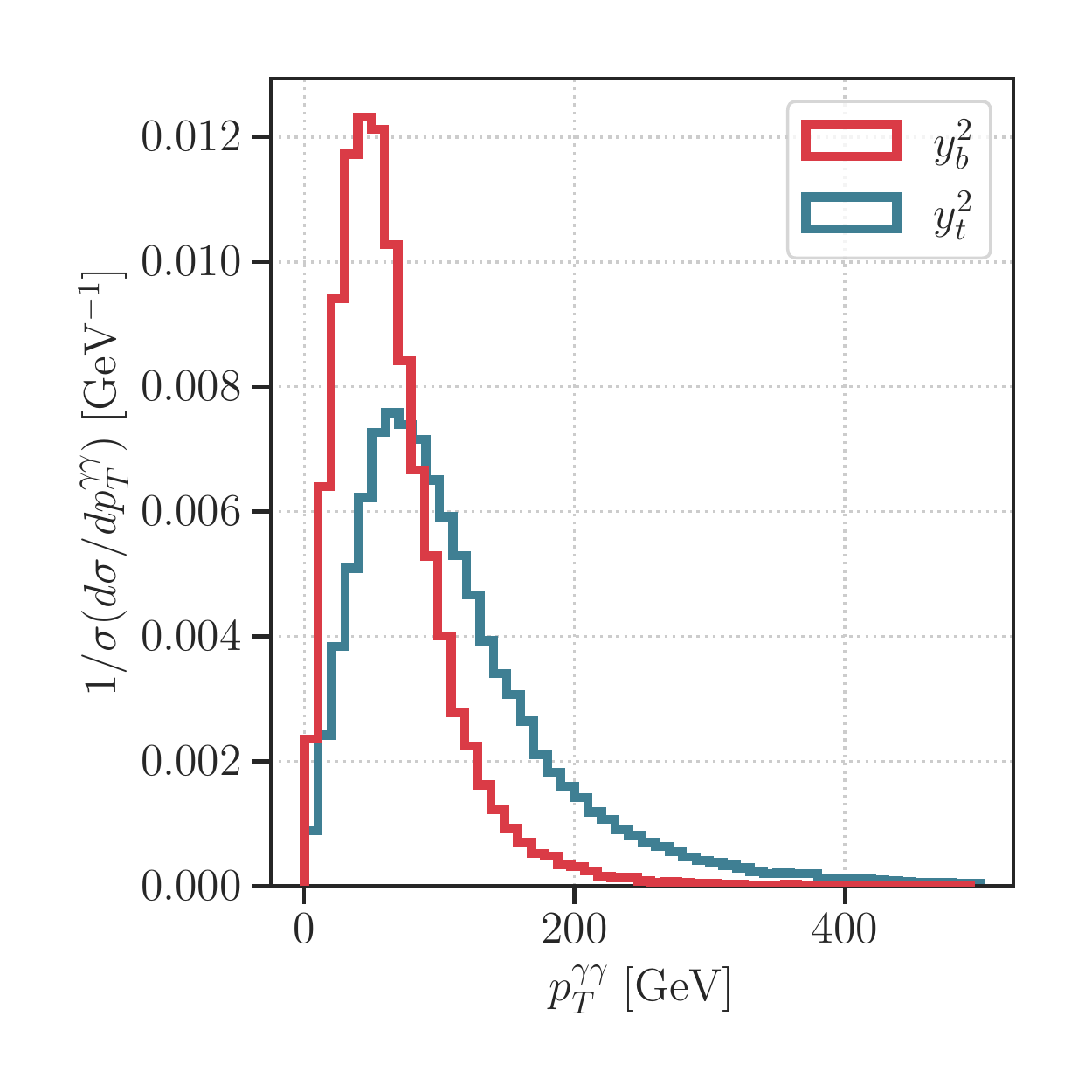}
    \includegraphics[trim=10 0 10 10,clip,width=0.32\textwidth]{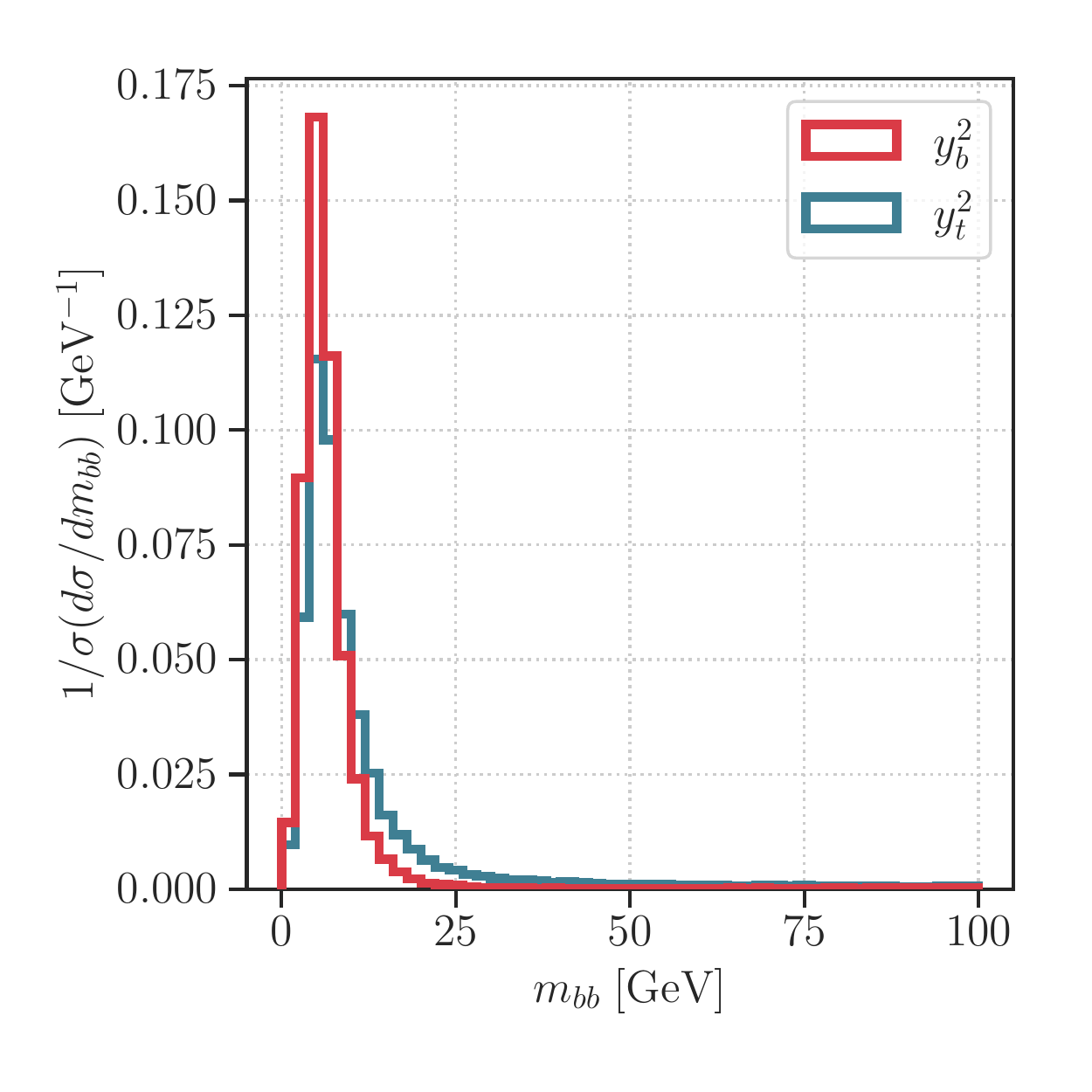}\\
    \includegraphics[trim=10 0 10 10,clip,width=0.32\textwidth]{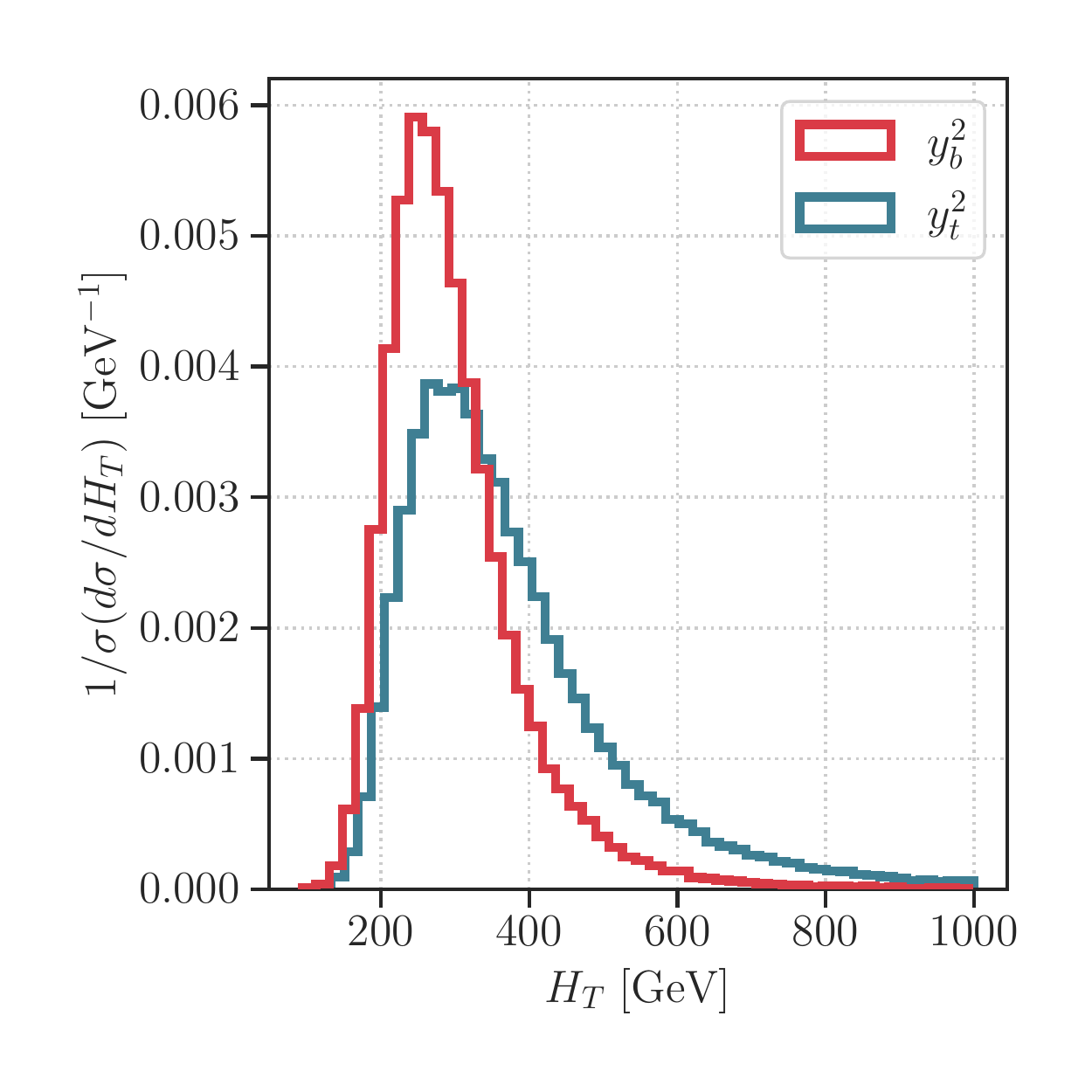}
    \includegraphics[trim=10 0 10 10,clip,width=0.32\textwidth]{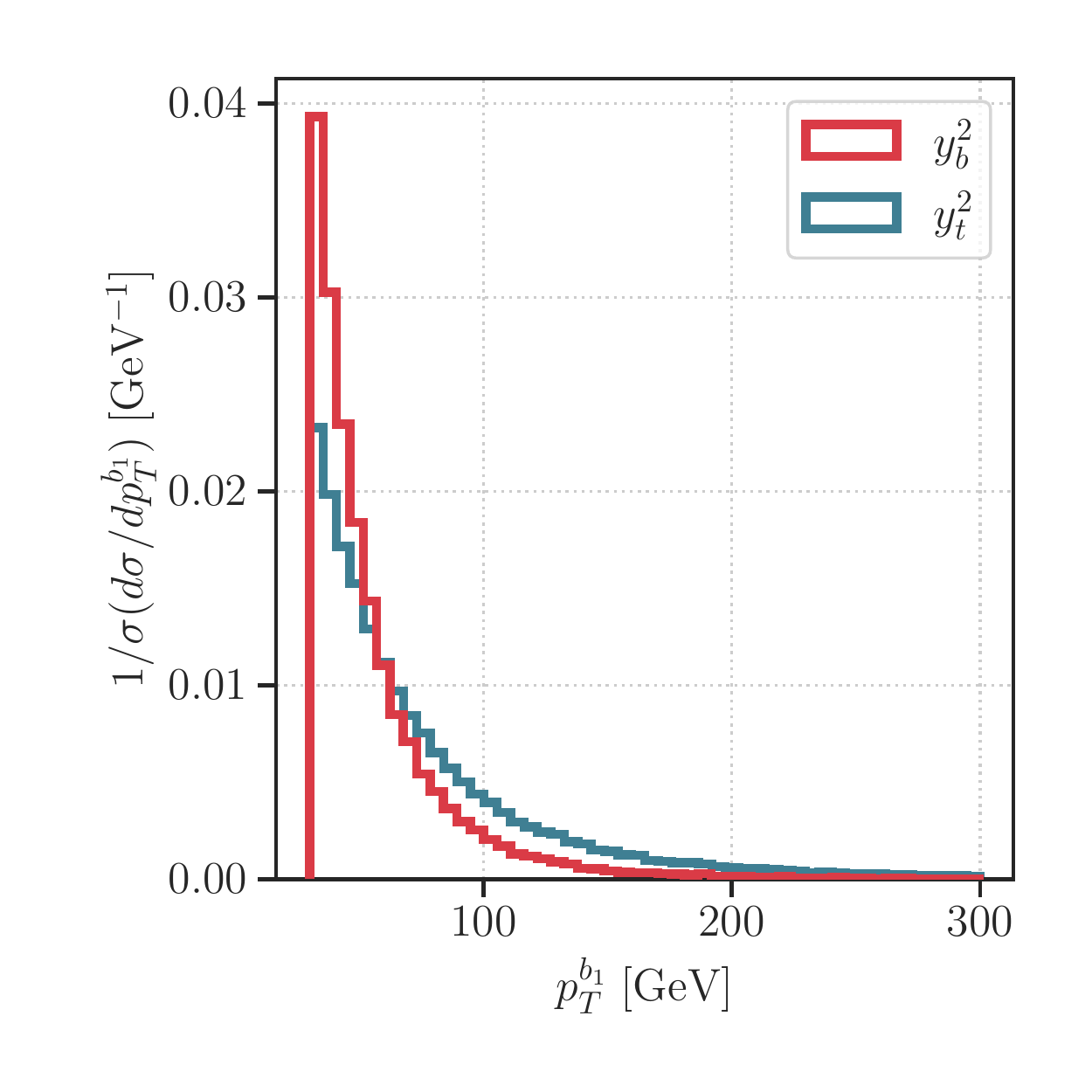}
    \includegraphics[trim=10 0 10 10,clip,width=0.32\textwidth]{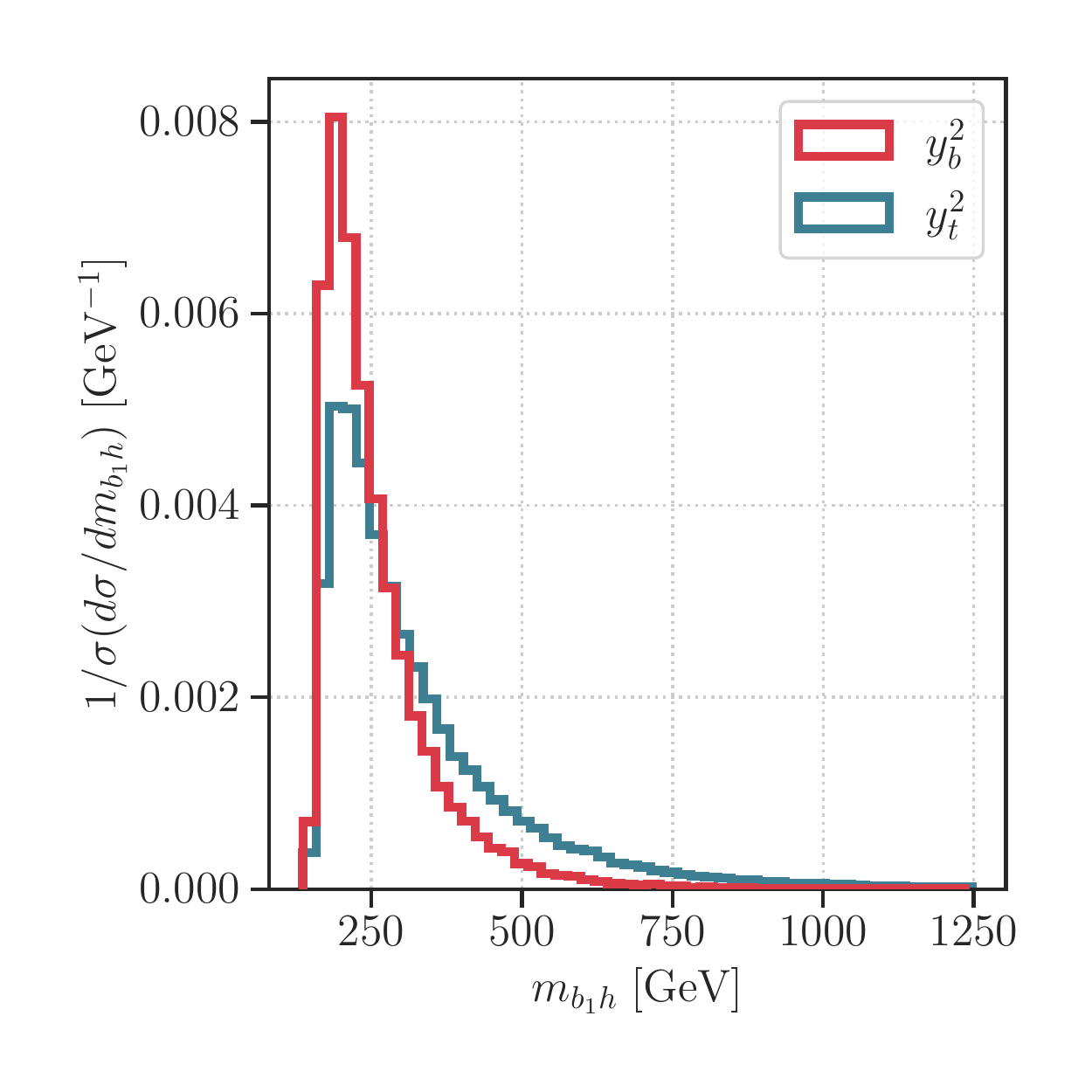}
    \caption{\it The Shapley values and differential distributions for the kinematic variables used to discriminate between the $y_b^2$ and $y_t^2$ contributions at HL-LHC, all \textbf{normalized to unity} for a clear comparison of shape. A SM signal is injected. \underline{Upper left corner:} The hierarchy of variable importance with increasing Shapley value denoting higher importance in the discrimination. The normalized distribution are those of the 5 most important kinematic variables as determined by their Shapley values.}
    \label{fig:yt2-yb2}
\end{figure}

Applying the methods that we used to show that the $y^2_b$-driven signal can be separated from the irreducible $Zh$ background in \autoref{sec:BDTybZh}, it can also be shown that $y^2_b$-driven signal can be separated from the irreducible $y_t^2$-driven background channel. However, the separation is not as clear as the separation between the $y^2_b$-driven signal and $Zh$. The top left plot in \autoref{fig:yt2-yb2} shows the hierarchy of importance of the kinematic variables for this classification, with $p_T^{\gamma\gamma}$ being, by far, the most important, and mostly corresponding to the $p_T$ of the Higgs. 

The $y_t^2$-driven channel tends to have a harder spectra in $p_T$, since majority of the events come from a Higgs recoiling against a gluon emission which further splits to the bottom pair, as can be seen in the representative Feynman diagrams \autoref{fig:feyndiag} and the kinematic distribution \autoref{fig:yt2-yb2}. The $p_T^{\gamma\gamma}$ distribution is shown in the top middle plot. The $y_t^2$-driven channel tends to have a harder $H_T$ spectrum too.

\begin{figure}
    \centering
    \includegraphics[width=0.495\textwidth]{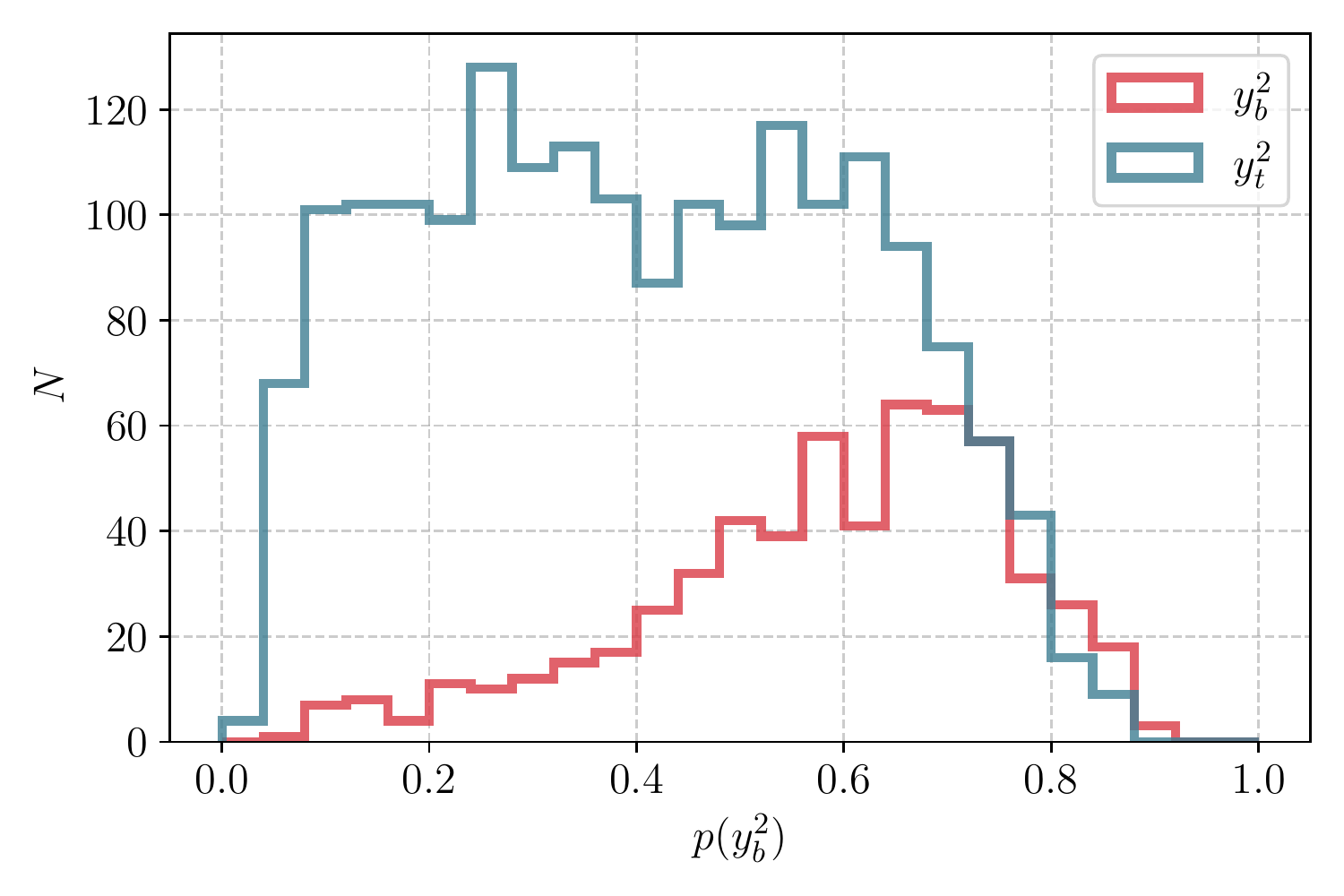}
    \caption{\it BDT discrimination for separating the contributions proportional to $y^2_b$ and $y^2_t$ at the HL-LHC with 6 ab$^{-1}$ of data. A SM signal is injected.}
    \label{fig:yb2-yt2-BDT}
\end{figure}

\begin{figure}
    \centering
    \figuretitle{\large\texttt{HL-LHC}}
    \vspace{0.1in}
    \includegraphics[trim= 10 10 10 20, clip, width=0.4\textwidth]{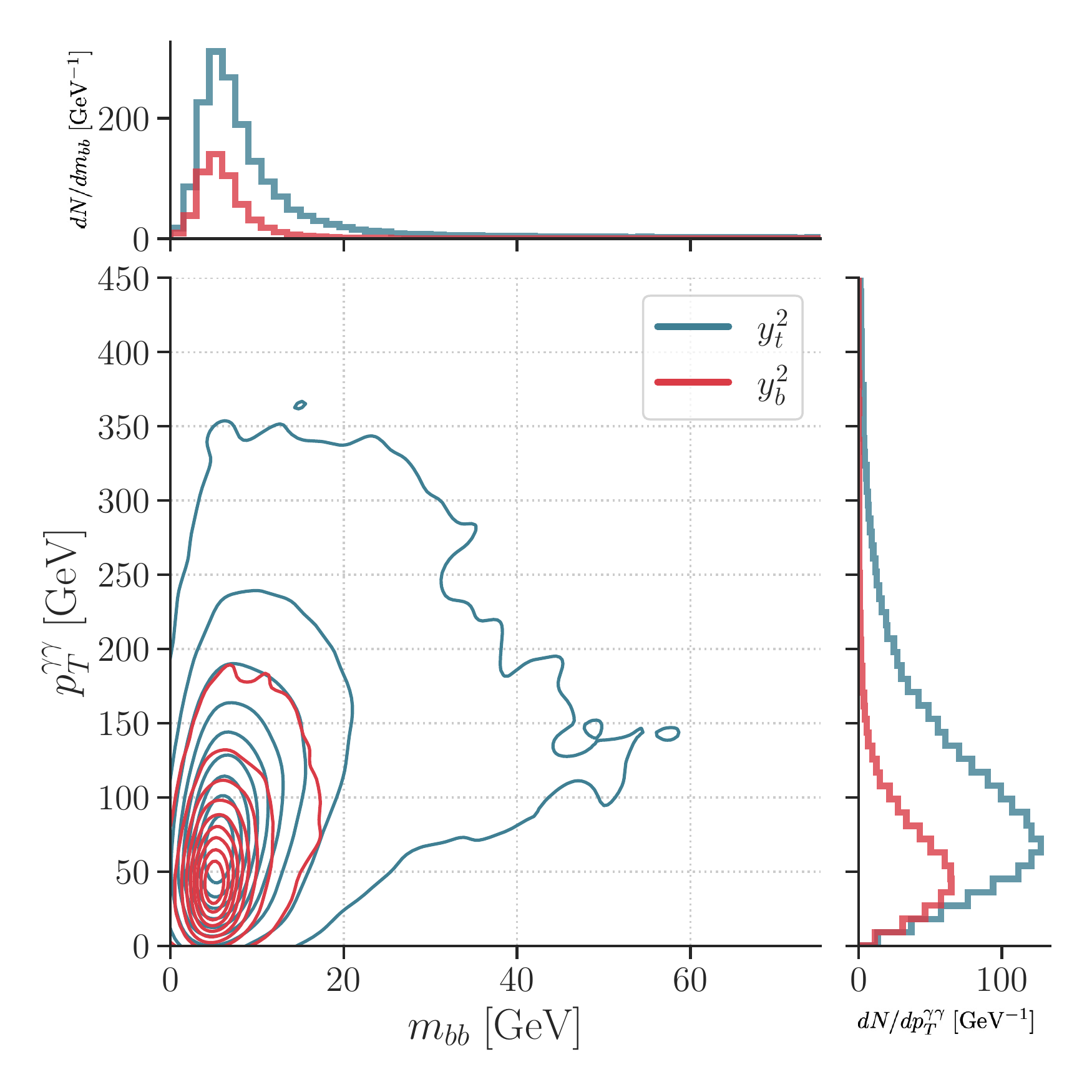}
    \includegraphics[trim= 10 10 10 20, clip, width=0.4\textwidth]{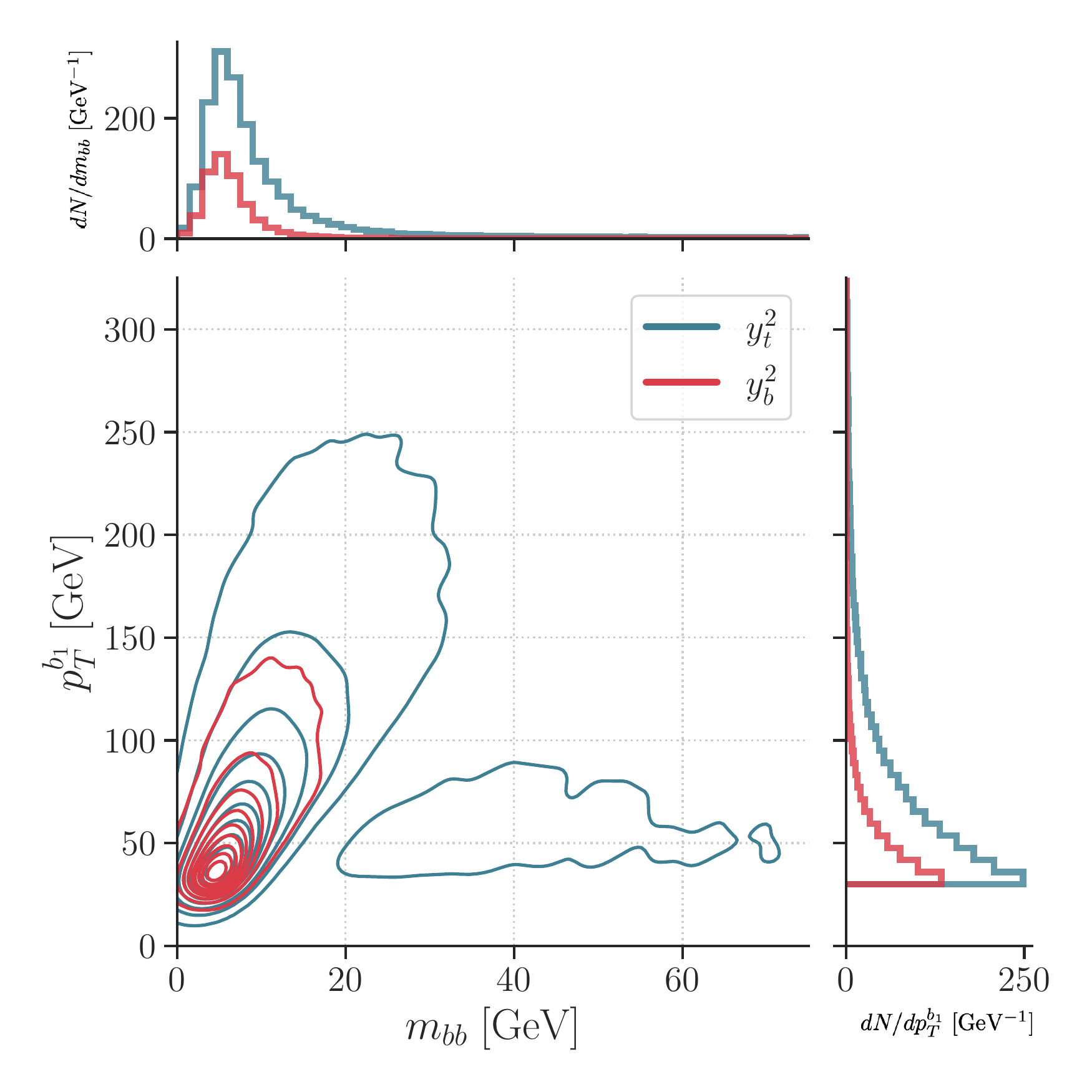}\\
    \includegraphics[trim= 10 10 10 20, clip, width=0.4\textwidth]{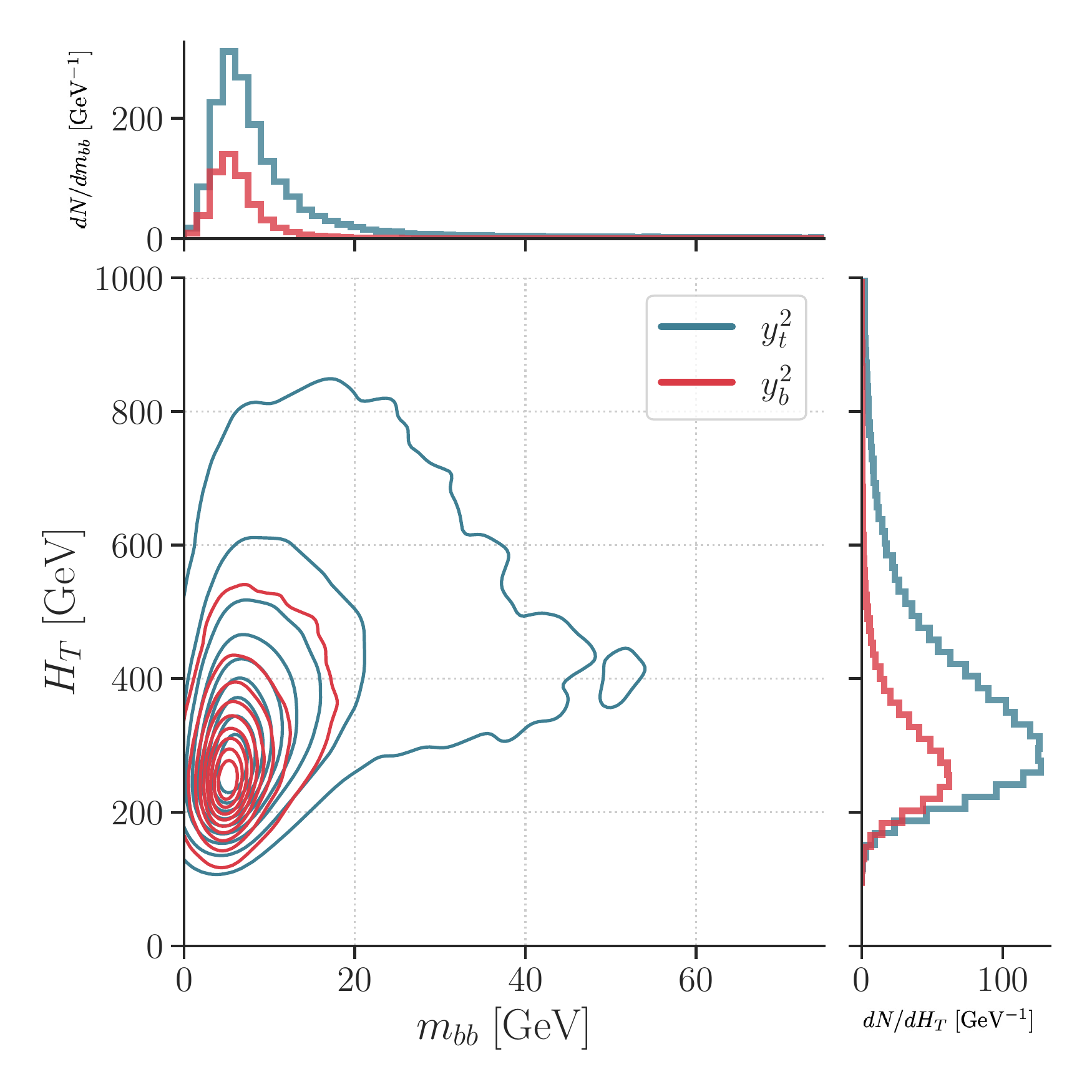}
    \includegraphics[trim= 10 10 10 20, clip, width=0.4\textwidth]{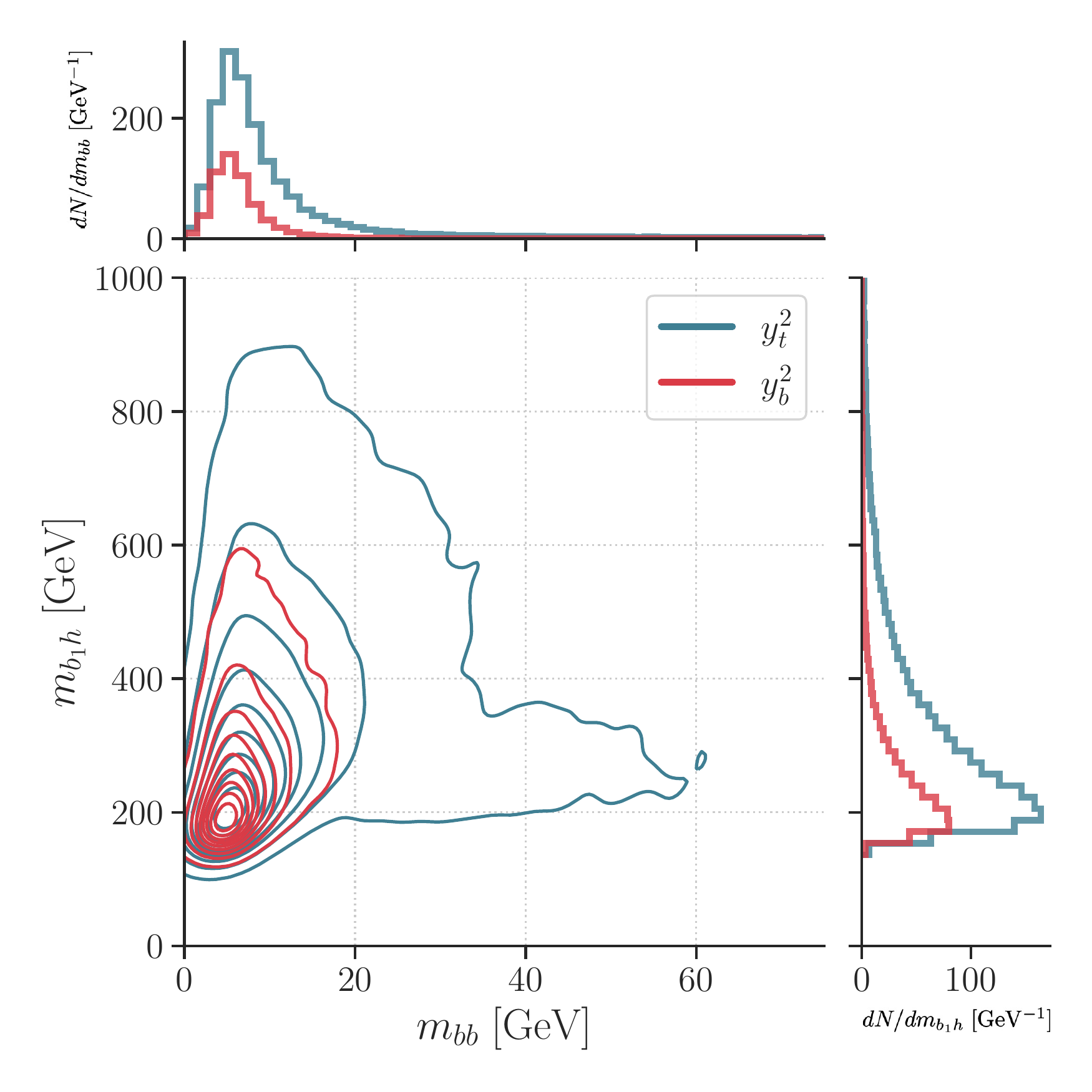}
    \caption{\it Joint \textbf{non-normalized} 1D and 2D plots, with SM signal, representing how kinematic shapes allow the BDT to discriminate between the $y_b^2$ and $y_t^2$ contributions at HL-LHC. While 1D distributions show that $m_{bbh}$ by itself cannot be used to discriminate between these two contributions, its correlation with other kinematic variables creates shapes that a BDT can use to discriminate between the two contributions.}
    \label{fig:shape-yt2-yb2}
\end{figure}

The plot in \autoref{fig:yb2-yt2-BDT} shows what classification would be possible in a pseudo-experiment with events from the $y^2_b$- and $y^2_t$-driven contributions only. This provides added confidence in the possibility of separating $y^2_b$-driven contributions in the full analysis. In \autoref{fig:shape-yt2-yb2} we show the joint distributions of the $y_b^2$- and $y_t^2$-driven channels. As can be seen, the shapes of the two channels are quite similar with the $y_b^2$-driven channel being overwhelmed by the $y_t^2$-driven channel. Suffice it to say, that this further highlights the ability of the BDT to separate two channels with very similar kinematic distributions and the added insight provided by the Shapley values in understanding why that is possible.


\section{Effective \texorpdfstring{$\kappa_g$}{kappag} and \texorpdfstring{$\kappa_\gamma$}{kappagamma}}
\label{app:kgkgamma}

Following the procedure in Ref. \cite{Brod:2013cka}, it is simpler to work out the constraints on $\kappa_b$ in terms of constraints on the effective $ggh$ and $\gamma\gamma h$ couplings which are defined as, 
\begin{equation}
    \mathcal{L}_{\rm eff}\supset c_g \frac{\alpha_s}{12\pi}\frac{h}{v}G_{\mu\nu}^a G^{\mu\nu,a} + \tilde c_g \frac{\alpha_s}{8\pi}\frac{h}{v}G_{\mu\nu}^a \tilde G^{\mu\nu,a},
\end{equation}
\begin{equation}
    \mathcal{L}_{\rm eff}\supset c_\gamma \frac{\alpha}{\pi}\frac{h}{v}F_{\mu\nu} F^{\mu\nu} +  \tilde c_\gamma \frac{3\alpha}{2\pi}\frac{h}{v}F_{\mu\nu} F^{\mu\nu}.
\end{equation}
The CP-even and CP-odd couplings are,
\begin{equation}
    c_g = \sum_{f=t,b} \kappa_f A(\tau_f),~~~\tilde c_g = \sum_{f=t,b} \tilde\kappa_f B(\tau_f),
\end{equation}
with $\tau_f = 4m_f^2/m_h^2$ and
\begin{equation}
    A(\tau) = \frac{3\tau}{2}\left[ 1+ (1-\tau)\arctan^2\frac{1}{\sqrt{\tau -1}} \right],~~ B(\tau) = \tau\arctan^2\frac{1}{\sqrt{\tau -1}}.
\end{equation}
Hence, $\kappa_g$ and $\kappa_\gamma$, the rescaling of the effective $ggh$ and $\gamma\gamma h$ couplings, can be written as functions of $\kappa_b$ and $\tilde\kappa_b$ as,
\begin{equation}
    \kappa_g = \frac{c_g}{c_{g}{\rm SM}} =  \frac{A(\tau_t) + \kappa_b A(\tau_b)}
    { A(\tau_t)+ A(\tau_b)}
\end{equation}
\begin{equation}
    \tilde\kappa_g = \frac{3}{2}\frac{\tilde c_g}{c_{g}^{\rm SM}} = \frac{3}{2}\frac{\tilde\kappa_b B(\tau_b)}{ A(\tau_t)+ A(\tau_b)}
\end{equation}
such that the modified inclusive gluon fusion rate has $\mu_g = |\kappa_g|^2 + |\tilde\kappa_g|^2$.
After plugging in the masses as defined in Section~\ref{sec:Sim}, these gives the numerical relations,
\begin{equation}
    \kappa_g \sim (-0.05 +0.8 i )\kappa_b + (1.05 - 0.08 i )
    \label{eq:kgkb}
\end{equation}
\begin{equation}
    \tilde\kappa_g \sim (-0.06 +0.08 i )\tilde\kappa_b 
\end{equation}
Similarly, the $h\gamma\gamma$ effective couplings can be expressed in terms of $\kappa_b$. In addition to the linear dependence on the Yukawa coupling, there is also a large constant term in the CP-even coupling from the leading $W$ and sub-leading top contributions at one loop in the SM:
\begin{equation}
    \kappa_\gamma = (0.996 + 0.005i) +  (0.004 - 0.005i)\kappa_b 
\end{equation}
\begin{equation}
    \tilde\kappa_\gamma = (0.004 - 0.005i)\tilde\kappa_b 
\end{equation}
We have used these expressions in \autoref{sec:YukConst} to set bounds on real and complex $\kappa_b$.

\section{Deep neural network analysis}
\label{app:dnn}

To construct the DNN~\cite{6472238,SCHMIDHUBER201585} we used TensorFlow~\cite{tensorflow2015-whitepaper}. The following are the details of the fully-connected DNN architecture:

\begin{itemize}
\itemsep0em
    \item depth: 12 layers
    \item width: 64 nodes in each layer
    \item optimizer: Adam 
    \item loss function: Sparse categorical crossentropy
    \item regularization: early stopping
    \item validation split: 0.2
    \item output layer: 5 nodes with softmax activation 
\end{itemize}

We used an early callback algorithm to stop the training of the network when the validation loss stopped decreasing and chose the epoch where the validation loss was the least. The loss function that we used was a simple mean square error function which measures the $L_2$ distance of the fit from the data. The training time was equivalent to that required for training a BDT to similar accuracies with the early stopping coming into effect at about 500 epochs. The trained DNN was used for constructing the confusion matrix shown in \autoref{tab:DNN_confusion}.

\begin{table}[t!]
    \centering
    \begin{tabular}{ll|rrrrr|r}
    \multirow{7}{*}{\rb{\bf Actual no. of events\hspace{0.6cm}}} & \multicolumn{7}{c}{\bf Predicted no. of events at HL-LHC}\\
    \cmidrule[\heavyrulewidth]{2-8}
    &Channel &      $y_b^2$ &     $y_by_t$ &      $y_t^2$ &         $Zh$ &  $bb\gamma\gamma$ &     total \\
    \cline{2-8}
    &$y_b^2$          &   167 &    770 &    320 &   1100 &        1970 &     583 \\
    &$y_by_t$         &    -4 &   -310 &    -30 &   -170 &        -400 &     -95 \\
    &$y_t^2$          &   268 &   1640 &   3810 &   5280 &        5120 &    1853 \\
    &$Zh$             &    24 &    380 &    160 &   4080 &        1640 &     650 \\
    &$bb\gamma\gamma$ &  2309 &  29660 &  11220 &  75940 &     1017800 &  115771 \\
    \cline{2-8}
    &$\mathcal{Z}_j$  &  3.18 &   0.55 &    9.68&   4.39 &          317&         \\
    \cmidrule[\heavyrulewidth]{2-8}
    \end{tabular}
    \caption{\it Trained DNN classification (confusion matrix) of the five channel contributions at HL-LHC with 6 ab$^{-1}$ luminosity (ATLAS+CMS), assuming SM signal injection. The right-most column gives the total number of events expected in each channel in the SM. This can be compared to \autoref{tab:HL-LHC-confusion} which was derived from training a BDT.}
    \label{tab:DNN_confusion}
\end{table}

The architecture described above was required for the five channel classification. To understand how the DNN scales with the number of channels, we tried a signal vs. background discrimination for $y_b^2$ vs. $Zh$ as described in \autoref{sec:BDTybZh}. To achieve accuracies comparable to the BDT implementation a DNN with 2 hidden layers with 16 nodes each was necessary. The rest of the architecture was the same as the larger DNN described above.

\bibliographystyle{JHEP-CONF}
\bibliography{bbh}

\end{document}